\documentclass[pdflatex,bst/sn-mathphys-num]{sn-jnl}

\usepackage{graphicx}%
\usepackage{multirow}%
\usepackage{amsmath,amssymb,amsfonts}%
\usepackage{amsthm}%
\usepackage{mathrsfs}%
\usepackage[title]{appendix}%
\usepackage{xcolor}%
\usepackage{textcomp}%
\usepackage{manyfoot}%
\usepackage{booktabs}%
\usepackage{algorithm}%
\usepackage{algorithmicx}%
\usepackage{algpseudocode}%
\usepackage{listings}%
\usepackage[numbers,sort&compress]{natbib}
\usepackage{fullpage}
\usepackage{graphics,graphicx}
\usepackage{hyperref}
\usepackage{tikz,pgfplots}
\usepackage{todonotes}
\theoremstyle{thmstyleone}%
\theoremstyle{thmstyletwo}%

\theoremstyle{thmstylethree}%

\begin{document}

\title[Article Title]{Models of Wildland Fire and Ember Spread}


\author*[1]{\fnm{Kevin} \sur{Speer}}\email{kspeer@ucsb.edu}

\author[2]{\fnm{Bryan} \sur{Quaife}}\email{bquaife@fsu.edu}

\author[3]{\fnm{Jie} \sur{Sun}}\email{jsun2@fsu.edu}

\affil*[1]{\orgdiv{Earth Research Institute}, \orgname{University of California Santa Barbara}, \orgaddress{\city{Santa Barbara},  \state{California}, \postcode{93106}, \country{USA}}}

\affil[2]{\orgdiv{Department of Scientific Computing}, \orgname{Florida State University}, \orgaddress{\city{Tallahassee},  \state{Florida}, \postcode{32306}, \country{USA}}}

\affil[3]{\orgdiv{Department of Earth, Ocean, and Atmospheric Science}, \orgname{Florida State University}, \orgaddress{\street{Street}, \city{Tallahassee}, \state{Florida}, \postcode{32306}, \country{USA}}}

\abstract{In this Chapter we explore the nature and theory of ember transport using observations from controlled laboratory settings, prescribed fire, and wildland fire. Specific examples and statistics from fires are used to gain insight and motivate a hierarchy of modeling approaches from simple idealized models to full-physics atmospheric boundary layer models. The emphasis is on the fundamental processes involved in moving embers away from their sources in vegetation and structures on and near the ground or in the atmospheric boundary layer winds and turbulent flows. We describe the problem first in terms of basic theory about the rate of spread of wildland fire, examine the physical principles at work in ember transport for two principle modes of transport near the ground and in the atmosphere well above the surface, and connect these to statistical models for transport.}




\maketitle

\newcommand{\pderiv}[2]{\frac{\partial #1}{\partial #2}}
\newcommand{\pderivtwo}[2]{\frac{\partial^2 #1}{\partial #2^2}}
\newcommand{\pig}{p_{\mathrm{ig}}}

\section{Introduction}
Wildland fires are complex, multiscale phenomena whose spread depends on
radiant and convective heat fluxes, combustion chemistry, smoke effects,
fuel type and geometry, terrain or topography, and atmospheric
conditions. Extreme atmospheric conditions with winds reaching hurricane
strength may be the progenitor of extreme wildfires or may be the result
of the fire. Fire-atmosphere coupling due to the high energy released in
fires can lead to unstable wind and fire spread conditions, as well as
intense coherent flow structures such as vortices and whirls. The heated
air produces buoyant plumes that are capable of modifying the larger
scale atmospheric conditions in the boundary layer and above, sometimes
producing clouds and precipitation, smoke related radiative effects and
strong associated downdrafts and updrafts. Examples of updraft
velocities in excess of 100~mph have been measured, and ember fall or
strikes at distances of tens of miles, well over the horizon from the
main fire front.

\begin{figure}[htp]
\centering
  \includegraphics[width=1.0\textwidth]{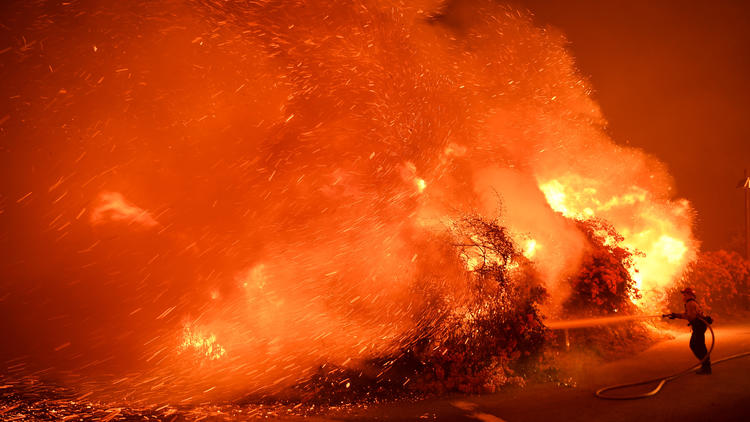}
  \caption{\label{fig:Firebrands} \em Embers are generated in trees and
  brush, lofted in the rising fire plume, with many falling immediately
  to the ground (Thomas Fire, La Conchita, California) Credit: Photo
  Wally Skalij, Los Angeles Times.} 
\end{figure}

What are embers? Usually we think of embers as pieces of vegetation,
such as bark, leaves, or twigs, but even whole plants---a good example
being tumbleweeds---can transmit fire. Built structures generate a large
amount of flying, burning debris not just from the building materials
but also from whatever is inside. As vegetation or structures burn, the
numerous small pieces of branches, leaves, and building materials detach
and either fall to the ground or are lofted by the strong vertical
velocities in the fire (Figure~\ref{fig:Firebrands}). These burning
embers or firebrands are liable to land on unburned fuels and cause
further ignitions, or spot fires, nearby or at great distances from the
main front of the fire. It is difficult to imagine the violent gusts of
wind that occur in large wildfires, shaking trees and brush and lifting
loose objects or burning embers from the ground. Short and long range
spotting is thought to be an integral part of fire spread in higher
winds across most fuel types. In built environments, fire spread can be
multiplied by the burning debris blown downwind and landing on
combustible fuels and other houses or buildings, effectively an ``urban"
regime of spotting and fire spread that has been extremely destructive
in mass fires and war.

Observations and models from simple to complex computational fluid
dynamic (CFD) simulations have been used to generate statistical
descriptions of ember flight. Ember landing patterns have been described
in terms of exponential, lognormal, and other more empirical patterns.
Some of these models are operational---used in real time to provide
support to ongoing fire incidents, or used by emergency management to
determine risk over communities, and many more models are used in the
research community to investigate various phenomenon in the wildland
fire system. We will be discussing a number of these in this Chapter,
with an emphasis on basic model concepts, data and data-models, and
physical constraints on ember transport.

\begin{figure}[htp]
\centering
  \includegraphics[width=1.0\textwidth]{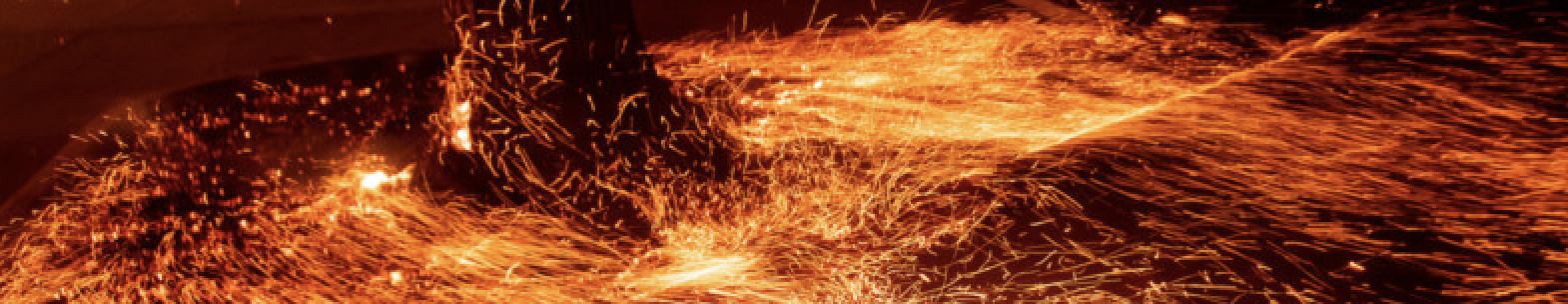}
  \caption{\label{fig:EmbersTree} \em Flow of embers around a tree and
  along the ground in the surface mode of fire spread. Credit: Detail of
  a Noah Berger/AP image of the Camp Fire in 2018.} 
\end{figure}

Some of the short range spotting is due to firebrands that do not rise
in the fire's plume---lofting---but instead spread nearly horizontally
in the surface boundary layer (Figure~\ref{fig:EmbersTree}). These and
the heavier embers and burning debris that emerge from the main fire
frontal region carried by strong winds are often partly responsible for
rapid movement and widening of the flaming front. We distinguish the
lofted form of ember transport from the surface mode and also use the
term ``ember wash" to denote this surface component of wildland fire
spread. Ejections of burning embers from the surface to higher levels by
boundary layer turbulence and eddies could also give rise to what would
effectively be a thicker, ``boundary layer" mode of near-surface ember
transport. Clearly, various transport processes may all be operating at
once in the complex 3D turbulent flow near the surface.

Numerous reports, anecdotal evidence, fire perimeter maps routinely
produced at large incidents, and diverse aerial and satellite images
have documented the impact of spotting on fire spread, and resulting
rapid movement and unpredictability of the location of the fire (see
e.g.~\citep{manzello2020role}). Consistent quantitative data on the {\em
spatial distribution} of spotting is difficult to produce due to the
wide range of observational techniques and gaps in coverage. Older
satellite products such as MODIS have been the backbone of many
large-scale studies of fire. In recent years, the growing capability of
satellite infrared camera images (e.g.~VIRS) are starting to provide
greater resolution of fires, both the main body and spot fires.
Nevertheless, the data still require careful analysis to avoid false
attributions. 

Two detailed studies based on traditional measurement techniques, in
particular, provide key data analyses that show the short-range behavior
and longer range distributions in diverse fuel and wind conditions based
on fires in Australia~\citep{storey2021experiments} and in the Rocky
Mountains, USA~\citep{page2019analysis}. Real fires do not, of course,
distinguish between surface and lofted modes of ember transport and both
intermingle especially at shorter range. We come back to these important
datasets in subsequent sections.

First, we briefly touch upon some of the key modeling concepts that have
been carried out to predict the mechanisms involved in the different
components of spotting, many of which grew from the early efforts of the
US Forest Service (see e.g.~\citep{albini2012mathematical}). Our aim here
is not to compile an exhaustive list of models and modeling ``schools of
thought", but rather to lay out the more fundamental aspects of fire
growth and ember transport modeling to enable new approaches. In common
with other geophysical areas of research there are two basic ways to
treat what is essentially a problem of Lagrangian transport: individual
embers or particles can be tracked in 3 dimensions, or the average
properties of embers can be calculated or parameterized. The latter
might involve physical or statistical models. A common challenge with
averaging is that it risks missing some of the important Lagrangian eddy
effects that transport particles in turbulent flow, making
parameterization of these effects crucial. An example relevant to embers
is the potential for waves in the atmosphere from various sources to
drive what is effectively a Stokes drift (see e.g.~\citep{far2024}) or in
general the role of fire plume turbulence magnifying ember
transport~\citep{thu-kep-tor-faw2017}.

Once released, the embers closest to the fire may be entrained or
directly injected into the buoyant plume and experience rapid lofting,
which is a fundamental aspect of spotting. Models have demonstrated the
important role of plume intensity and turbulence in the degree and range
of spotting, and suggested ``short-range'' lognormal and long-range
Poisson distributions~\citep{sar-con-kai-fer-por2008} or
exponential-like distributions of spotting downwind of the
fire~\citep{martin2016spotting, bhutia2010comparison}. A great deal of
effort has gone into a better understanding of the detailed mechanisms
and mechanical forces that are applied to embers during this
process~\citep{sardoy2007modeling, albini2012mathematical,
tohidi2017stochastic}.

Along with these aspects, the combustion characteristics of embers in
flight, their shape, the mechanical forces and stresses from drag, their
changing composition as they pyrolyze, and associated mass loss are
important to consider for their ultimate role in igniting vegetation and
structures~\citep{manzello2020role, dos-yag2023}.
\citet{koo-pag-wei-woy2010} provide an excellent overview of spotting
processes. Fewer studies have been made of the surface mode of ember
transport, though the fact that embers travel over the ground and ignite
fires is well-known in the fire community. \citet{qua-spe2026} use an
idealized, statistical approach to model ember transport processes that
control fire area growth from spotting and surface modes of ember
transport.

A host of factors influencing spotting were analyzed
by~\citet{storey2020analysis}, and they concluded that fire area was the
most important predictor of spotting, with weather, topography, and
fuels making secondary contributions. These results were based on
Australian fires and may not carry over exactly to other landscapes, such as the mountainous western US, dense forests, prairies, etc., but the basic results are robustly related to key wildland fire variables. Another interrelated predictor is the source term. How many embers are produced in a given area of fire, given vegetation or fuel, and given weather conditions? An ember source term represents all the factors that release burning materials from vegetation or structures and inject them into the surrounding environment. An ember transport term naturally depends on the type of fuel and the wind, but also obstacles and terrain. This factor represents the complex fluid dynamics of the problem and interactions within the atmospheric boundary layer. A further factor is a probability of ignition ($\pig$), reflecting a standard quantity in forestry that depends on temperature, humidity, time-of-day, aspect, shading, fuel class, and so on. It is clearly a dynamic quantity in its own right, evolving rapidly under the changing fire conditions as well as the local weather.

\citet{martin2016spotting} derived an ember transport model using a
similar framework leading to a specific form for the final spotting
distribution. Moreover,~\citet{sto-bed-pri-bra-sha2021} derived a
probabilistic model for fire spread taking into account real wildland
fire spread rates. These key studies envelop, in a sense, the collection
of model design and details we discuss here, and could be starting
points on their own for statistical models of fire spread including
ember effects. 

To frame the subject we can view the process schematically as a chain of
``models" from an initial wildland fire generating an ember to the ember
landing in a receptive fuel and causing a fire:
\begin{align}
  \boxed{\begin{array}{c}
    \text{\bf Fire}\\ \text{\bf Area}
  \end{array}}
  \;\times\;
  \boxed{\begin{array}{c}
    \text{\bf Ember}\\ \text{\bf Source}
  \end{array}}
  \;\times\;
  \boxed{\begin{array}{c}
    \text{\bf Ember}\\ \text{\bf Transport}
  \end{array}}
  \;\times\;
  \boxed{\begin{array}{c}
    \text{\bf Landing}\\ \text{\bf Distribution}
  \end{array}}
  \;\times\;
  \boxed{\begin{array}{c}
    \text{\bf Probability}\\ \text{\bf of Ignition}
  \end{array}}
  \label{eqn:Chain}
\end{align}
Each of the factors in equation~\eqref{eqn:Chain} is challenging and
amounts to its own field of study; each has been the subject of modeling studies from simple idealized cases to extremely complex dynamics and thermodynamics with many physical processes and variables. Our aim here is to strive to present a reduced form of the problem, focusing on the minimal models capable of reproducing key facets of observed fires, spread, ember transport, and ignition. 

First, in Section~\ref{sec:spread}, we review models for the rate of
spread from an empirical, idealized, deterministic, and stochastic point of view. Then, in Section~\ref{sec:physics} we examine the fundamental physical principles at work in ember transport, including combustion, lofting, settling (terminal velocity and drag), and the turbulent wind carrying the embers. These principles are illustrated with a fully 3D coupled fire-atmosphere model. Next, Section~\ref{sec:ember_stats} describes a statistical approach to ember transport, aiming to produce models for the ember landing distributions. The nature of the landing pattern is explored further in Section~\ref{sec:fire_model} in the context of a surface transport mode of ember transport that moves embers around on or near the ground in the boundary layer, with and without structures. This mode may be the only ember transport mode if no lofting occurs initially at the source, due to heavy ember production or intermittent turbulence. To compare with more realistic and complete models, Section~\ref{sec:WUI} shows examples of ember trajectories and the landing distributions are presented from computational fluid dynamics models, and a brief discussion of Wildland-Urban Interface (WUI) effects is given. Finally, Section~\ref{sec:discussion} summarizes
the Chapter and outlines future directions of ember research.

\section{Fire Spread and Rate of Spread}
\label{sec:spread}
Predicting the evolution of wildland fires is a central challenge for
both operational forecasting and scientific modeling. Wildland fire
growth is commonly characterized in terms of motion normal to the fire
front, with a rate of spread (RoS) that is parameterized in terms of
empirical evidence, physics, or data-driven approaches. To give some intuitive feel for just how fast large fires spread on average, a comprehensive synthesis of observations suggests that the downwind front spread rate is about 10\% of wind speed in forests and shrublands, with a much larger, but not well determined rate in grasslands~\citep{cru-ale2019}. Strong winds can easily drive fires faster than people can walk or run. 

The RoS depends on many factors, with wind, topography, and fuels being
the most important. Sullivan published a series of articles that
reviewed different fire spread models developed between 1990 and
2007~\citep{sul2009a, sul2009b, sul2009c}. Semi-empirical models have
been developed from field measurements, with two of the most
foundational models being Rothermel~\citep{rot1972}, developed primarily
in the context of US wildland fires, and~\citet{mca1966} developed for
Australian bushfires. These semi-empirical RoS closure models remain
central to many modern modeling systems, including
FARSITE~\citep{fin1998}, CAWFE~\citep{cla-coe-lat2004},
WRF-Fire~\citep{coe-cam-mic-pat-rig-yed2013},
WRF-SFire~\citep{man-bee-koc2011}, Behave~\citep{and1986},
SPARK~\citep{mil-hil-sul-pra2015}, and
Phoenix~\citep{tolhurst2008phoenix}. Computationally, these models
advance the fireline in time using level set methods, cellular automata, or by representing the fireline as a polygon, and advancing its vertices.

\begin{figure}[htp]
\centering
\includegraphics[width=0.9\textwidth]{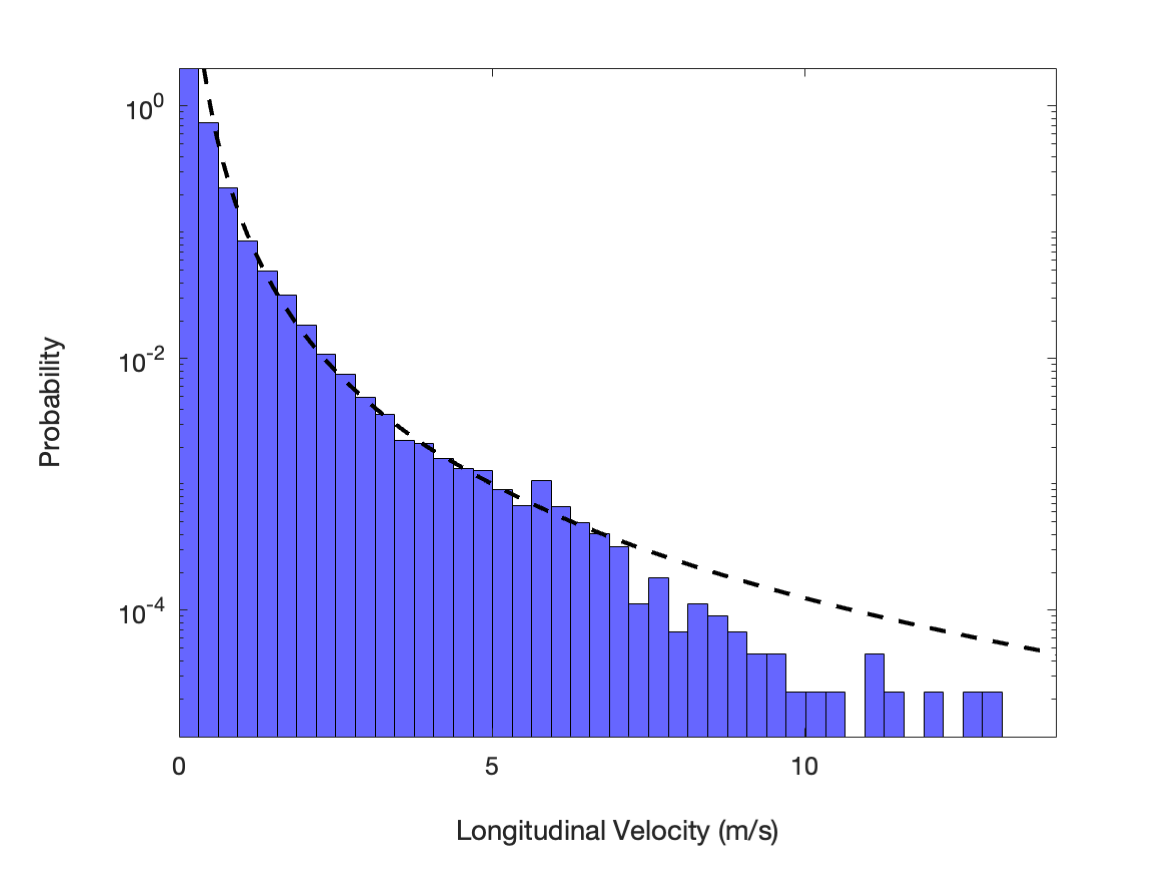}
  \caption{\label{fig:paugam} \em A distribution of the RoS from data
  collected by~\citet{pau-woo-rob2013}. The small values for the RoS
  follow a Pareto distribution with a scale parameter of $x_m =
  0.25$~m/s, and a shape parameter of $\alpha = 2$, resulting in an
  expected value of 0.5~m/s. The tail of the distributions contains rare
  RoS values that exceed 10~m/s.}
\end{figure}

While these semi-empirical models have provided the engine for many fire
spread models, it has become clear from coupled model studies that the
RoS of the fire front in any direction is not a single number, but a
stochastic process resulting from multiple feedbacks on wind and fire
intensity~\citep{sag-spe-pok-qua2021, sim-sha-eva2016}. In an outdoor
laboratory setting with imposed wind, \citet{sag-spe-pok-qua2021} found
that the RoS (forward component of velocity) followed an exponential distribution, roughly consistent with an approximately laminar or weakly turbulent fixed wind. An example of a broader range of spread rates becomes apparent in high-resolution analyses of experimental prescribed fires conducted outdoors in realistic conditions~\citep{pau-woo-rob2013} (Figure~\ref{fig:paugam}). Here, the wind field is stronger and much more turbulent, including the natural atmospheric boundary layer effects. The lower-RoS portion of the distribution is approximately a Pareto distribution, while the tail contains rare, much higher spread rates. This observation suggests a lack of a characteristic spread rate scale and aggregation or multiplicative effects, possibly better described as a cascade process.

While the detailed RoS offers crucial insight into the local behavior in the fire spread, large observational datasets as a whole are more readily
interpreted through integral measures, such as the total burned area,
$A(t)$, and the rate of growth, $dA/dt$, the latter being a primary
metric to assesses fire severity and risk. \citet{coe-cru-ros-spe2022}
describe fire spread and numerical experiments with coupled
fire-atmosphere models, and provide examples of other large-scale
metrics that measure distinctive characteristics of the fire perimeter
including reduced area (burned area divided by burned perimeter
squared), which measures the dispersion or ``fingering" of the fire
front, and the forward RoS divided by bulk rate of spread (RoS/BRoS),
which indicates the amount of forward spread compared to spread in all
directions. These metrics give a numerical value, for example, to
behavior associated with classic point-shaped and parabolic-shaped
fire~\citep{che-gou-cat1993}. It is important to understand that
apparently straightforward measurements of fire spread become extremely
challenging in realistic conditions of turbulent boundary layer winds
with complex arrangements of fuels and topography.
\citet{coe-cru-ros-spe2022} emphasize that this makes true model
validation challenging and great care must be taken interpreting
observations and experimental results and comparisons to models. 

In the rest of this section we attempt to gain an integrated view of
fire spread with the aim to link observations to simple growth models.
These reference or benchmark observational ensembles are also meant to
be used to develop consistent statistical models of fire behavior.

\subsection{Geometric Models for the Rate of Spread}
An early kinematic model of fire growth is the ellipse model. The
elliptical fire shape is embedded in the traditions of fire modeling for
good reason---early experimental work showed that fires spreading in
simple uniform fuels rapidly develop an elliptical shape with expanding
radii and could be modeled by a Huygens' principle approach with a fire
origin at the focal point of ellipses positioned on the perimeter of a
fire~\citep{and-cat-dem-par1981, and1986, and2014}. Present day
operational fire spread models use these principles (see
e.g.~\citep{fin-mca-gru-for2021}). Crucially, for an ellipse, the fire
can spread outward with a variable rate along the perimeter, controlled
by observable parameters for the forward and backward movement along the
major axis and widening along the minor axes (for a full implementation
of this model see for example \citet{lau2013} or the operational model
FARSITE~\citep{fin1998}). In the case of constant spread rates the area
growth from a single origin is always quadratic---based on an assumed
linear spread rate of the ellipse parameters. Quadratic fire area growth
implies that the rate of growth $dA/dt$ is increasing in time, a
behavior that is observed and we shall explore further below. It does
not appear to be well recognized that an {\em elongating} ellipse, when
the ratio of the major and minor axes increases in time, area grows
faster than $t^2$, for example as $t^3$ for linear elongation. Such
elongation will occur in the model when the wind itself increases over
time since the ellipse axes ratio is an increasing function of wind.
These details are important for the interpretation of fire area growth
as a measure of model success and validation.

To illustrate the insight provided by the elliptical model, consider a
purely artificial example that might seem reasonable at first, a fire in
which lateral spread is in one direction, downwind, with no spread
laterally or upwind. A model for this could be a rectangle with the
downwind side (head) at the fire front moving steadily and the other
sides (flanks and rear or back) stationary. In this case, the burned
area growth $A(t) \sim t$ is linear meaning that the rate of growth of
the burned area is constant. Of course, in real fires the other sides do
expand as flanking and backing fires. Assuming a fire grows in all
directions at a constant rate of spread, then the length and width, for
the case of a rectangle, or diameter, for the case of a circle, of the
burn scar will grow linearly with time, and result in the area growth
scaling as $A(t) \sim t^2$. However, it is well known that the three
basic fire fronts: head, flank, and back, typically move at different
speeds, and moreover fires do not expand as simple rectangular shapes,
even from line sources, because fires drive winds that focus spread at
the head and suppress spread at the flank and
rear~\citep{coe-cam-mic-pat-rig-yed2013}, producing more of an
elliptical shape. Thus these artificial examples are poor
representations of actual fire growth.

For fixed environmental conditions and constant spread rates, the
elliptical model results in quadratic fire growth. We will see that
observations suggest that a class of large fires grows linearly in area.
How can this be reconciled with the elliptical model? Characterizing
fire spread by a simple continuous expansion normal to the perimeter is
appealing but another paradigm for spread arises from the complex
physics and random variations in fuels, moisture content, etc., or the
random convective motion of flames and hot gases at the fire front,
igniting fuels ahead of the front in apparent stochastic jumps. These
approaches can be represented by models with partial differential
equations (PDE).

\subsection{PDE Models for the Rate of Spread}
A physics-based approach to model fire spread and growth rate is to
begin with the full set of equations for the mass, chemical species,
momentum, and energy or thermodynamics of the solid and gas phases. An
excellent survey of the physical modeling of wildfires and comparisons
with experimental spread rates was given by \citet{mor2011}, and another
by \citet{sul2009b}. Both these articles discuss very advanced PDE
models including FIRETEC~\citep{lin-cun2005},
WFDS~\citep{mel-jen-gou-che2007}, and
FIRESTAR~\citep{morvan2009physical}. These are all very mature and
sophisticated models that include subgrid-scale models for complex
processes including combustion, canopy drag, and fire-induced winds. 

Alternatively, by interpreting diffusion as a representation of the
turbulent convection of flame and hot gases in and above the fuel bed,
PDE models can represent the back-and-forth motion of flaming gases that
impinge on unburned fuel, ignite the leading edge, and carry fire
forward. This idea of turbulent heat diffusion as a primary driver of
RoS leads to a diffusive fire spread model that can produce realistic
spread rates~\citep{beb-oli-qua-sko-hei-spe2020}. The scaling becomes
strongly linked to the turbulent diffusion rate as $K^{1/2}$ where $K$
is the (now) turbulent diffusion coefficient extending down through the
canopy and into the fuel bed. In many geophysical conditions, the
turbulent diffusion is related to variance and roughly proportional to
the mean wind. Hence this model explains the result that large fire
spread rates are tightly linked to the mean wind. Since the fire itself
produces turbulent motion this model provides for spread even in the
absence of background wind, or, in other words, upwind spread as a
backing fire. By itself, a purely diffusive model for fire spread has an
approximately linear area growth scaling at small times. In order to
better represent turbulent effects in an idealized framework with
uniform background wind, \citet{qua-spe2021} implemented a 2D diffusive
spread model to drive spread physically in low or no wind conditions.
Then, by modifying the turbulent diffusion coefficients to represent
stronger, inhomogeneous boundary layer wind conditions, greater fire
growth in higher winds may be realized, with various quasi-elliptical
shapes. This interpretation lends itself, in principle, for more complex
transport processes with super-diffusion in self-avoiding random walks
across a spatial domain relevant to bimodal burn states. Nonlinear
diffusion effects, where the spread rate depends on fuel density, fire
intensity, or other dynamical quantities, may prove to be a useful
extension as well.

At this point we focus on simpler systems, with various degrees of
idealization, with the advection-diffusion equation for heat, including
a reaction term for the heat produced by combustion, and a radiation
term for the heat gained or lost by radiation. A mathematical
description of solutions to this equation for a wildland fire
application, focused on the role of wind, is given by
\citet{bab-bou-hil2009}. The simplified 1D version of the local heat
balance and fuel consumption equations are
\begin{align}
  \pderiv{Y}{t} &= 
  -\frac{1}{\tau}\,Y\,H\!\left(T-T_{\mathrm{ign}}\right), \\
  \rho C\!\left(\frac{\partial T}{\partial t} + U\,\pderiv{T}{x} \right)
  &= k\,\pderivtwo{T}{x} + Q\,\rho\,\frac{1}{\tau}\,Y\,H\!
    \left(T-T_{\mathrm{ign}}\right) - \gamma\,(T-T_{a}),
  \label{advdiff}
\end{align}
where $Y(x,t) \in [0,1]$ is the dimensionless fuel mass fraction,
$T(x,t)$ is absolute temperature, $\rho$ is fuel density, $C$ is the
specific heat, $U$ is the wind velocity, and $k$ is the thermal
conductivity. The reaction term consumes fuel only when $T$ exceeds an
ignition temperature $T_{\mathrm{ign}}$, with $\tau$ being the reaction
timescale. In some cases this simple Heaviside form $H(\cdot)$ is
replaced by a smoother, exponential Arrhenius form. The term
$Q\rho\tau^{-1}YH(T-T_{\mathrm{ign}})$, which has units W $\cdot$
m$^{-3}$, is the heat release when the fuel mass is burning, and $Q$ is
the heat released per unit fuel mass. The term $-\gamma(T-T_{a})$ is
heat loss to the ambient environment to represent radiation with
$\gamma$ a heat-loss coefficient, and $T_{a}$ the ambient temperature.
The addition of an oxygen conservation equation can be very important in
some circumstances, particularly where fuels are highly confined, and
leads to additional combustion regimes that we ignore here.

Diffusion of heat occurs in this system only as a thermal conductivity in fuels, and for systems with this version of diffusion, various forms
of~\eqref{advdiff} have been used very successfully to model fire spread
in experimental fuel beds (e.g.,~\citep{mor-sim-san-bal2005}) and are
able to reproduce observed spread rates in the laboratory setting. Air
flow convergence and divergence are allowed via a boundary flux term.
This convergence provides a dynamical feedback between the fire strength
and air flow, which in turn affects the fire intensity
\citep{beb-oli-qua-sko-hei-spe2020}.

In a very different, purely mathematical context,
\citet{bab-bou-hil2009} find traveling wave solutions of
equation~\eqref{advdiff}. They arrive at analytical relationships for
spread rate with different (constant) winds and ignition temperatures,
with the RoS scaling as $U + \sqrt{1/(\tau T_{ign})}$. In this system,
entirely within the fuel bed without a direct connection to the air
layer above, the forward fire front travels faster than the air flow
within the fuel bed, while the backing fire travels at a similar speed
as the air flow. The RoS increases with wind but decreases as the
non-dimensional reaction time or ignition temperature increases. Higher
fuel moisture content is one way these two quantities can be effectively
increased. This result also highlights the limitations of the closed 1D
model that are apparent since in wildland environments the wind blows
over the fuel bed at much higher speeds and much of the ignition and
rapid combustion takes place on the fuel bed surface before propagating
down into the interior of the fuel bed.

A major limitation of deterministic PDE models with a given wind is that
they primarily capture averaged or continuum fireline motion, all the
while missing important stochastic behaviors. This is appropriate to
capture broad fireline motions, such as an average RoS. However, as we
saw in Figure~\ref{fig:paugam}, the RoS varies over several orders of
magnitude---a consequence of the stochasticity present in the wind
field, fire intensity, fuel structure, and more. Moreover, the data
suggests various ember generated effects, near and far from a fire
front, leading to rapid, stochastic jumps in area growth that continuous
deterministic models do not naturally capture. Therefore, in the next
section, we discuss models for the RoS that include varying levels of
stochasticity.

\subsection{Stochastic Models for the Rate of Spread}
Continuous thermodynamic and geometric models can struggle to capture
the turbulent nature of real-world wind fields, motivating the use of
stochastic processes to represent the RoS. In addition to the
turbulence-induced randomness, fire spread also has non-local spatial
behavior caused by new ignition sites. These new ignitions are most
commonly caused by embers, but they can also be caused when smoldering
combustion occurring below the surface in the duff gains access to
oxygen, thereby transitioning into flaming combustion and emerging as
new fires.

Randomness in fireline patterns is often modeled with stochastic
differential equations. The Kardar-Parisi-Zhang (KPZ) equation is a
nonlinear stochastic partial differential
equation~\citep{kar-par-zha1986} that has been used to describe the
fireline shape. It is an example of a surface growth model, and these
models are often analyzed to study the statistics of the rough surfaces,
such as firelines, and how they relate to key parameters such as the
amount of randomness. A clever example of non-flaming (smoldering)
combustion frontal movement as a stochastic process on a micro-scale was
described by~\citet{zha-zha-als-lev1992}, who analyzed a burning front
on paper fuel using the KPZ model. In this model there is no wind and
the front advances in the normal direction by accretion with an added
stochastic noise term, and diffuses or smooths laterally. With front
position or height $h$ in 1D (similar in 2D) the model is
\begin{align}   
  \pderiv{h(x,t)}{t} = \nu \nabla^2 h(x,t)
    + \frac{\lambda}{2}\,\lvert\nabla h(x,t)\rvert^2
    + \xi(x,t),
\end{align}
with white noise statistics
\begin{align}
  \langle\xi(x,t)\rangle = 0,\qquad
  \langle\xi(x,t)\,\xi(x',t')\rangle = 2D\,\delta(x-x')\,\delta(t-t').
\end{align}
Here, $\nu$ quantifies small-scale diffusion, $\lambda$ quantifies the
nonlinear coupling due to motion normal to the front, and $D$ quantifies
the noise amplitude. A scaling for the mean spread rate in 1D
is~\citep{sas-spo2010}
\begin{align}
  \langle h\rangle \sim v_{\infty}\,t + \mathcal{O}(t^{1/3}), 
    \qquad v_{\infty}=\frac{\lambda^{2}D}{24\,\nu^{2}},
\end{align}
but closed-form solutions generally do not exist in 2D. It is difficult
to relate these model parameters to wildland fire quantities, but this
scaling nevertheless provides intuition about the direct role of random
jumps versus smoothing or merging in frontal motion, and suggests an
asymptotically linear net spread rate along the perimeter. On the other
hand, this type of model may be more appropriate for the microscale
spread process within a fuel bed rather than a typical flaming
combustion fire.

Natural wildland fire fronts have both width and depth and their
propagation vertically as well as horizontally is a 3D process, which
among other things makes measurements of spread at the fuel scales
challenging. Moving to the case with flaming combustion, frontal
velocity and displacements in wildland and experimental fires have been
observed at fine resolution using various techniques and suggest that a
fire proceeds across a fuel bed with exponentially distributed motion,
at least at low to moderate wind
speed~\citep{joh-whe-woo-pau-dav-deb2018, pau-woo-rob2013,
sag-spe-pok-qua2021} (see Figure~\ref{fig:paugam}). This is thought to
be linked to fine scale wind turbulence on the fuel bed surface and
interior, as well as fuel characteristics, and implies an independent
arrival time process of ignition. Subsequent fuel burn time (or burnout
time) following the frontal passage exhibits various distributions
depending to a large extent on the nature of the
fuels~\citep{cur-spe-hie-obr-goo-qua2018}. These detailed statistics are
very promising for the development of more sophisticated models of fire
spread, and point to the importance of making observations that can
resolve the fine scale behavior of fire in the turbulent atmospheric
boundary layer. These results also demonstrate that stochasticity is an
essential component of local RoS models. 

Looking ahead, we shall see that ember transport also shows exponential
distributions that may relate back to the underlying physical context of
a turbulent wind field. Additional statistical models that focus on
ember dynamics will be presented in Section~\ref{sec:ember_stats}.
However, in the next section, we first provide observational evidence
that spotting caused by embers alters statistical behaviors of the RoS
by considering progression maps of burn scars from the western US.

\subsection{Observations of Fire Area Growth}
To motivate the need for stochasticity for RoS models, and to relate the
RoS to ember dynamics, we conclude this section by turning to
observations of area growth in historical wildfires. The dataset is of
western US wildland fires, and we use them to provide a reality check of
some of these statistical modeling approaches to fire spread. 

\begin{figure}[htp]
  \centering
  \includegraphics[height=5cm,trim=2cm 0cm 0.3cm 0cm,clip=true]{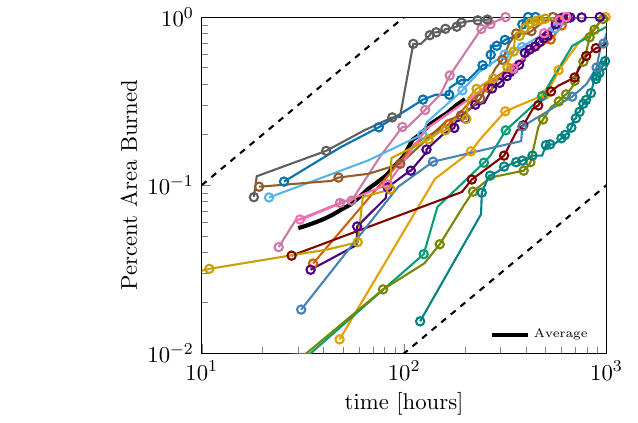}
  \hspace{40pt}
  \includegraphics[height=5cm,trim=0.9cm 0cm 0.1cm 0cm,clip=true]{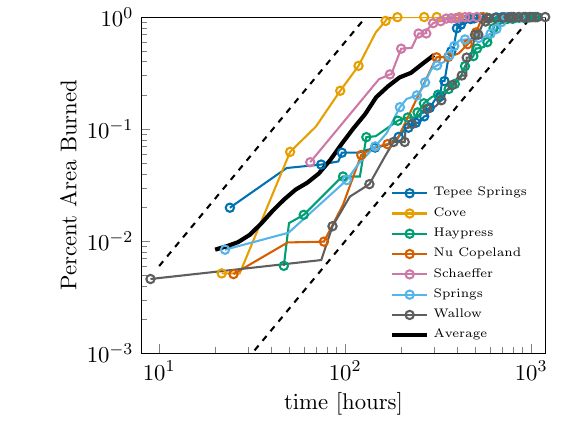}
  \caption{\label{fig:historical1} \em Left: The burnt area of 15 (left)
  and 7 (right) wildfires in the Western USA. All areas are normalized
  according to their final size. The slope of the dashed black lines is
  1 (left) and 2 (right), indicating a linear (left) and quadratic
  (right) growth in the fire's area.}
\end{figure}

\begin{figure}[htp]
  \begin{center}
    \fbox{\includegraphics[width=0.225\textwidth,trim=7cm 7cm
    7cm 7cm,clip]{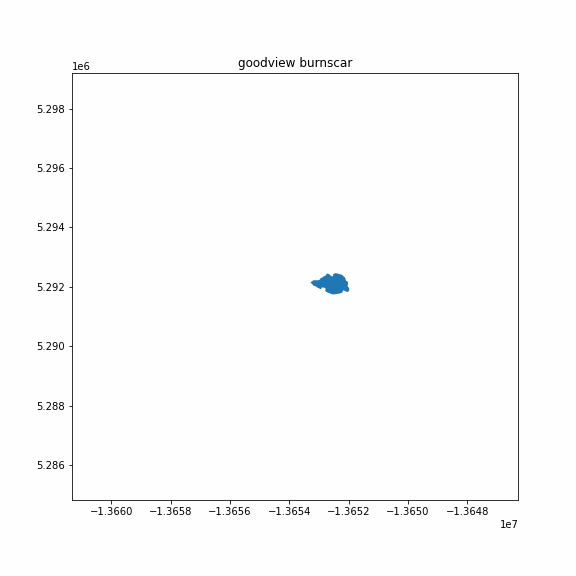}}
    \fbox{\includegraphics[width=0.225\textwidth,trim=7cm 7cm
    7cm 7cm,clip]{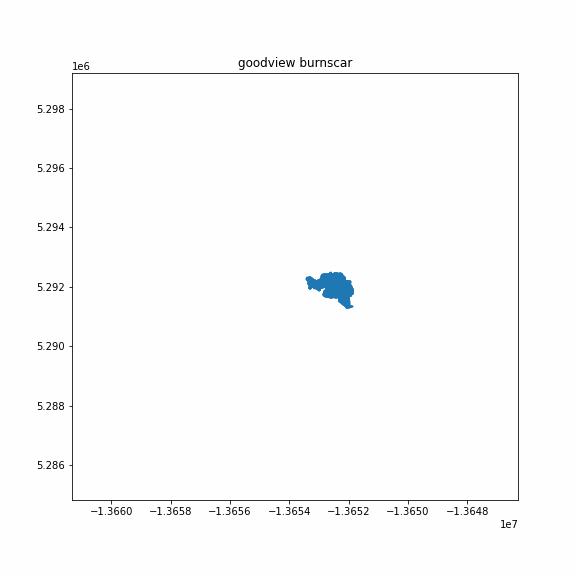}}
    \fbox{\includegraphics[width=0.225\textwidth,trim=7cm 7cm
    7cm 7cm,clip]{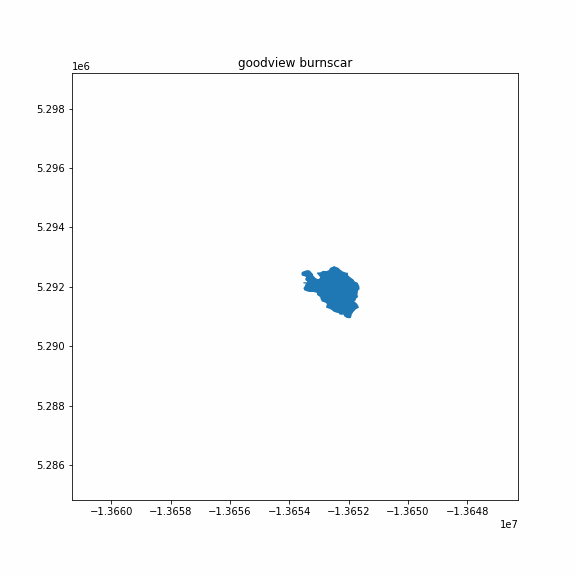}}
    \fbox{\includegraphics[width=0.225\textwidth,trim=7cm 7cm
    7cm 7cm,clip]{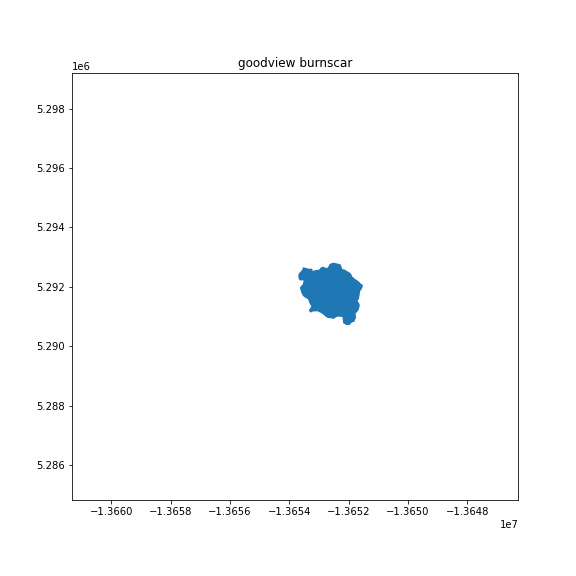}} \\
    \fbox{\includegraphics[width=0.225\textwidth,trim=7cm 7cm
    7cm 7cm,clip]{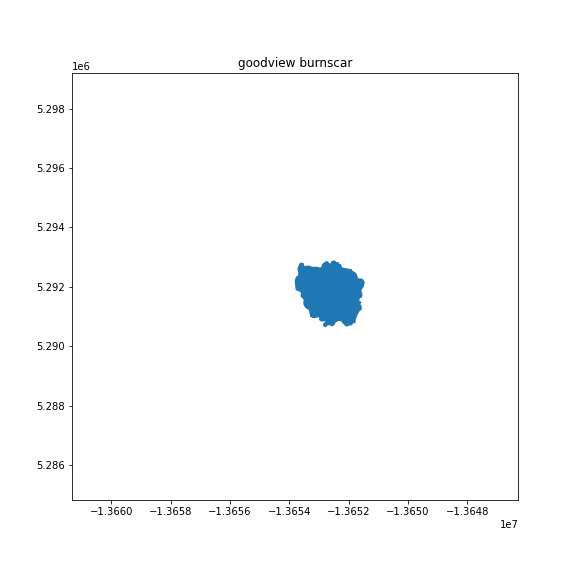}}
    \fbox{\includegraphics[width=0.225\textwidth,trim=7cm 7cm
    7cm 7cm,clip]{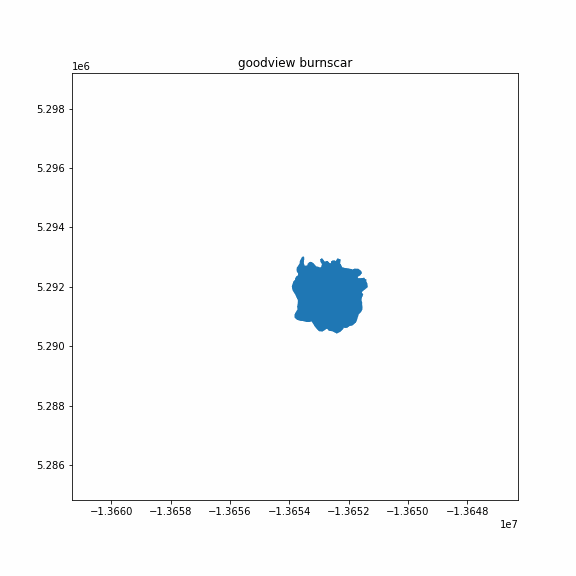}}
    \fbox{\includegraphics[width=0.225\textwidth,trim=7cm 7cm
    7cm 7cm,clip]{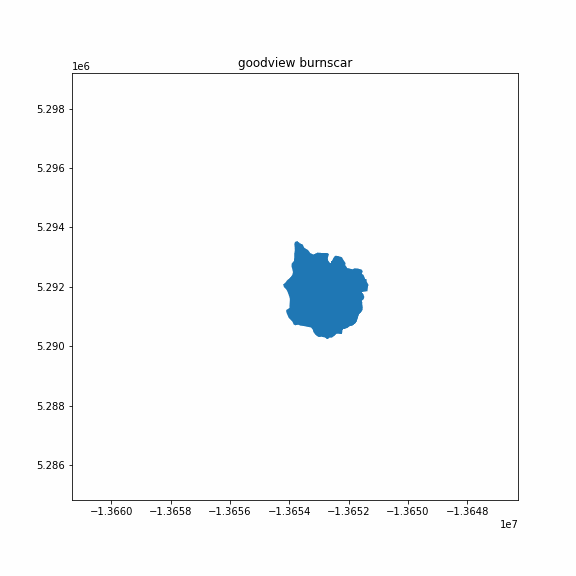}}
    \fbox{\includegraphics[width=0.225\textwidth,trim=7cm 7cm
    7cm 7cm,clip]{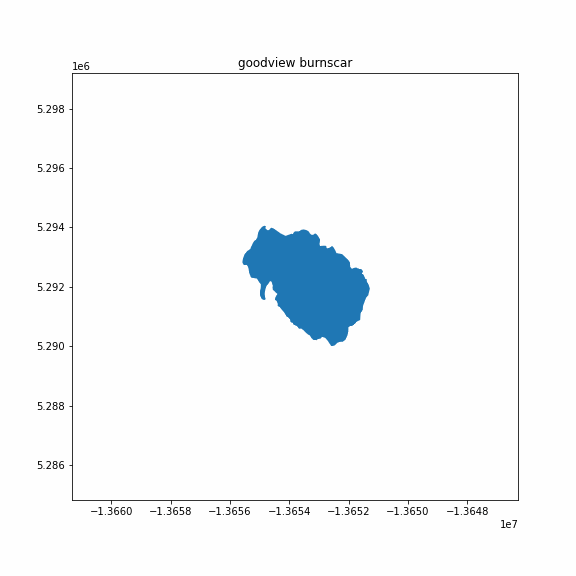}} \\
    \fbox{\includegraphics[width=0.225\textwidth,trim=7cm 7cm
    7cm 7cm,clip]{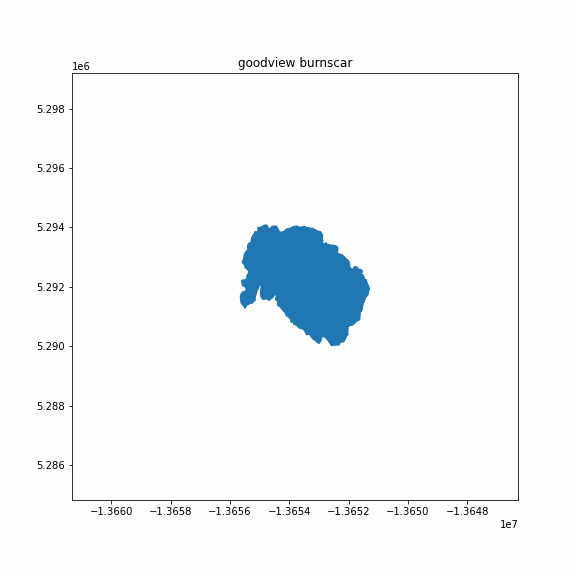}}
    \fbox{\includegraphics[width=0.225\textwidth,trim=7cm 7cm
    7cm 7cm,clip]{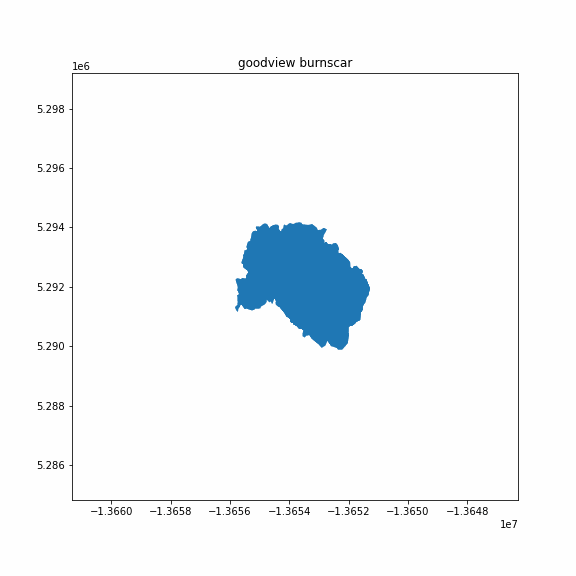}}
    \fbox{\includegraphics[width=0.225\textwidth,trim=7cm 7cm
    7cm 7cm,clip]{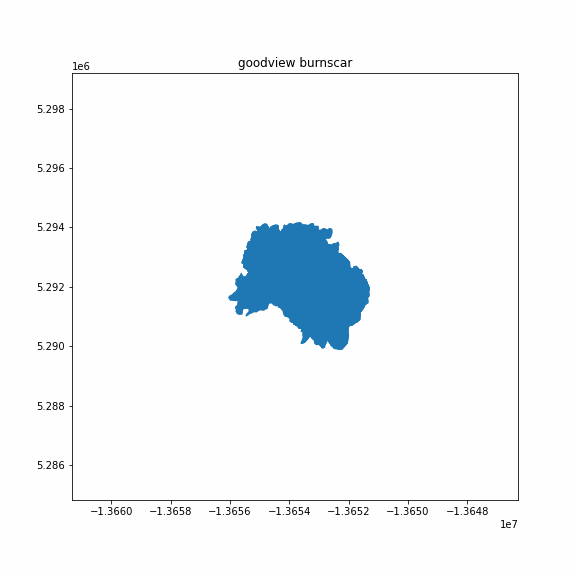}}
    \fbox{\includegraphics[width=0.225\textwidth,trim=7cm 7cm
    7cm 7cm,clip]{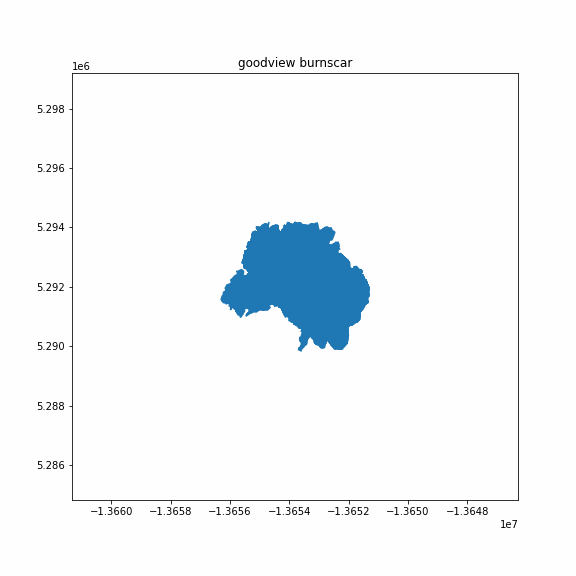}}
  \end{center}
  \caption{\label{fig:goodview} \em The Goodview fire exhibits the
  linear area scaling $A(t) \sim t$. Notice the lack of spotting.}
\end{figure}

To investigate the asymptotic growth of burn scar area from historical
fires, GIS data of burn scars of over 500 fires that occurred in the
western US between 2015 and 2021 were processed~\citep{qua-spe2026}.
Many datasets have short or only crudely sampled area making the
behavior of the growth rate difficult to determine. We used 22 of these
fire incidents that have enough samples to be able to characterize an
asymptotic trend in the area growth. The left plot in
Figure~\ref{fig:historical1} shows the area as a function of time of 15
wildland fires from the western US. The black dashed lines have slope 1,
indicating that the area of these fires is growing linearly with time.
The solid black line is an average over these 15 fire events. One
particular fire event is the Goodview fire, and GIS frames of the burn
scar at different times are provided in Figure~\ref{fig:goodview}. One
characteristic that we can observe is the regular spreading and lack of
physically separated fires indicating weak or absent long range
spotting.

\begin{figure}[htp]
  \begin{center}
    \fbox{\includegraphics[width=0.225\textwidth,trim=5cm 5cm
    5cm 5cm,clip]{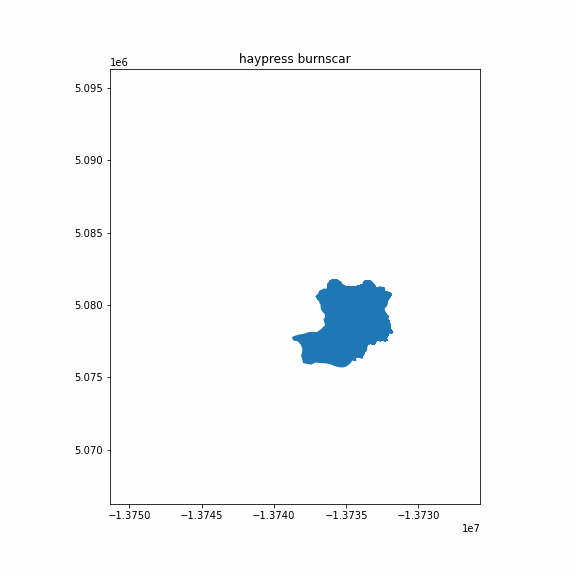}}
    \fbox{\includegraphics[width=0.225\textwidth,trim=5cm 5cm
    5cm 5cm,clip]{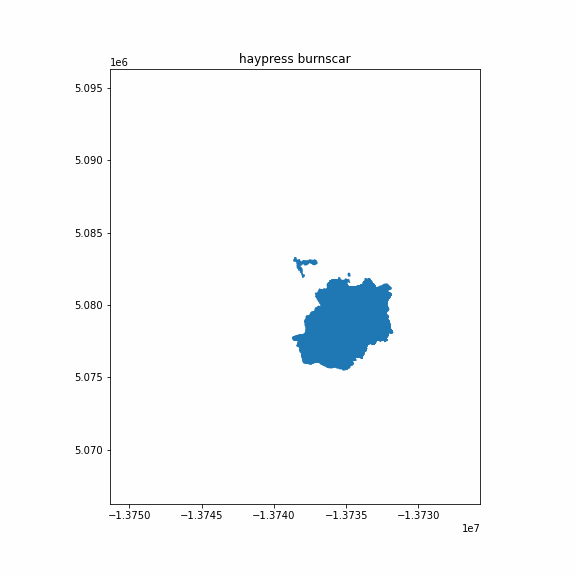}}
    \fbox{\includegraphics[width=0.225\textwidth,trim=5cm 5cm
    5cm 5cm,clip]{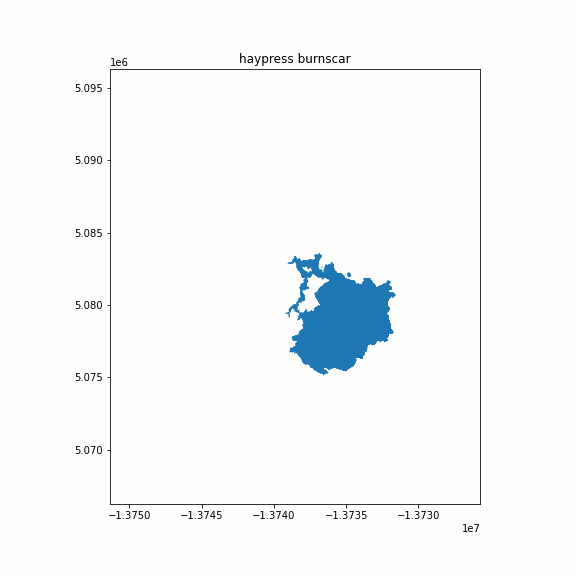}}
    \fbox{\includegraphics[width=0.225\textwidth,trim=5cm 5cm
    5cm 5cm,clip]{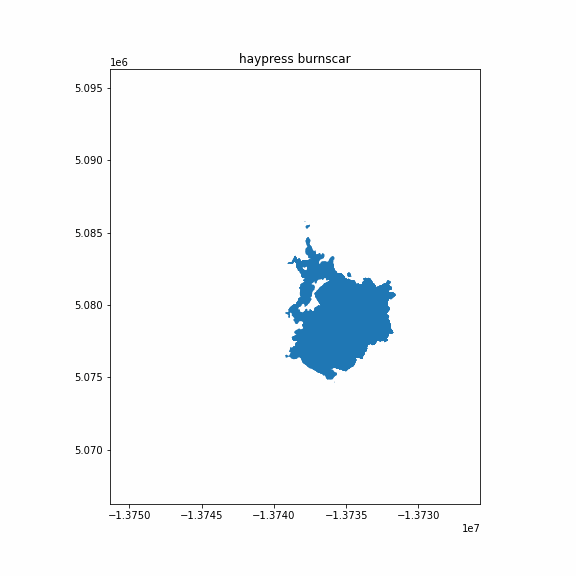}} \\
    \fbox{\includegraphics[width=0.225\textwidth,trim=5cm 5cm
    5cm 5cm,clip]{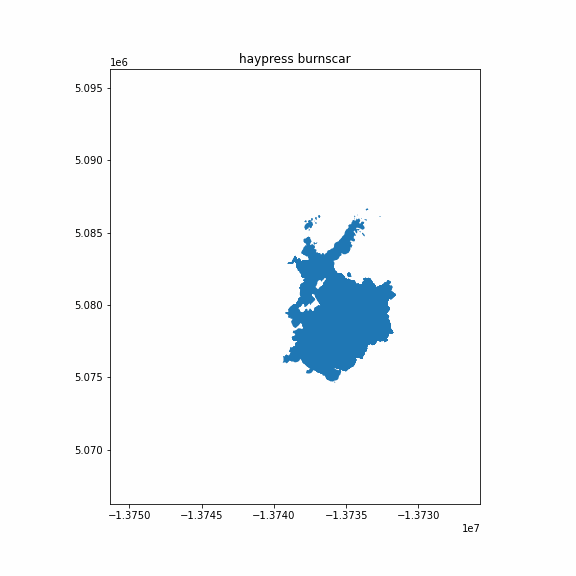}}
    \fbox{\includegraphics[width=0.225\textwidth,trim=5cm 5cm
    5cm 5cm,clip]{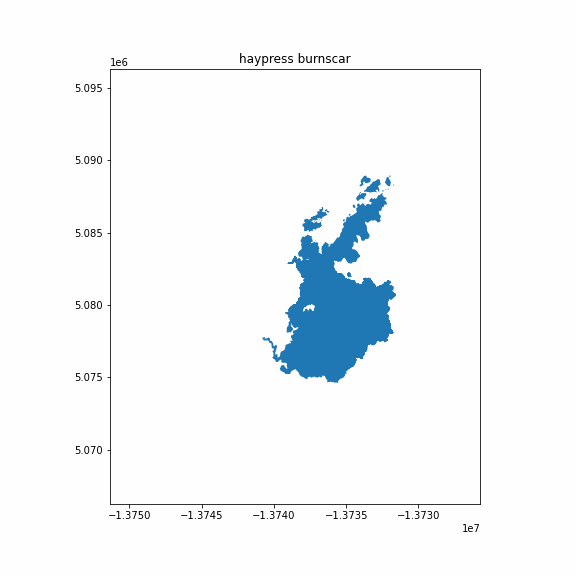}}
    \fbox{\includegraphics[width=0.225\textwidth,trim=5cm 5cm
    5cm 5cm,clip]{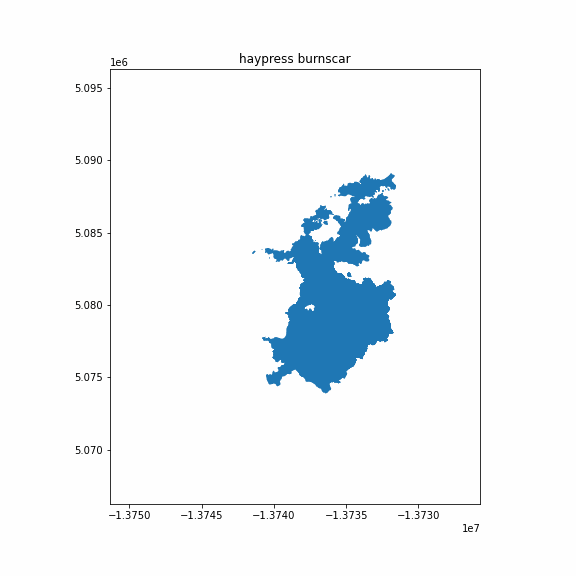}}
    \fbox{\includegraphics[width=0.225\textwidth,trim=5cm 5cm
    5cm 5cm,clip]{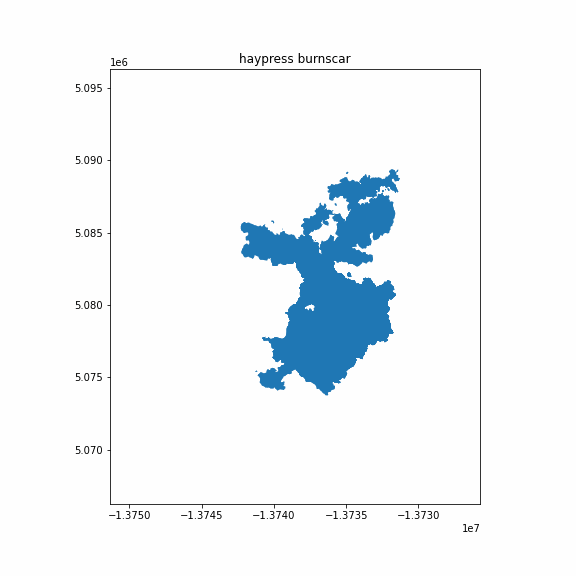}} \\
    \fbox{\includegraphics[width=0.225\textwidth,trim=5cm 5cm
    5cm 5cm,clip]{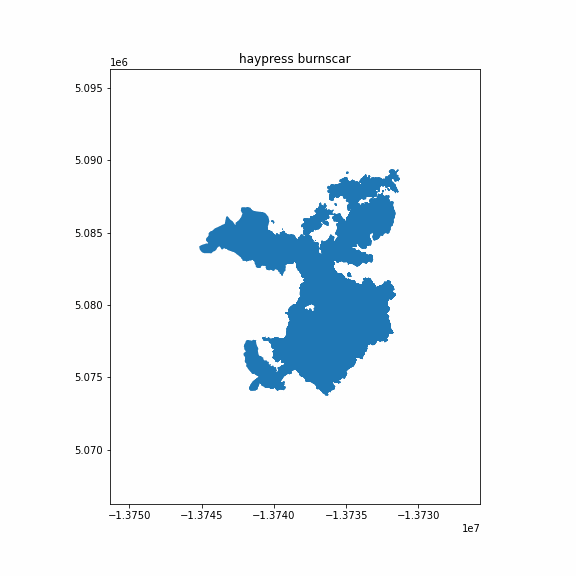}}
    \fbox{\includegraphics[width=0.225\textwidth,trim=5cm 5cm
    5cm 5cm,clip]{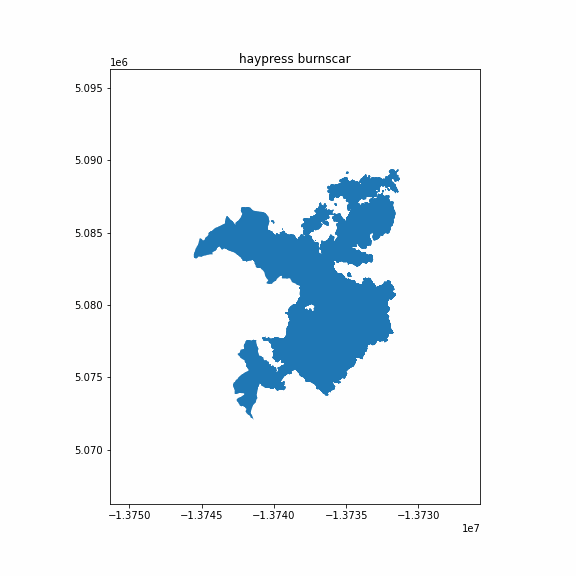}}
    \fbox{\includegraphics[width=0.225\textwidth,trim=5cm 5cm
    5cm 5cm,clip]{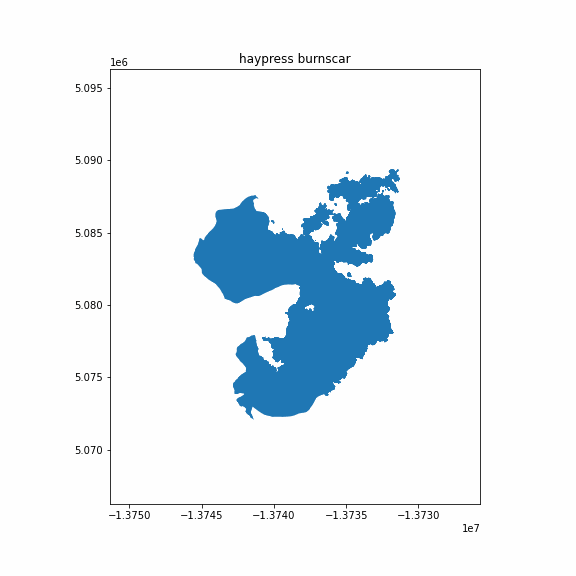}}
    \fbox{\includegraphics[width=0.225\textwidth,trim=5cm 5cm
    5cm 5cm,clip]{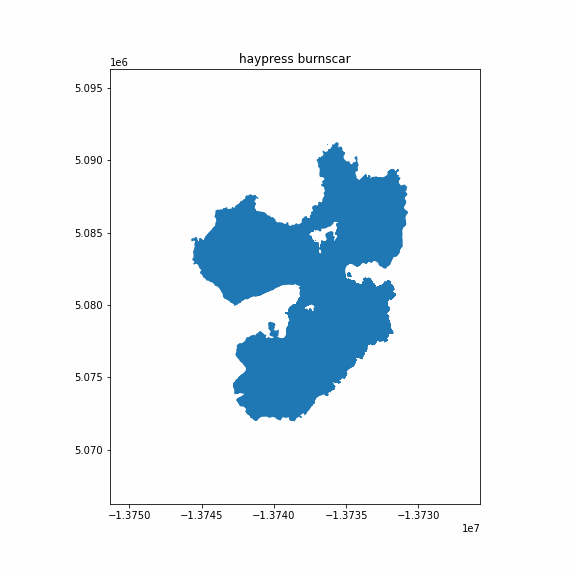}}
  \end{center}
  \caption{\label{fig:haypress} \em The Haypress fire exhibits the
  quadratic area scaling $A(t) \sim t^2$. Notice the abundance of
  spotting.}
\end{figure}

The right plot of Figure~\ref{fig:historical1} shows that the area
growth can also be quadratic for other large fires in the western US.
The black dashed lines have slope 2, and the solid black line is an
average over the seven fire events. One particular fire event is the
Haypress fire, and GIS frames of the burn scar at different times are
provided in Figure~\ref{fig:haypress}. Note that the fire growth is
clearly not regular or isotropic, and isolated neighboring fires are
evident, indicating that another process is responsible for the
quadratic growth. The notion here is that nonlocal ember spotting, which
is observable in these images, drives the quadratic growth.

An even wider range of growth rates in fire area has been inferred for
western US fires between 2001 and 2020 by
\citet{jua-wil-aba-bal-hur-mor2022} [Figure 2]. They analyzed a large
number of fires to find a statistical relation between the increase in
the size of the fire, $\Delta A$, and the fire size $A$. That is, $dA/dt
\sim A^{\beta}$. So long as $\beta \neq 1$, this implies
\begin{align}
  A(t) \sim t^{\frac{1}{1-\beta}}.
\end{align}
Averaging over 2,352 measurements, they calculate an average value of
$\beta = 0.46$, suggesting that the average area scaling result is $A(t)
\sim t^{1.85}$, however, their results are so broadly distributed in
$\beta$ that pretty much any slope between linear and quadratic scaling
is also equally as likely, probably due to the inclusion of noisier
samples.

These observations raise a central modeling question: which physical
mechanisms control the transition between a linear and quadratic growth
in a fire's area? While plume-driven \textit{long-range} spotting has
been studied extensively, near-field spotting and ember wash has not,
despite copious anecdotal evidence for ember transport close to the
front and near the surface. The GIS data are suggestive of various
ember generated effects, near and far from a fire front, leading to
rapid growth. The GIS-derived fire growth behavior in
Figure~\ref{fig:historical1}, and observations discussed by
\citet{jua-wil-aba-bal-hur-mor2022} suggest that fire areas can grow
quadratically with time, with this asymptotic growth lasting as long as
three weeks. As noted earlier, typical operational modeling choices
disallow linear growth in the overall fire area.

Does fire area growth at very large scales follow a diffusive process?
It is clear that the movement of the front is strongly controlled by
turbulence in the boundary layer wind field, due to ambient and
fire-generated winds. This could imprint a more diffusive, random-walk
character to frontal progression at the largest scales of the fire.
Rather than using a model that kinematically results in linear or
quadratic area growth, it is important to allow physical processes to
play a significant role in the asymptotic area growth. In particular, by
comparing Figures~\ref{fig:goodview} and~\ref{fig:haypress}, many more
spotting-like events near and far from the front can be seen in the
Haypress fire whose area grew quadratically with time.

Having described observations of fire spread, highlighting the scaling
behavior, we next return to the basic physical processes responsible for
ember transport. Our goal is to outline the fundamental framework of
physics-based models and provide justification for simpler models of
transport.

\section{The Physics of Ember Transport}
\label{sec:physics}
Embers are inertial, burning particles that are responsible for both
short- and long-range spotting. Their trajectories depend on
particle-scale properties, including drag, shape, density, and mass. At
the same time, ember trajectories are strongly controlled by the
atmospheric environment through which they travel, including both
background winds and fire-induced plume dynamics. In this section, we
first describe reduced models for individual ember motion and then
discuss how plume-scale atmospheric transport modifies ember
trajectories.

Physics-based ember models range from idealized trajectory calculations
to coupled fire-atmosphere simulations. \citet{sar-con-kai-fer-por2008}
considered landing distributions of disk-shaped embers under different
wind speeds and fire intensities, finding that the number of firebrands
landing in a flaming state correlates with flight time. More recently,
\citet{lop-tru-mor-fio-car-pag2024} coupled a cellular automata fire
spread model with ember spotting models. Broader reviews of ember
generation and transport are provided by \citet{koo-pag-wei-woy2010} and
more recently by \citet{wad-sul-wic-kyn-kha-moi2022}.

\subsection{Ember Equations of Motion}
The equations of motion for an ember carried by atmospheric boundary
layer winds $U_w$ of density $\rho_a$ are complex due to time-evolving
mass, drag, smoldering combustion induced flow, buoyancy, and a highly
variable Reynolds number in turbulent atmospheric motion. There have
been a variety of models that make different assumptions about key
parameters including the ember's shape, its combusting behavior, and its
density. One simplified set of equations that captures the most
important factors that transport an ember is
from~\citep{Tohidi_Kaye_2022}
\begin{subequations}
\begin{align}
  \frac{dX_e}{dt} &= U_e, \label{eqn:ember_location} \\
  M_e \frac{dU_e}{dt} &= \frac{1}{2}\rho_aA_e(t)C_d(U_e-U_w)^2,
  \label{eqn:ember_horizontal} \\
  M_e \frac{dW_e}{dt} &= \frac{1}{2}\rho_aA_e(t)C_d(W_e-W_w)^2 -
  M_e(t)g, \label{eqn:ember_vertical} \\ 
  A_e &= \pi (D/2)^2, \label{eqn:ember_area} \\
  M_e &= \rho_e (4/3)\pi (D/2)^3, \\
  M_e(t) &= M_{e0}/(1+at^2). \label{eqn:ember_mass}
\end{align}
\end{subequations}
where $U_e$, $W_e$ are the ember velocity components in the
$(x,z)$-coordinate system, $A_e$ is the ember area, $D$ is the ember
diameter, $M_e$ is the ember mass, $g$ is gravity, $\rho_e$ and $\rho_a$
are the densities of the ember and air, respectively, and $a$ is a
combustion rate parameter. The drag coefficient $C_d$ dictates the rate
at which the wind $(U_w, W_w)$ transfers momentum to the ember.
Equations~\eqref{eqn:ember_location}--\eqref{eqn:ember_vertical}
describe the particle dynamics, while
equations~\eqref{eqn:ember_area}--\eqref{eqn:ember_mass} provide
geometric and mass-loss closures. 

More complete models can update the ember diameter, density, or both,
rather than prescribing a phenomenological burnout law.
\citet{koo-pag-wei-woy2010} used the combustion model
\begin{align}
  \frac{d}{dt}D = -A_e(W_e-W_w)
    \frac{\rho_{\mathrm{air}}}{\rho_{\mathrm{ember}}},
\end{align}
in which the ember diameter, $D$, evolves according to the relative
vertical wind velocity ($W_e-W_w$). Various forms for the drag and
settling terms are relevant that depend on geometry, combustion,
composition, and flow regime. Typically the wind comes from an
atmospheric model, which may be more or less complete in terms of
parameterized subgrid scale processes, thermodynamics, vertical extent
and horizontal domain. In some applications the wind may be decomposed
into a mean, obeying a simplified model, and fluctuating parts $U_w =
\overline{U}_{w} + U_w'$ with the latter obeying turbulent statistics.
Depending on the model, the fluctuating velocity may be represented
using Gaussian components, Weibull or Rayleigh speed distributions,
intermittent bursts, or non-Gaussian statistics associated with coherent
structures.

\subsection{A Large-Eddy Simulation of Embers in a Plume}
\label{sec:LES}

Numerous studies in the atmospheric literature have used numerical
simulations to generate the 4D (space and time) wind fields that advect
particles of diverse composition, typically to represent hydrometeors
(rain, snow, hail, etc.). This area of research translates directly into
the wildland fire environment for the evolution of embers lofted into
the atmosphere by the plume. In fact, it is not uncommon for fires to
create clouds (pyrocumulus) and even rain and localized thunderstorms
under certain conditions. To set the stage for idealized modeling
discussed later on in this Chapter, and to build some intuition about
ember behavior in more realistic, yet simulated conditions, we use a
cloud-resolving model called Cloud Model 1 (CM1)~\citep{bry-fri2002} to
simulate a wildland plume with embers.

The model is what is termed a ``large-eddy simulation" or an LES,
meaning that while eddies and other flow structures are resolved at
scales from several grid lengths to the domain size, the grid is still
much larger than the dissipation scales of several centimeters for air.
Hence the flux of energy to those subgrid scales has to be
parameterized. It is important to recognize that individual realizations
of LES turbulence fields contain substantial stochastic variability.
Instantaneous flow features, particularly the timing, exact position,
and detailed morphology of individual vortices or eddies will vary from
one simulation to another.

The horizontal domain size in the initial reference simulation used here
is 4~km $\times$ 4~km with a horizontal resolution of 20~m. The vertical
domain is 0 to 12.5~km, with a stretched vertical resolution of
2.5-250~m~\citep{wilhelmson1982simulation}. The heat source of the fire
is represented as a constant value of sensible heat flux at the surface
of the domain ($z = 0$) throughout the entire simulation period. We
impose a fuel condition and reference value of the heat flux for a
typical large prescribed fire~\citep{frankman2013measurements}.
Specifically, a heat flux of 30~kW\.m$^{-2}$ is used in the control run
to mimic a typical prescribed fire. The fixed location of the added heat
source in the model has dimensions of 40~m $\times$ 2000~m in the $x$
and $y$ directions, respectively. All simulations employ the equilibrium
Smagorinsky subgrid-scale turbulence closure scheme (see
\citet{stevens1999large}).
Zero-flux boundary conditions are applied at both the lower and upper
boundaries for all scalars, including temperature and passive tracers.
Lateral boundaries are treated with open-radiative boundary conditions.

\begin{figure}
  \centering
  \includegraphics[width=0.8\linewidth]{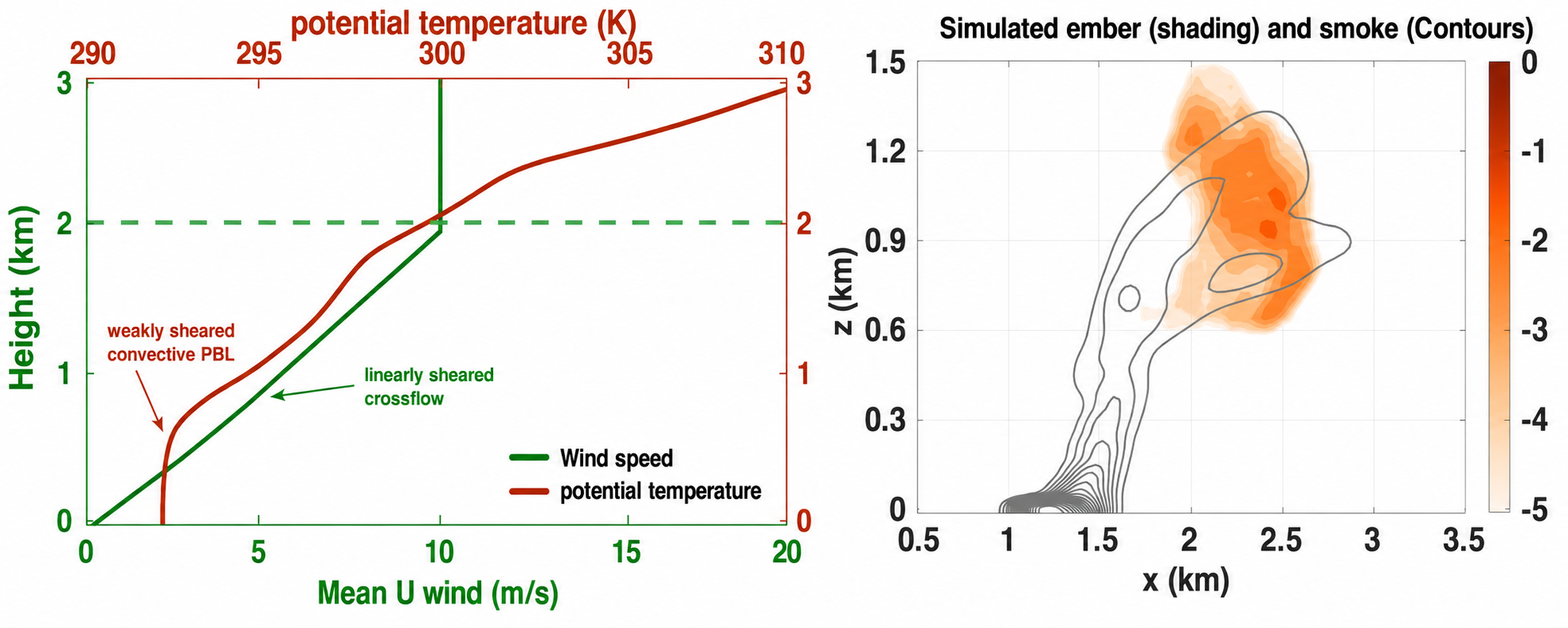}
  \caption{\label{fig:Model1} \em Background conditions used in CM1
  (left panel), and the simulated smoke concentration (contours;
  logarithmic scale) and ember particle density distribution (shading;
  logarithmic scale) (right panel).}
\end{figure}

The specified background atmospheric conditions are the primary control
on plume development and ember transport. For the example used here, the
wind is linearly increasing from the surface to 2~km, with a constant
value above 2~km. In this case, the observed and simulated plumes are
well below 2~km. A well-mixed boundary layer condition (close to neutral
stratification) is applied for the lowest 1~km, and the atmosphere above
1~km is stable. The left panel of Figure~\ref{fig:Model1} shows the
experimental setup, including the linear crossflow shear, a well-mixed
boundary layer, and then a stable boundary layer.

Embers are released in the plume at each gridpoint if the vertical
velocity exceeds a threshold, simulating the natural release and rise
from strong updrafts. The threshold is 2~m/s in these simulations. A
passive tracer representing smoke is released continuously from the heat
source.

The simulated smoke concentration and ember particle density fields
displayed in the right panel of Figure~\ref{fig:Model1} indicate that
embers closely follow the atmospheric flow represented by the smoke
plume. Owing to the strong vortical circulation within the plume head,
the smoke concentration exhibits vortex-like structures, which are also
reflected in the spatial distribution of embers. The similarity between
the simulated smoke and ember patterns illustrates the dominant role
played by plume dynamics in governing ember transport. Overall, the
simulated smoke concentration and ember distribution are consistent with
the observed plume structure.

\begin{figure}
  \centering
  \includegraphics[width=0.95\linewidth]{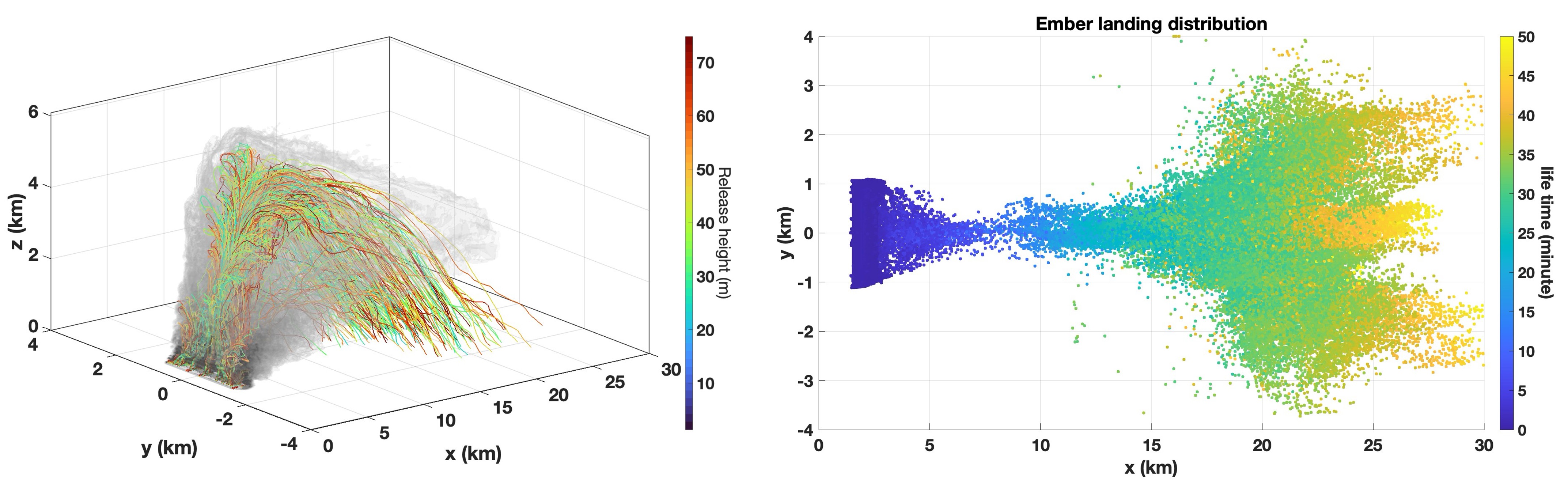}
  \caption{\label{fig:Model4} \em Simulation of a large western
  U.S.~wildland fire (the 2021 Dixie Fire; 13 July–25 October 2021). The
  left panel shows the simulated smoke concentration field together with
  three-dimensional trajectories of selected ember particles. Trajectory
  colors indicate the ember release height (m). The right panel shows
  the spatial distribution of ember landing locations, with colors
  representing the residence time (min) of each ember in the atmosphere
  prior to landing.}
\end{figure}

Additional simulations were run in larger domains to gain insight to the
larger scale and longer range ember fallout patterns of real wildland
fires. An example meant to represent the 2021 Dixie Fire shows the
complex turbulent evolution of ember trajectories
(Figure~\ref{fig:Model4}). The trajectories are lofted and intertwine
due to advection by chaotic and coherent structures generated by the
plume. The horizontal distribution of the ember landing pattern reflects
these structures and resembles a hourglass with higher and lower
concentrations depending on ember path and distance. Far-field embers
have likely exceeded typical burnout times of several minutes, but which
time varies widely for different vegetation and wind conditions. 

\begin{figure}
  \centering
  \includegraphics[width=0.95\linewidth]{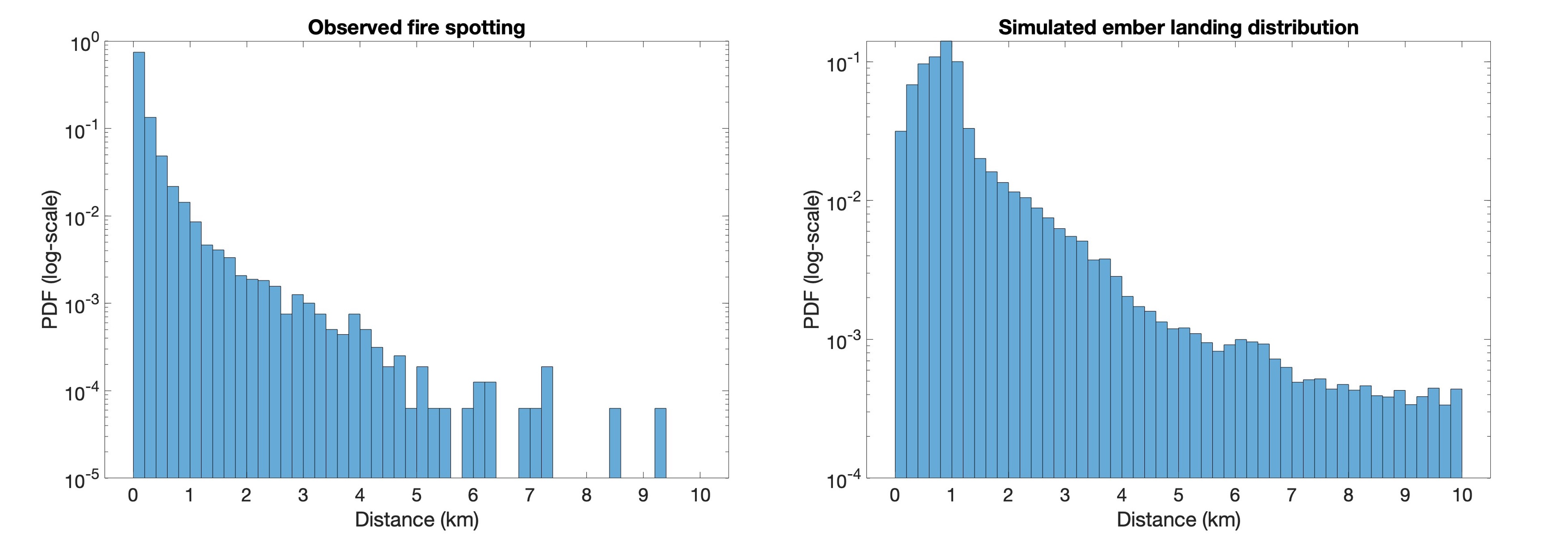}
  \caption{\label{fig:Model5} \em Comparison of the probability density
  functions (PDFs) of observed fire-spotting distances (left), derived
  from the observational studies of \citep{page2019analysis} and
  \citep{storey2020analysis}, and simulated ember landing distances from
  the example above (right). Only embers with atmospheric residence
  times shorter than 10 min are displayed in the simulated
  distribution.}
\end{figure}

A close up of the near field spotting distribution shows similar
behavior as observed distributions from wildland fire
(Figure~\ref{fig:Model5}). Note that the observed distribution is based
on actual spotting events and fires whereas the simulation is based on
landing without consideration of ignition. Ignition would only occur in
some fraction of embers but the distribution shape would not be
significantly affected in the near field. Taken together the overall
picture, even without the complication of terrain or topography, is one
in which atmospheric conditions, surface boundary layer processes,
background and plume-induced turbulence all play a strong role in ember
transport and landing patterns.

The physics-based approach provides a solid base to move forward with
simpler models. It provides statistical distributions which can be used
and tested to infer net ember transport in a wide variety of situations.
We go on to make use of some of the intuition built from the observed
and modeled statistical behavior to try to construct simplified or
idealized models of ember transport.

\section{A Statistical Approach to Ember Transport}
\label{sec:ember_stats}

Our aim now is to introduce relatively simple statistical distributions
that connect ember transport, landing distributions and fire-area
growth. Rather than resolving every ember trajectory, these approaches
treat ember locations, landing distances, or fire-front displacements as
random variables subject to physical constraints. These constraints may
arise from landing distributions generated by physics-based simulations
or observations, or from model parameters tied to physical quantities
such as background wind, turbulent diffusion, and ember burnout. The
remainder of this section presents complementary statistical viewpoints
of ember transport.

\subsection{A Stochastic Jump Model for Ember Transport}
Ember transport is strongly influenced by the background wind, but
turbulence, combustion-driven mass loss, lofting, settling, and
interactions with obstacles introduce substantial uncertainty in ember
location. This motivates the use of a stochastic model, such as a
stochastic differential equation (SDE). We let $X_t \geq 0$ be the
downwind displacement of an ember from its source at time $t$. A simple
multiplicative-noise SDE model is
\begin{align}
  dX_t = \mu X_t \, dt + \sigma X_t \, dW_t,
  \label{eqn:sde}
\end{align}
where $W_t$ is standard Brownian motion, $\mu$ is an effective drift
parameter representing systematic downwind growth of displacement, and
$\sigma$ controls the unresolved turbulent variability. The drift term
should not be interpreted as the local wind velocity, but rather as an
effective transport parameter that represents the systematic tendency of
an ember to move away from its source. The multiplicative noise implies
that uncertainty in ember position grows with displacement from the
source.

The It\^o solution of equation~\eqref{eqn:sde} is
\begin{align*}
  X_t = X_0 \, \exp\left(\left(\mu - \frac{\sigma^2}{2}\right)t +
  \sigma W_t\right).
\end{align*}
Thus, at fixed time, $X_t$ is lognormally distributed, and this provides
one possible explanation for lognormal-like ember displacement
statistics. Similar distributions have been observed in other stochastic
geophysical processes~\citep{and2021}. However, final landing distances
also depend on stopping mechanisms, including burnout, settling, and
trapping by obstacles. These mechanisms can transform the instantaneous
displacement distribution into a landing-distance distribution,
including one with exponential-like decay. Sections~\ref{sec:FP}
and~\ref{sec:survival} describe complementary mechanisms that lead to
ensemble probability densities and exponential survival laws.

\subsection{A Fokker-Planck Approach to Ember Transport}
\label{sec:FP}
Another method to model ember transport is to use the Fokker-Planck
equation. This equation has been used in many different applications and
can be used to describe the evolution of a probability distribution of
particle positions in the presence of a time and state dependent drift,
and a stochastic Gaussian diffusion that also depends on time and
state~\citep{Pavliotis2014}. Assume that a normalized ember
concentration is advecting in the positive $x$ direction with a drift
velocity $u$, there is an additive turbulent spread $D \geq 0$, and a
per-time burn-out or termination rate $\lambda$. The probability of
observing a particle at location $x$ and at time $t$ satisfies the
Fokker-Planck PDE
\begin{align}
  \pderiv{p}{t} = -u \pderiv{p}{x} + D \pderivtwo{p}{x} - \lambda p.
\end{align}
In this application it has the same form as the usual
advection-diffusion equation with decay. Both $u$ and $D$ can be
generalized to time and space dependence. Introducing a length scale
$L$, velocity scale $U$, and the time scale $T = L/U$, the dimensionless
equation is 
\begin{align}
  \pderiv{p}{t} = -\pderiv{p}{x} + 
    \frac{1}{\mathrm{Pe}} \pderivtwo{p}{x} - \Lambda  p,
\end{align}
where $\mathrm{Pe} = UL/D$ is the Peclet number and $\Lambda = \lambda
L/U = \kappa L$ is a dimensionless decay rate. 

Instead of considering the time-dependent probability density, much can
be learned by considering the steady-state of the distribution. In
nondimensional variables, the steady state solution takes the form $p(x)
= \exp(\mu x)$ where 
\begin{align}
  \mu = \frac{\mathrm{Pe}}{2} \pm \frac{\mathrm{Pe}}{2} \sqrt{1 +
  \frac{4\Lambda}{\mathrm{Pe}}}.
\end{align} 
Since $p$ is a probability density, it must decay as $x > 0$, so we
require $\mu < 0$. Therefore,
\begin{align}
  p(x) \propto \exp\left(\left(\frac{\mathrm{Pe}}{2} -
  \frac{\mathrm{Pe}}{2} \sqrt{1 +
  \frac{4\Lambda}{\mathrm{Pe}}}\right)x\right),
\end{align}
where the constant of proportionality is chosen so that $p$ is a
probability density function. An interesting point is that we have
recovered an exponential distribution. If advection dominates, then
$\mathrm{Pe} \gg 1$, and $\mu \approx -\Lambda$, so the nondimensional
decay length is $1/\Lambda$. Since $x$ was scaled by $L$, the
corresponding dimensional decay length is $L/\Lambda = U/\lambda$. In
contrast, if diffusion dominates, then $\mathrm{Pe} \ll 1$, and $\mu
\approx -\sqrt{\Lambda\mathrm{Pe}}$, and the nondimensional decay length
is $(\Lambda \mathrm{Pe})^{-1/2}$, corresponding to the dimensional
scale $\sqrt{D/\lambda}$.

Under more general conditions and with time dependence the solutions can
take a variety of forms, including error functions with various decay
rates. For example, given the linear advection-diffusion equation with
the point source initial condition, $p(x,0) = \delta(x)$,
the solution is
\begin{align}
  p(x,t) = e^{-\lambda t} \frac{1}{\sqrt{4\pi Dt}} 
    \exp\left(-\frac{(x-ut)^2}{4Dt} \right).
\end{align}
The term $e^{-\lambda t}$ captures the decay of the concentration over
time while the remaining part of $p(x,t)$
corresponds to the diffusion of the initial point source into a Gaussian
profile, and translated downwind by the advection term $ut$. This
solution might represent an initial localized source of embers that move
and spread away from the fire front. It could be used to see the time
dependent effects of a diffusive ``wave" of embers moving away from a
fire front, while the front may eventually overtake the ember cloud.
Different wind fields and boundary conditions can produce accumulation
regions, depletion zones, or more complex landing distributions.


\subsection{Ember Spread as Survival}
\label{sec:survival}

We turn to an application that focuses on ember transport near and on
the surface or ground. This can arise from the redistribution of embers
that are initially lofted, subsequently land and are then carried around
by surface transport processes. This situation will be described later
in the Chapter pertaining to structural elements blocking the wind. Or,
embers may fall directly to the ground and remain in the near-surface
environment their whole lifetime. As they are carried or advected by the
wind, ember motion is intermittent and repeatedly interrupted by
obstacles and traps of various sorts, including dynamic flow generated
stagnation or attraction manifolds. Assuming these mechanisms act
stochastically along an ember's path, this motivates a probabilistic
description of ember travel distance rather than a deterministic
transport law~\citep{qua-spe2026}.

Let $R$ be the random variable of the travel distance of an ember in an
ember storm. Then, 
\begin{align}
  S(r) = P(R > r),
\end{align}
is the probability that an ember travels farther than distance $r$, and
this is the ember's survival function. It is related to the cumulative
density function $F_R$ through the identity $S = 1 - F_R$. Next, define
the hazard of an ember ignition per unit distance 
\begin{align}
  \kappa(r) = \lim_{\Delta r \rightarrow 0} 
    \frac{P(r<R<r+\Delta r \mid R>r)}{\Delta r}.
\end{align}
Physically, $\kappa(r)$ is the instantaneous risk per unit distance that
an ember ceases to be transported due to extinction, trapping, or loss
of mobility. Taking the limit, we have
\begin{align}
  \kappa(r) = -\frac{1}{S(r)}\frac{dS}{dr},
\end{align}
resulting in a governing equation for the survival function
\begin{align}
  \frac{dS}{dr} = -\kappa(r) S(r), \quad S(0) = 1.
\end{align}
The survival function depends on how we model the hazard function
$\kappa$. If $\kappa$ is independent of distance, corresponding to the
assumption that each additional meter of travel carries the same
probability of ember termination, then the survival function is
exponential:
\begin{align}
  \label{eqn:exponential}
  S(r) = \exp(-\kappa r) \Rightarrow 
    R \sim \texttt{Exp}(\kappa).
\end{align}
If $\kappa$ depends on $r$, then
\begin{align}
  S(r) = \exp\left(-\int_{0}^r \kappa(s)\, ds \right),
\end{align}
and if $\kappa$ varies slowly with $r$, then $R$ can be approximated
with an exponential distribution.

Given the hazard function, we convert the per distance hazard ($\kappa$)
to a per time hazard ($\lambda$), which depends on the fuel type,
trapping of embers, and environmental conditions. We define
\begin{align}
  \lambda = \kappa u_{\mathrm{emb}},
\end{align}
where $u_{\mathrm{emb}}$ is comes from a simple saltation
model~\citep{qua-spe2026} for ember transport. This conversion allows
the distance-based survival model to be coupled consistently to the
time-stepping fire spread model.

The total distance traveled can be viewed as the accumulation of many
small spatial increments. The key assumption is not the detailed
statistics of individual increments themselves, but that each additional
meter of travel carries the same probability of ``survival" independent
of the distance already traveled. This corresponds to a constant hazard
per unit distance, meaning risk accumulates additively. A constant
spatial hazard uniquely leads to the exponential survival
law~\eqref{eqn:exponential}, so the total distance $R$ is exponential.
Next this exponential survival model is coupled to an idealized
fire-atmosphere solver to examine how near-surface ember wash modifies
fire geometry, first-arrival times of the fire front, and area-time
scaling.

\section{A Coupled Fire-Ember Model}
\label{sec:fire_model}
We carry on to couple the statistical model from
Section~\ref{sec:survival} with an idealized 2D model for near-surface
wind in the presence of fire~\citep{qua-spe2021}. For added background
and intuition to the development here, we briefly summarize the original
idealized model, which we previously demonstrated various effects of
fire-atmosphere interactions, but in the absence of embers. 

\subsection{The Baseline Fire-Atmosphere Model}
\begin{figure}[htp]
\centering
\includegraphics[width=0.7\textwidth]{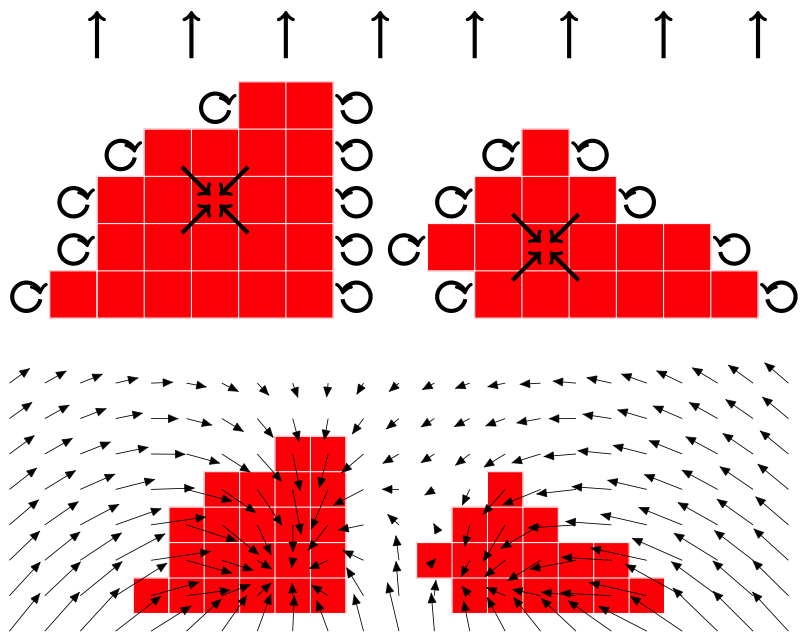}
\caption{\label{fig:smokeparcels} \em Model wind field around combusting
  elements (indicated by red cells). The atmospheric wind velocity is
  the sum of three components: a background flow, a pyrogenic potential
  or divergent term, and a vorticity component predominantly along the
  flanks.} 
\end{figure}

The model uses a 2D rasterized geometry, and each cell is always in one
of three states: actively burning; unburnt with available fuel; and,
burnt with no remaining fuel. Given a fire configuration, contributions
in the near-surface flow include a constant background wind, a divergent
flow induced by the buoyant fire plume, and a turbulent diffusion term.
The model for the divergent flow term was first introduced by
\citet{hil-sul-swe-sha-tho2018}, and they call this term the pyrogenic
potential. We also introduce positive and negative vertical vorticity
sources reflecting the buoyancy driven uplift of strong horizontal
vorticity induced by the background wind over a rough
surface~\citep{cla-jen-coe-pac1996b}. This allows for modeling
vorticity-driven effects and local fire front dispersion by coherent
structure in the boundary layer. The pyrogenic potential and vorticity
are depicted in the top of Figure~\ref{fig:smokeparcels}.

Computing the flow field requires solving the two-dimensional Poisson
equation with a known (and evolving) forcing function. We use the
standard second-order central difference formula to discretize the
Laplacian, and we impose a constant Neumann boundary condition an solve
the resulting sparse linear system with Matlab's backslash operator.
The velocity field is depicted in the bottom of
Figure~\ref{fig:smokeparcels}. The wind in our CA model is kinematic,
and satisfies mass conservation but not a full set of momentum
equations. Its role is to act as a ``convective agent'' for the
probability of fire in a cell.

Once the total velocity field is calculated, new cells are ignited by
applying Bresenham's line algorithm to carry the fire downwind. In
addition, we stop igniting cells along the line if it reaches a cell
that is either combusting or has no remaining fuel. Turbulent diffusion
is incorporated into fire spread to represent the fluctuations in hot
gases and flames at the near-surface boundary due to small-scale
horizontal turbulent motion of the wind within and above the fuel. This
stochastic term models the statistics of a multitude of processes and
interactions in the surface boundary layer ahead of the fire front.
Finally, burning cells are extinguished after they have combusted for a
specified amount of time.

In our setup, the cell size is 1~m$^{2}$, and the time evolution occurs
over minutes to hours. The full model domain is flat and 200~m on each
side. To minimize edge effects, we remove the first 5~m of fuels around
the entire perimeter of the domain. The only fuel parameter in the model
is a constant burn time which depends on the fuel and combustion process
and is regarded here as an observed parameter. We use a value of 15~s
which we estimated from
observations~\citep{cur-spe-hie-obr-goo-qua2018}. While the range of
this quantity can be large, mean values of a few seconds to minutes are
typical in light to moderate fuels. The strength of the pyrogenic source
is related to fire intensity by standard plume scaling. A uniform
distribution with mean value 0.4 is used to defined the stochastic term
that allows cells to ignite neighboring cells the time step before the
original cells extinguish. This diffusion-based ignition is independent
of the convection-based ignition due to wind. 

A suite of runs that demonstrate various effects of fire-atmosphere
interactions on spread rate and geometry of the fire has been carried
out~\citep{qua-spe2021}. Standard phenomena such as a parabolic fire
front shape, accelerated merging of flank fires, fingering, and other
effects are well represented in the idealized model.

\begin{figure}[htp]
  \centering
  \includegraphics[width=0.45\textwidth]{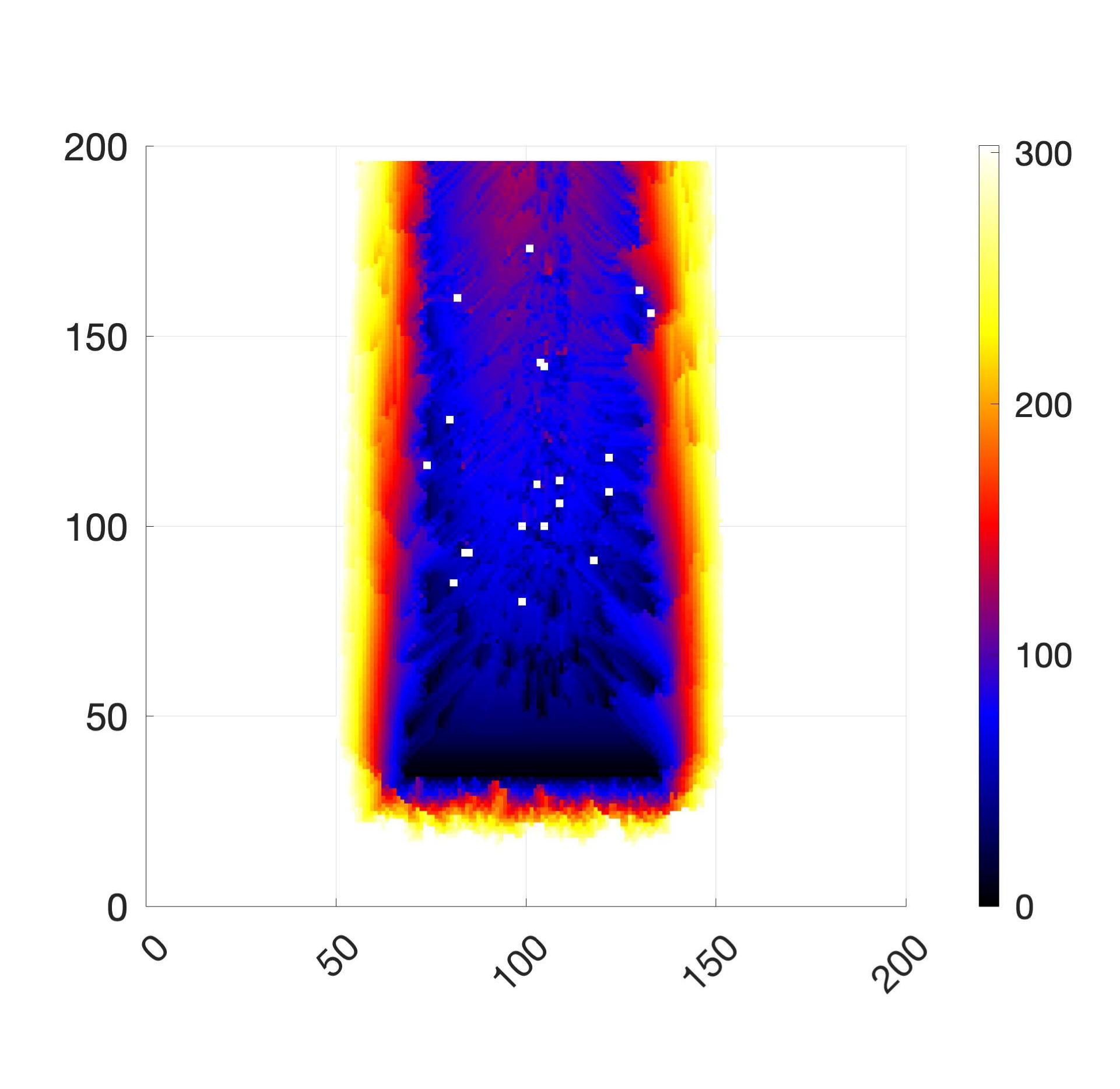}
  \includegraphics[width=0.45\textwidth]{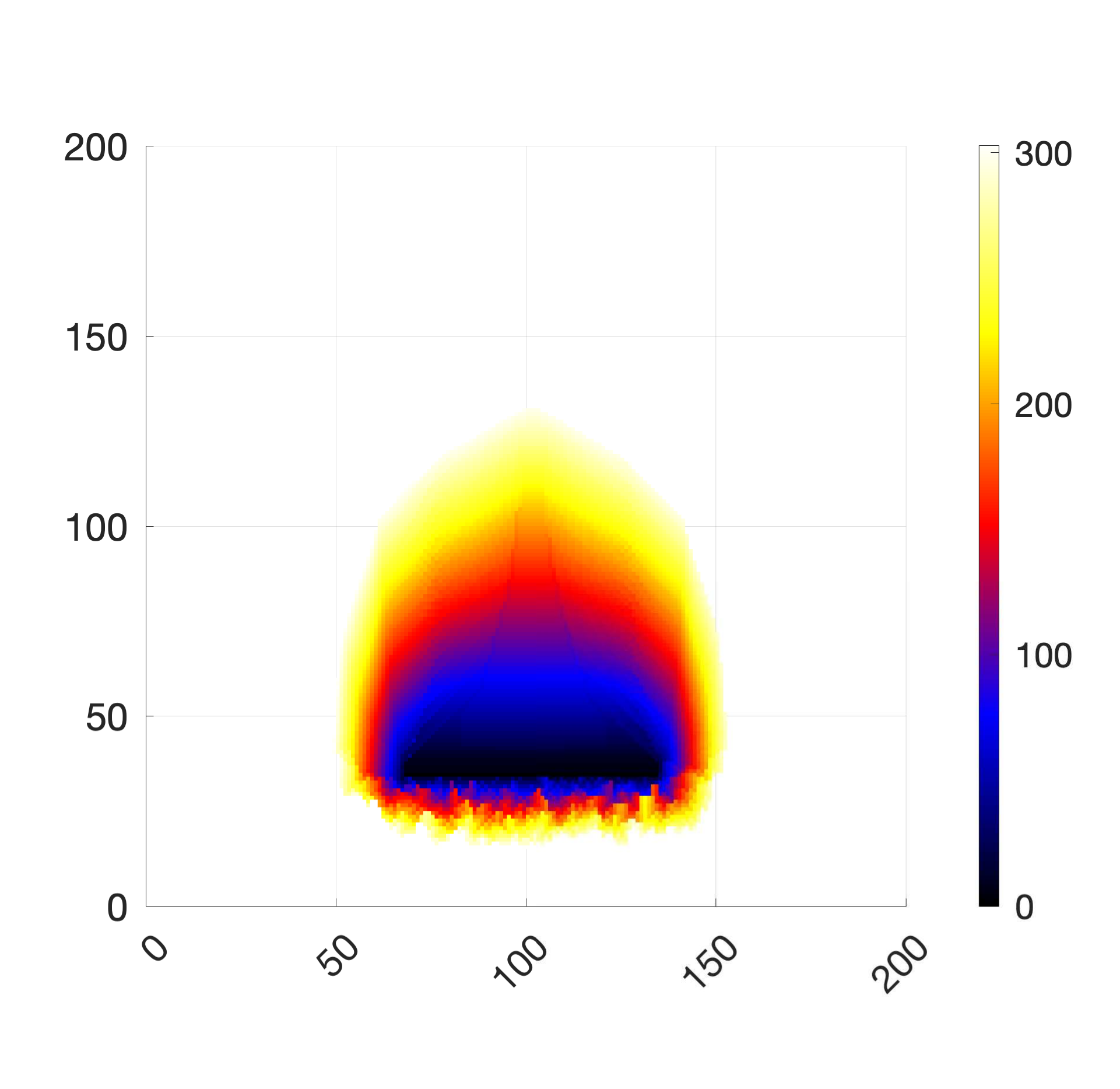}
  \caption{\label{fig:spotting_FAT} \em A representation of fire
  behavior with (left) and without (right) embers as a first arrival
  time map.} 
\end{figure}

There are several ways that the model can be used to characterize fire
behavior. For example, the first arrival time provides a spatial
description of the first time that each pixel first sees fire. This
first arrival time can also be parameterized using neural network
architectures~\citep{ton-qua2025}. Examples of first arrival time maps
both with and without ember spread are shown in
Figure~\ref{fig:spotting_FAT}. However, to draw comparisons to
observations, we study the behavior of the burn scar's area. To
nondimensionalize the burn scar's area, we consider the ratio of $A(t)$
to the total amount of available fuel. Instead of introducing a new
variable, we simply understand that $A(t) \in [0,1]$, with $A(t) = 1$
meaning that all fuel has been consumed. Other ways to normalize area
and compare RoS and area to models have been
proposed~\citep{coe-cru-ros-spe2022, cru-ale2013}. 

We next relax the assumption that fire spread occurs only through local
convective and diffusive propagation and extend the baseline model by
adding ember transport. The presence of embers allows burning cells to
ignite new cells downwind through ember spotting
(Section~\ref{sec:EmberSpotting}) and ember wash
(Section~\ref{sec:wash}).

\subsection{Extension to Lofted Spotting}
\label{sec:EmberSpotting}

To begin, in order to understand the effects of long-range spotting, we extend this model with a simple statistical model for embers. We consider an idealized spotting model to illustrate how embers may be distributed down the background or ambient wind when they are lofted, and the resulting flow complexity induced by the fire. The development is
similar to that of~\citep{sch1969}, who used forms of probability
distributions for ignition based on firebrand size, moisture, excess
heat to build an index of spotting fire risk. Spotting in this form produces distinct fires well-separated from the source fire.

Spot fires are introduced downwind using the background flow based on a
Gaussian-distributed lifetime~\citep{tohidi2017stochastic} with mean
$\mu$ and standard deviation $\sigma$. In all simulations we use $\mu$ =
20~s and $\sigma$ = 5~s. Because embers responsible for long-range
spotting are lofted high in the atmosphere, they are transported by the
background flow and not the total surface wind that includes
fire-induced flow. The reason for this is that a fully 3D model
(discussed in Sections~\ref{sec:LES} and~\ref{sec:WUI}) would be
required to capture the atmospheric boundary layer effects including
plume lofting and turbulence, whereas in the 2D model the background
wind is used to represent flow aloft. The background winds may be as
complex as desired but for present purposes are represented by uniform
flow. The background wind speed is 4~m/s in this example to match ember
flight excursions to the domain size, but can be scaled up arbitrarily to larger domains. Here, the ember trajectories are determined by the background wind, with the plume effects secondary once the embers have been released by the plume into the surrounding environment, well above the surface boundary layer.

\begin{figure}[htp]
  \centering
  \includegraphics[width=0.243\textwidth,trim=0.5cm 0.5cm 1.5cm
  1.0cm,clip=true]{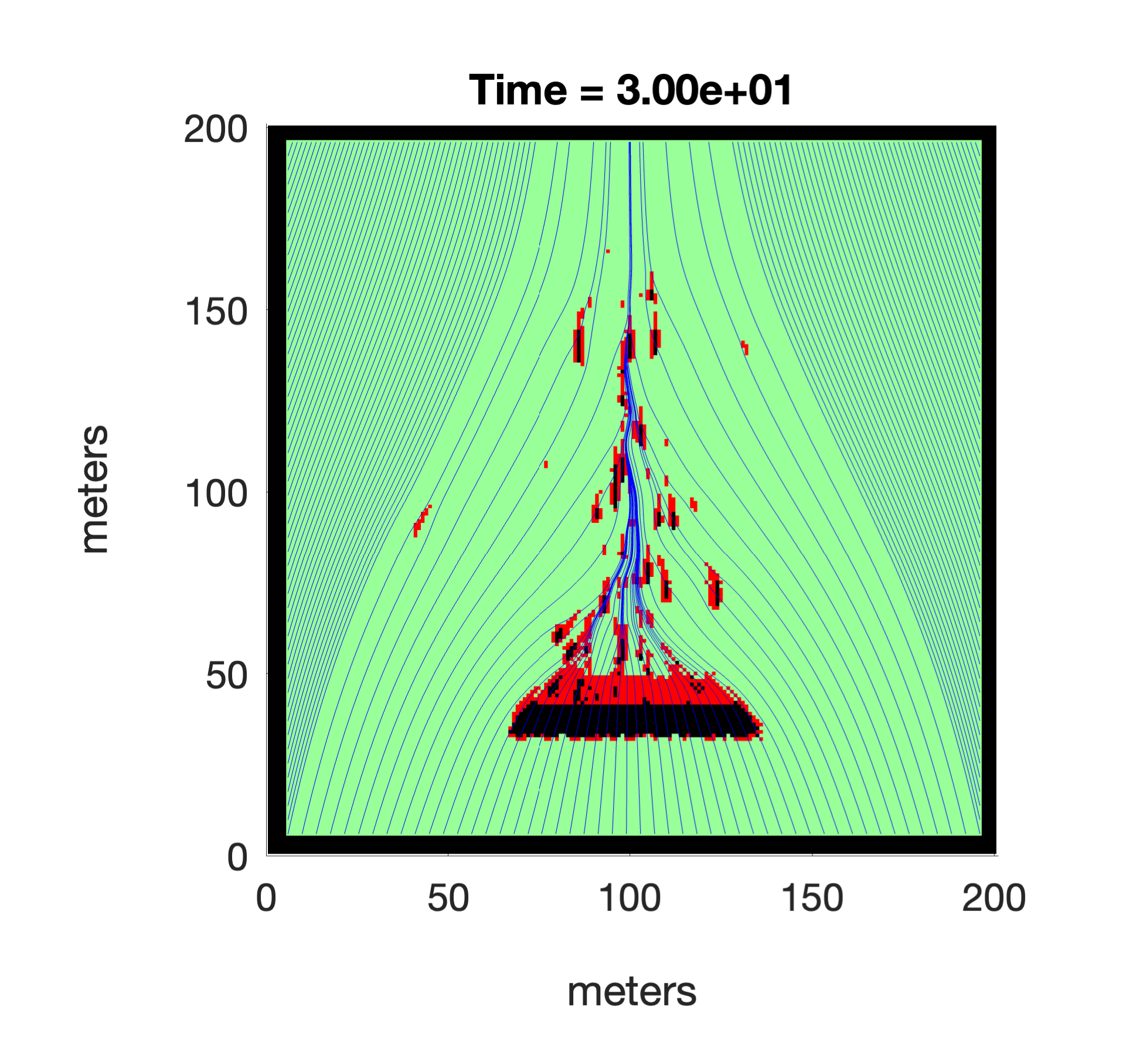}
  \includegraphics[width=0.243\textwidth,trim=0.5cm 0.5cm 1.5cm
  1.0cm,clip=true]{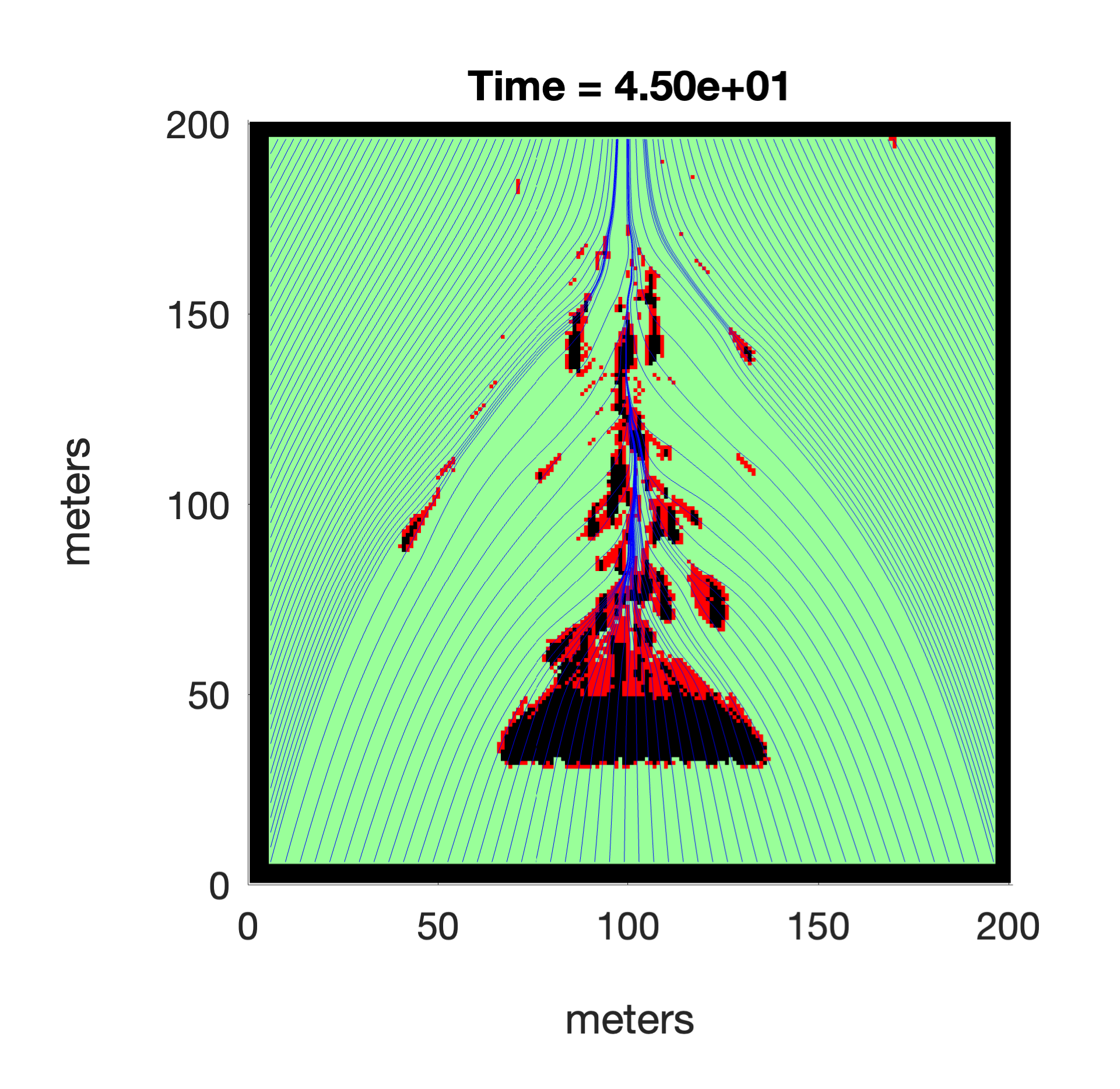}
  \includegraphics[width=0.243\textwidth,trim=0.5cm 0.5cm 1.5cm
  1.0cm,clip=true]{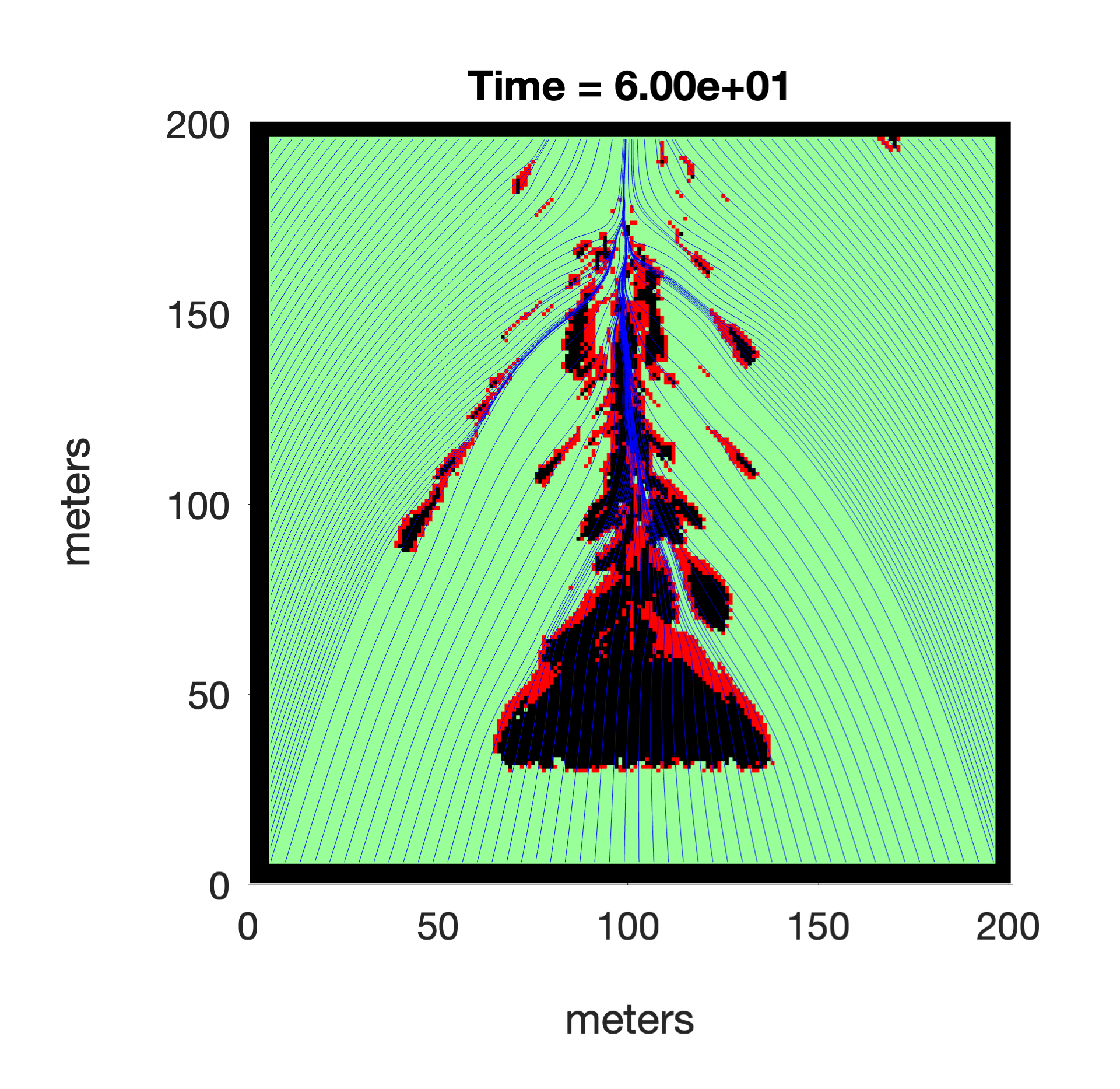}
  \includegraphics[width=0.243\textwidth,trim=0.5cm 0.5cm 1.5cm
  1.0cm,clip=true]{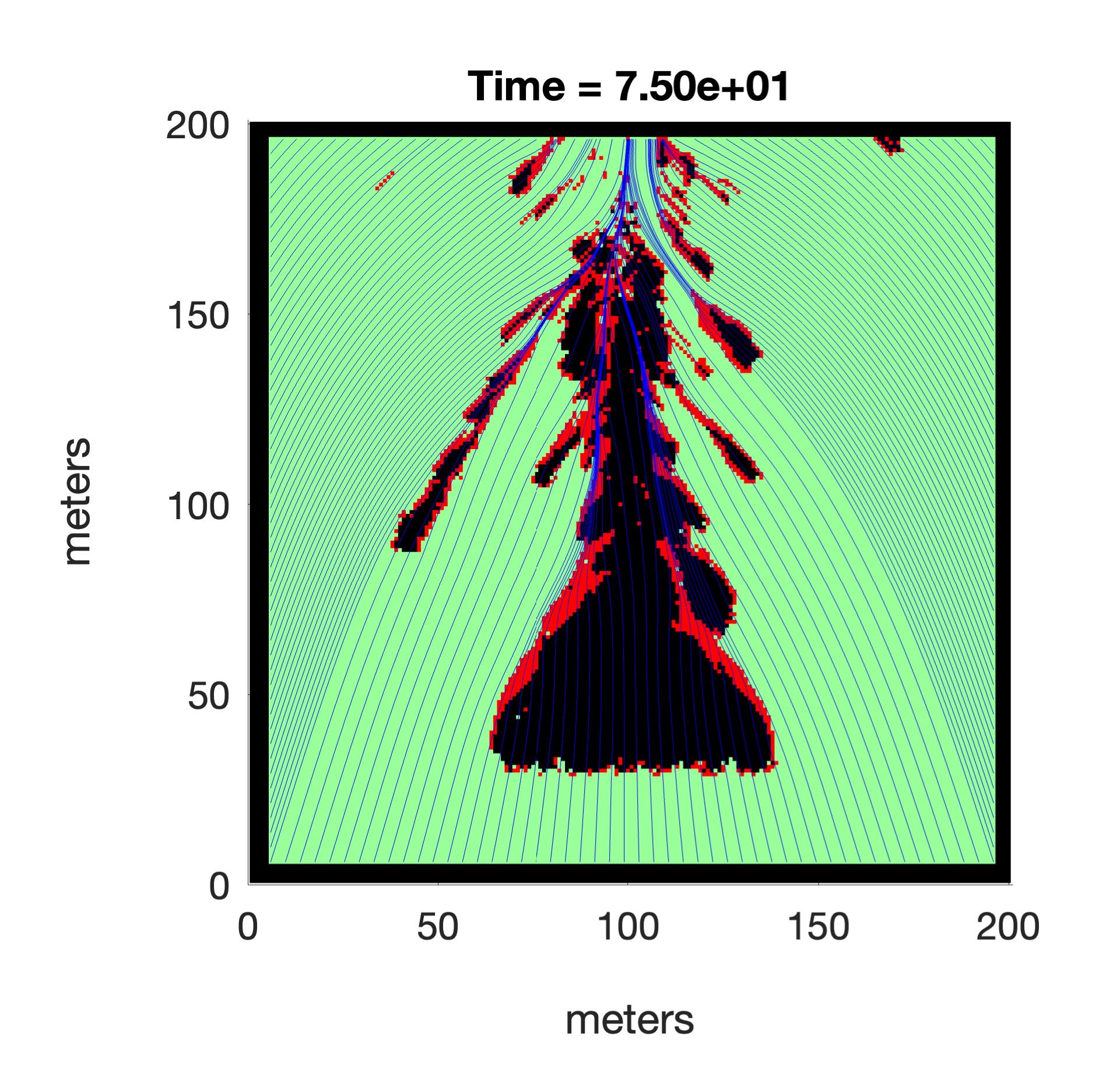}
  \\
  \includegraphics[width=0.243\textwidth,trim=0.5cm 0.5cm 1.5cm
  1.0cm,clip=true]{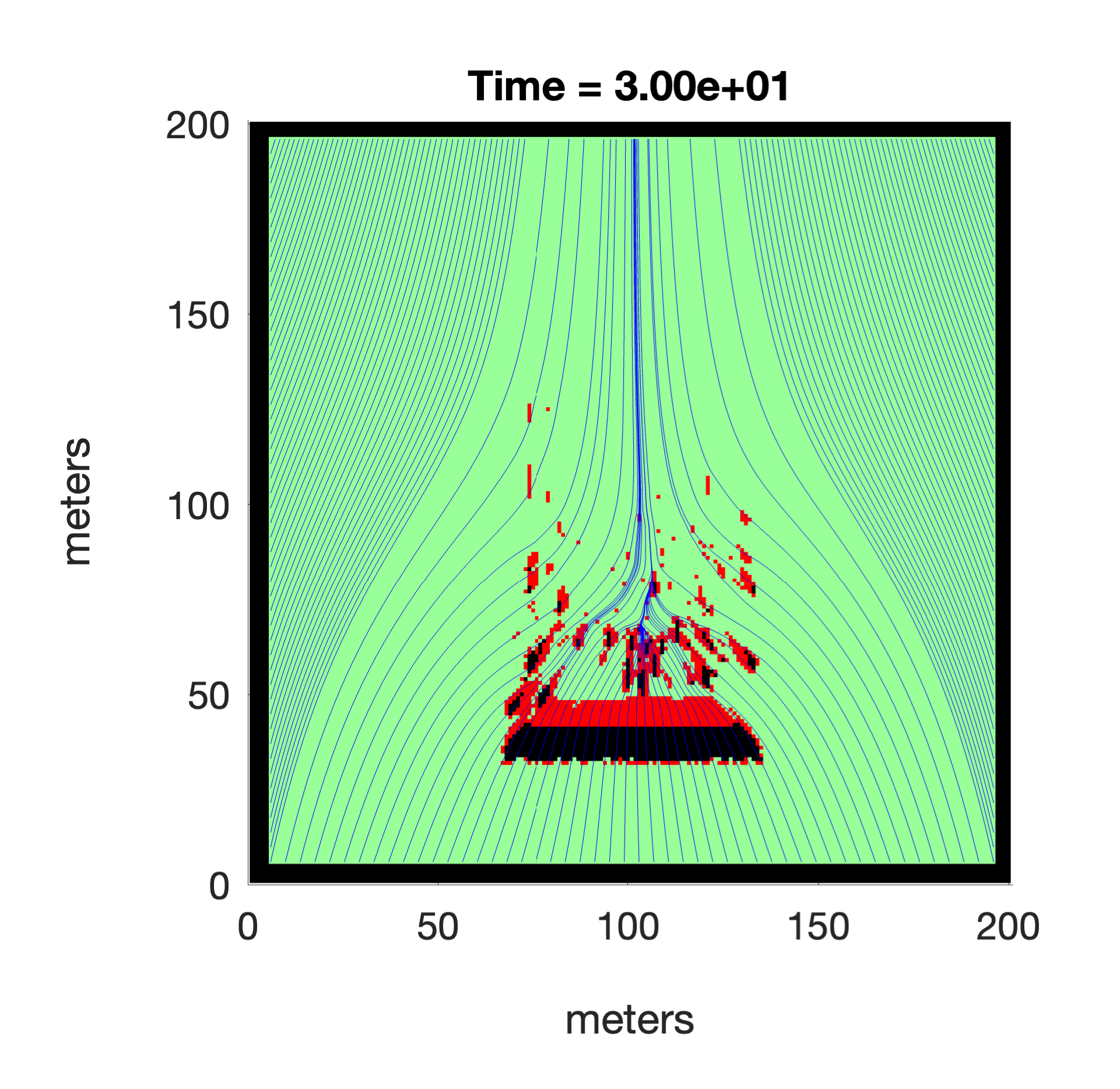}
  \includegraphics[width=0.243\textwidth,trim=0.5cm 0.5cm 1.5cm
  1.0cm,clip=true]{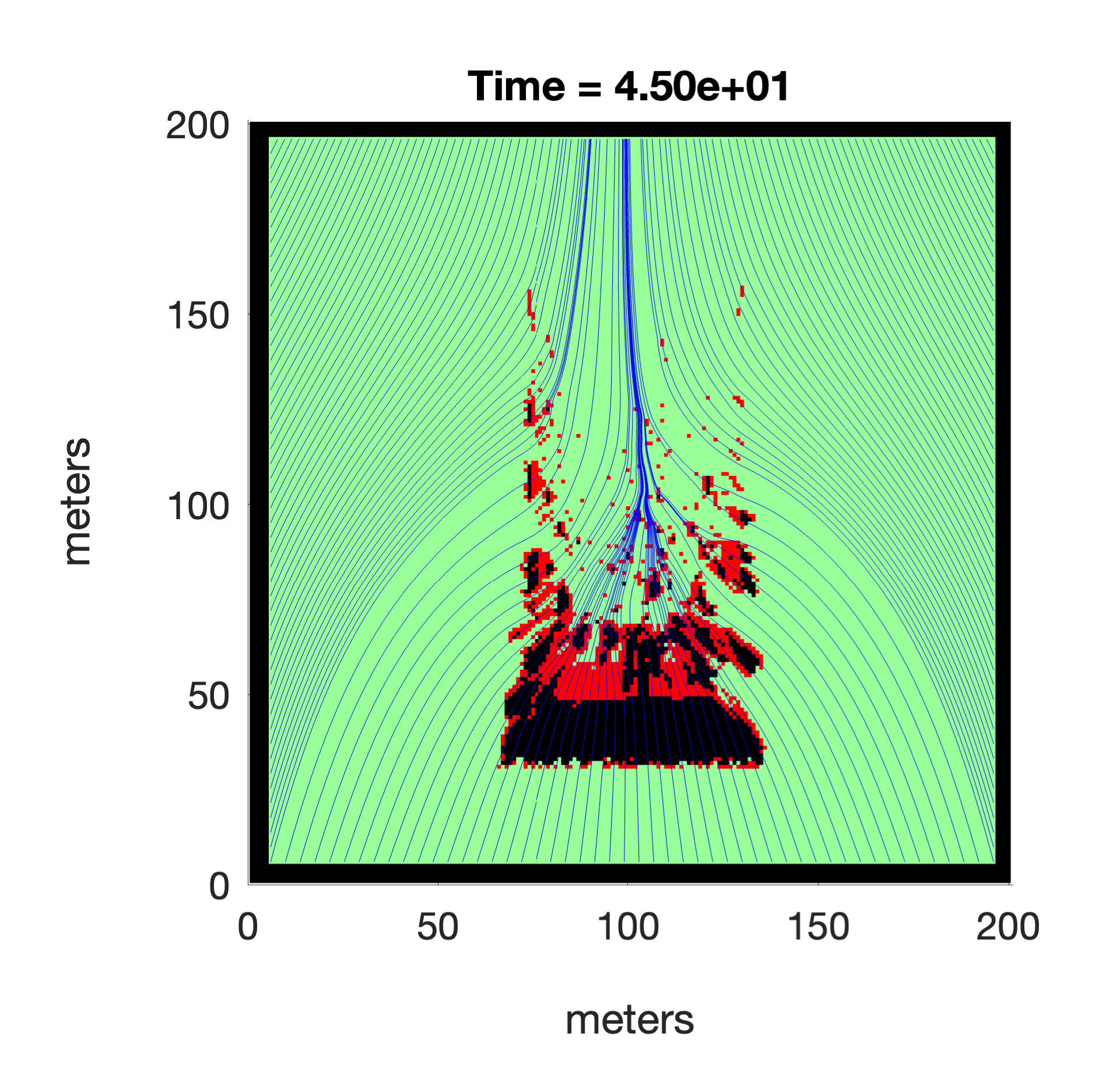}
  \includegraphics[width=0.243\textwidth,trim=0.5cm 0.5cm 1.5cm
  1.0cm,clip=true]{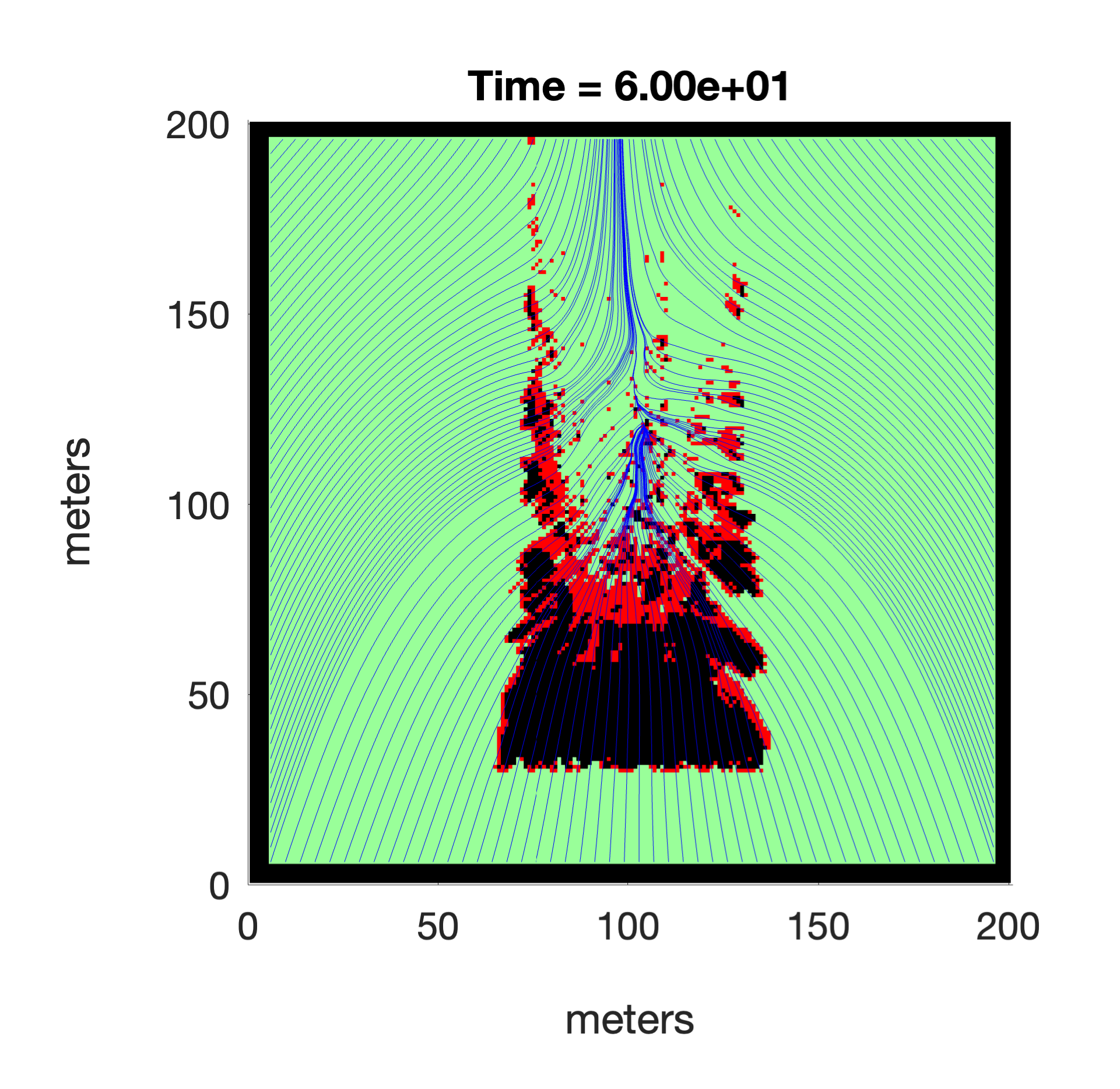}
  \includegraphics[width=0.243\textwidth,trim=0.5cm 0.5cm 1.5cm
  1.0cm,clip=true]{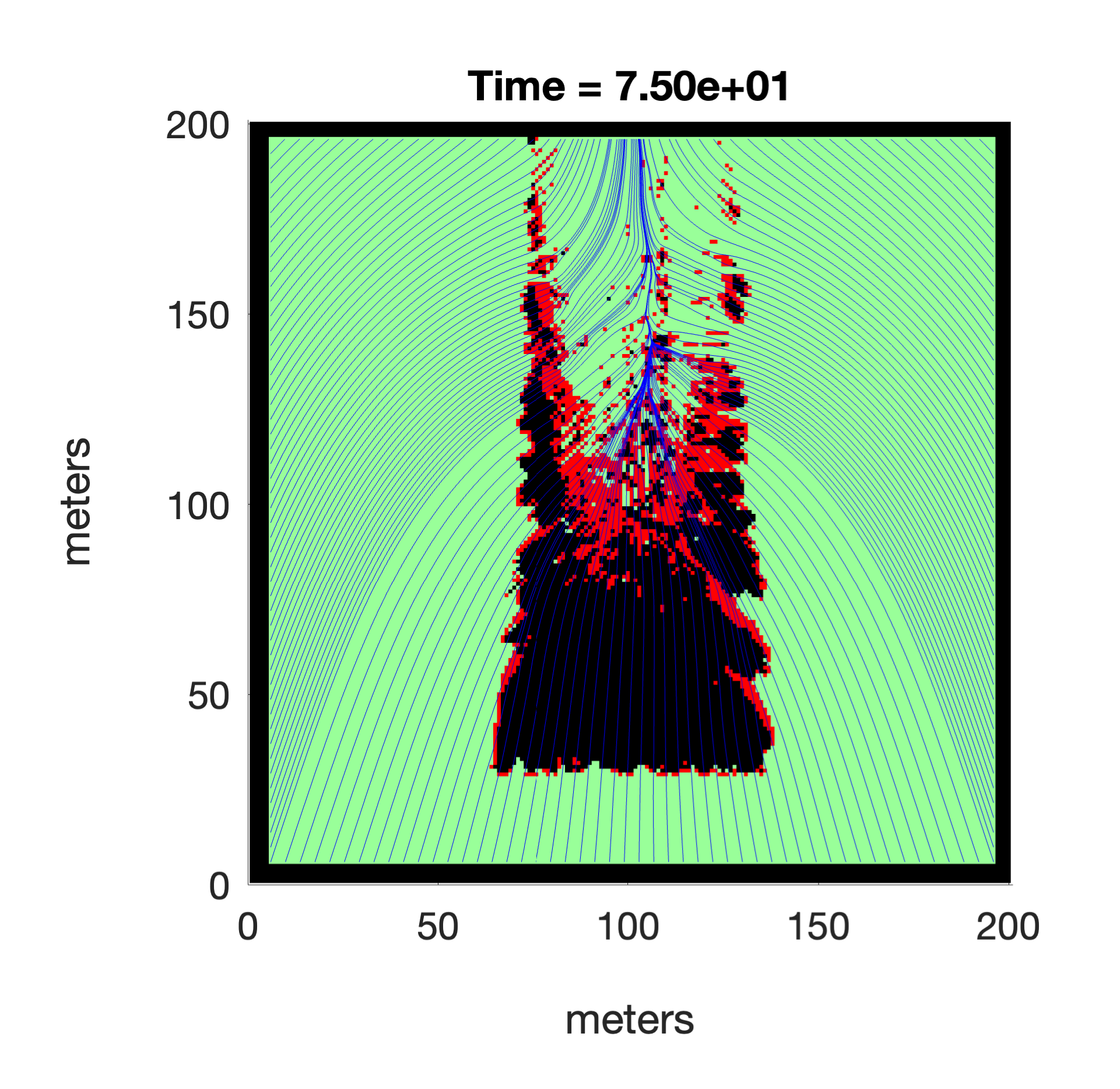}
  \\
  \includegraphics[width=0.243\textwidth,trim=0.5cm 0.5cm 1.5cm
  1.0cm, clip=true]{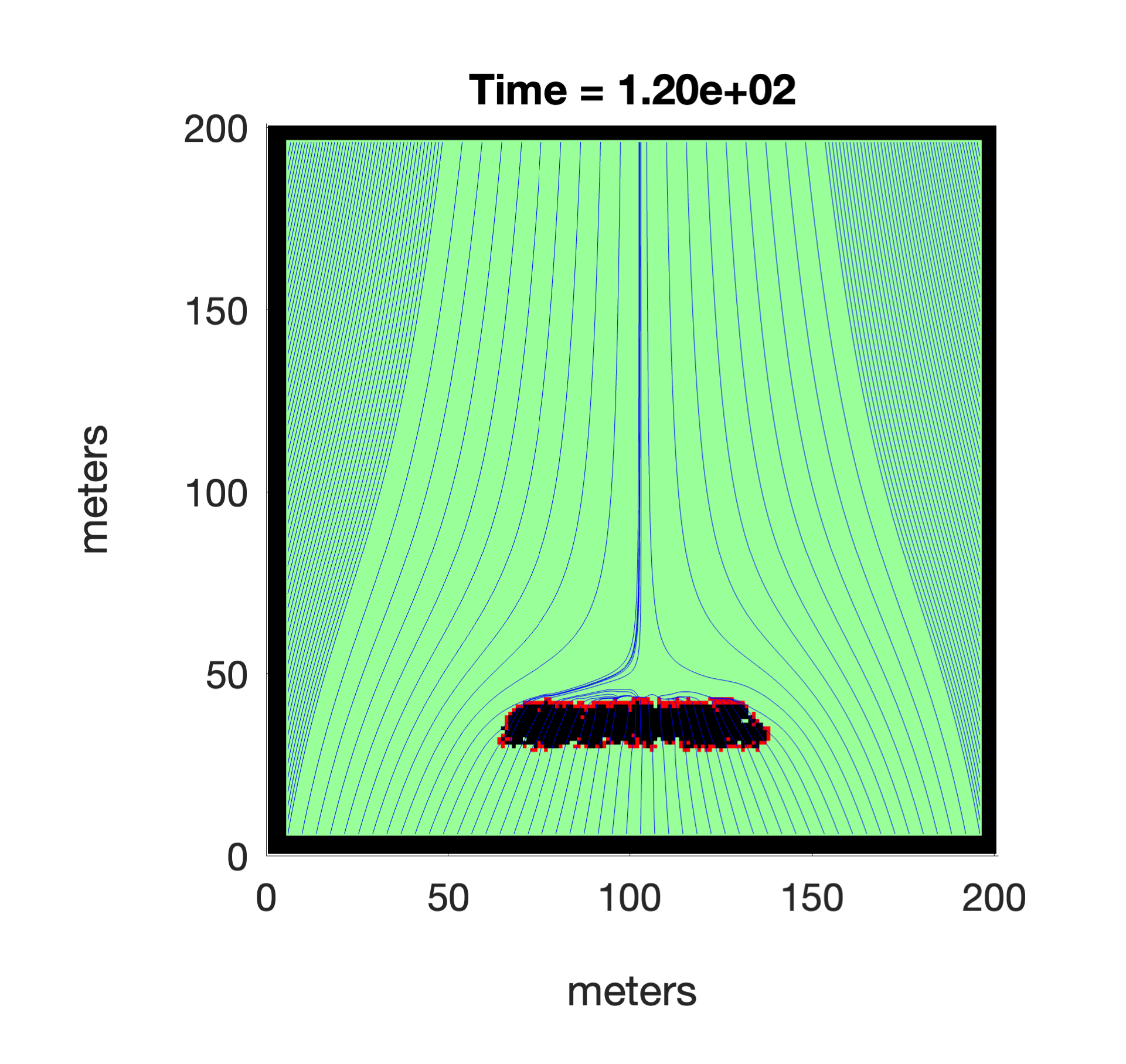}
  \includegraphics[width=0.243\textwidth,trim=0.5cm 0.5cm 1.5cm
  1.0cm, clip=true]{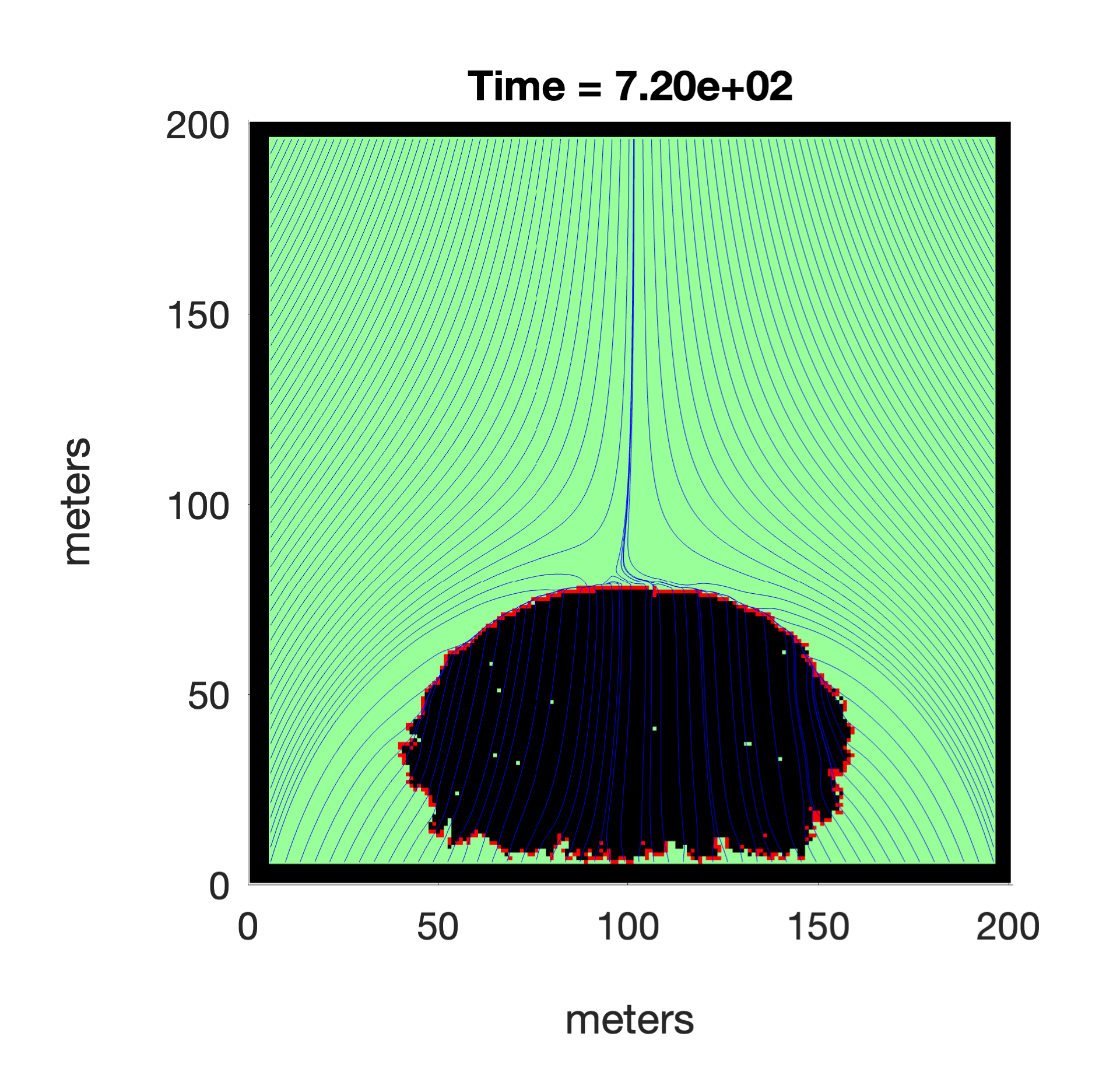}
  \includegraphics[width=0.243\textwidth,trim=0.5cm 0.5cm 1.5cm
  1.0cm, clip=true]{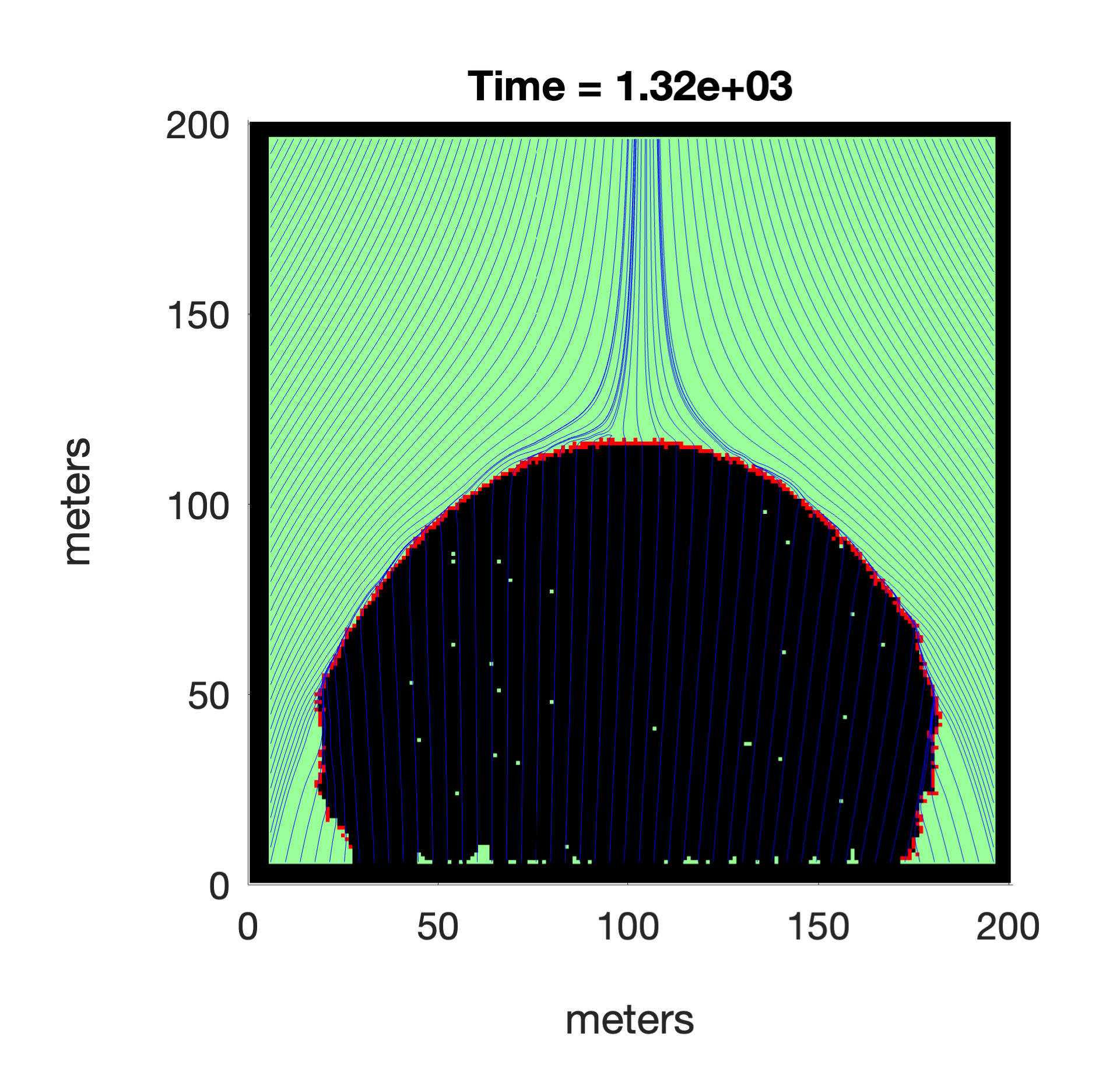}
  \includegraphics[width=0.243\textwidth,trim=0.5cm 0.5cm 1.5cm
  1.0cm, clip=true]{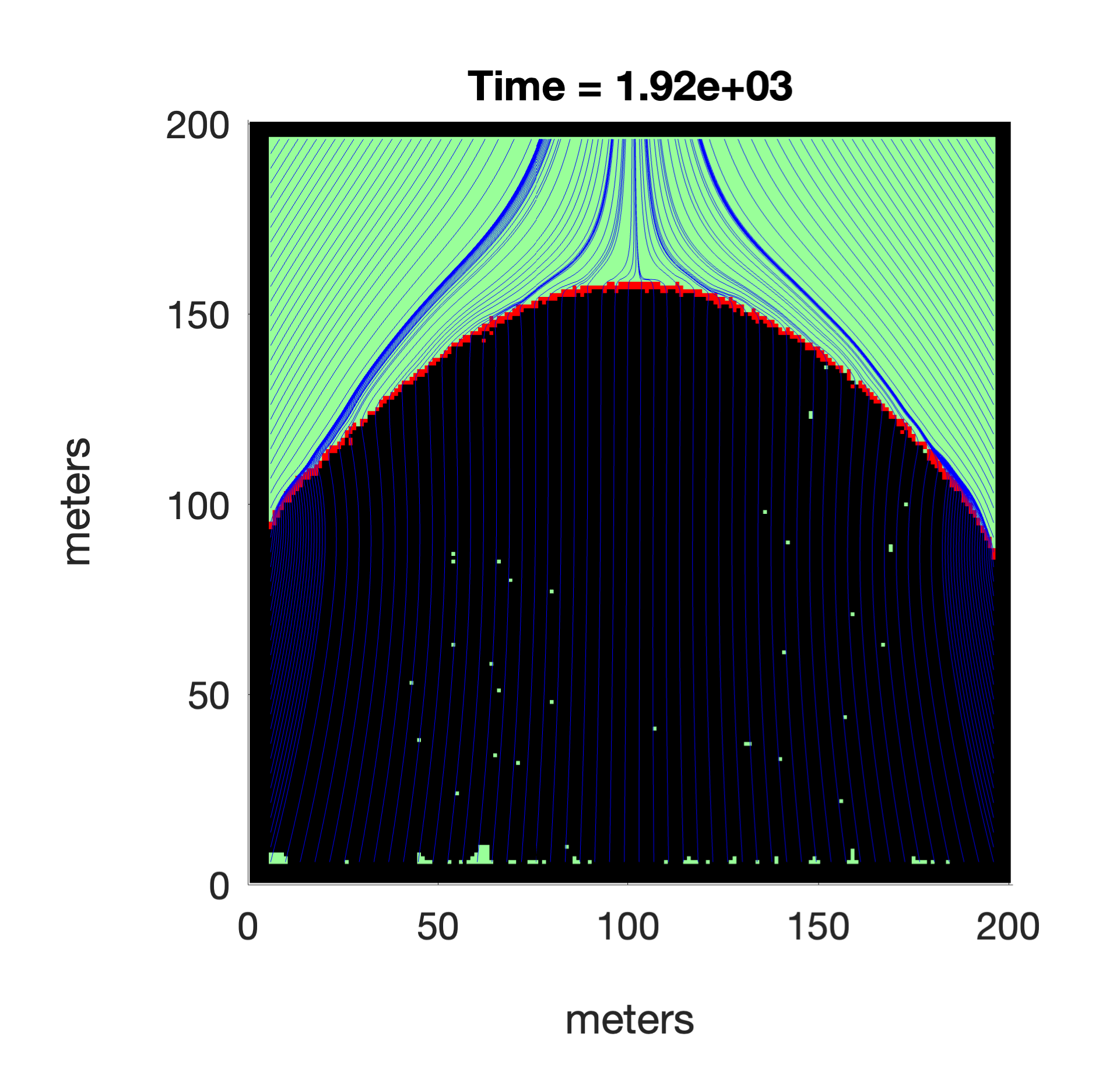}
  \caption{\label{fig:visualsegment2} \em An idealized example of fire
  spread due to spotting (top two rows) and without ember-driven spread
  (bottom row). The ember flight time is normally distributed with a
  mean of $\mu$ = 40~s (top row), $\mu$ = 20~s (middle row), both with a
  standard deviation of $\sigma$ = 5~s, and a probability of ignition of
  $\pig = 0.1$. The ignition pattern is a head fire. Streamlines of the
  wind (solid curves) indicate the interaction of the constant
  background wind from the south at 4~m/s with the fire-induced wind.
  Note that the ember velocity is the constant background velocity.}
\end{figure}

Once the ember ``lands" it may or may not start a fire. The ignition
probability of the spot is given by a Bernoulli distribution with a
probability of ignition of $\pig$. This statistical representation is a
useful idealization of the overall spotting process that reflects actual
physical processes in an idealized manner. The subsequent spread of the
original fire is strongly influenced by the presence of spotting, which,
as the spot fire grows, draws air in toward the new fire and
dramatically changes the local flow. Figure~\ref{fig:visualsegment2}
illustrates the fire progression both with spotting (top two rows), and
without spotting (bottom row). The lifetimes of the embers are normally
distributed with a mean of 40~s (top row) and 20~s (middle row), both
with a standard deviation of 5~s. In both cases, the probability of
ignition is $\pig = 0.1$. Higher probabilities of ignition produce too
much fire with new fronts emerging rapidly mirroring the initial front;
lower probabilities produce infrequent spotting and longer runs are
needed in this case to determine the effects. Because of the significant
difference in the spread rate, the time stamps of the ember and no ember
simulations are not aligned.

\subsection{Extension to Surface Ember Transport}
\label{sec:wash}
Next we model the surface ember transport as movement directed down the
local near-surface wind. The physical transport process is substantially
different from spotting. In contrast to the typical longer range
spotting process, ember wash as described here spreads near the surface
rather than aloft. Some of this spreading occurs in a flow-following
inertial particle mode, but also in a tumbling mode of transport with
embers and debris released and falling, bouncing, and rolling along the
ground, and then being resuspended and carried rapidly forward again in
wind gusts. A wind turbulence-driven suspended component probably
includes a mix of very short range spotting and resuspended embers that
have fallen near the fire front.

Several new parameters are used to introduce ember wash into the model.
First, embers require sufficiently large surface velocity to break free
from vegetation. Since our model calculates the near-surface horizontal
velocity at each cell in the grid, we use a single threshold of 0.2~m/s
for this threshold. Smaller thresholds result in all combusting cells
launching embers, while larger thresholds result in no ember generation.
Inspired by the survival function~\eqref{eqn:exponential}, once an ember
is created, we assume it travels with the near-surface velocity for a
time $\tau$, where $\tau$ is exponentially distributed with a mean value
of $\mu$. To model this transport, we modify the Bresenham ignition
line. The direction of the ignition line is now oriented along the
surface flow; however, the distance $r$ at which cells are ignited is
determined by the spatial distribution function described above.
Finally, once the ember lands, a Bernoulli probability of ignition,
$\pig$, determines if the ember ignites the new cell.

The number of firebrands launched, or the firebrand generation rate, is
somewhat arbitrarily chosen to be 1/s/cell (cells here are 1~m $\times$
1~m). For comparison, \citet{koo-pag-wei-woy2010} describe modeling
studies with a 1~ember/s/cell where the cell size is 2~m $\times$ 2~m.
Relating these numbers to generation rates in large fires is
challenging. \citet{man-mar-mel2007} determined ember generation rates
from real trees, and using their results one may roughly infer rates of
ember generation in heavy fuel model loads of 2~embers/s/m of linear
fire front perimeter, depending on the fuel load, type, ember density,
rate of spread, and other factors not well determined instantaneously in
actual fires such as local turbulence levels.

We also note that actual observations of spotting contain modes of
exponential-like behavior~\citep{storey2020analysis} with measured mean
spotting distances within 600~m from the main fire front. They pointed
out that the exponentially distributed behavior was not usually
associated with the long distance spotting, which is consistent with a
different mechanism for transport in long and short-range spotting.
\citet{page2019analysis} also found exponential behavior within a few
hundred meters of 48 analyzed Rocky Mountain fires. As described
earlier, the distinction between ember wash and spotting becomes fuzzy
at close distances hence the spot fire analyses likely have an ember
wash component close to the fire front. \citet{page2019analysis} also
discussed how the spotting distance analysis itself has limits of
resolution since the main fire front may overtake the spot fire in the
time between infrared images used to determine the presence of spot
fires. They added a correction factor (ranging from about 50~m to 400~m)
to try to account for this effect, but this did not modify their results
significantly. Higher time resolution observations will provide
greater insight to the close-range nature of spotting.  Eventually, the
mean flight time parameter $\mu$ may be estimated from such high
resolution ``spotting'' data, but for now the results from the large
scale analysis are very helpful to provide bounds on this parameter, at
least until more appropriate data become available.

\begin{figure}[htp]
  \centering
  \includegraphics[width=0.243\textwidth,trim=0.5cm 0.5cm 1.5cm
  1.0cm, clip=true]{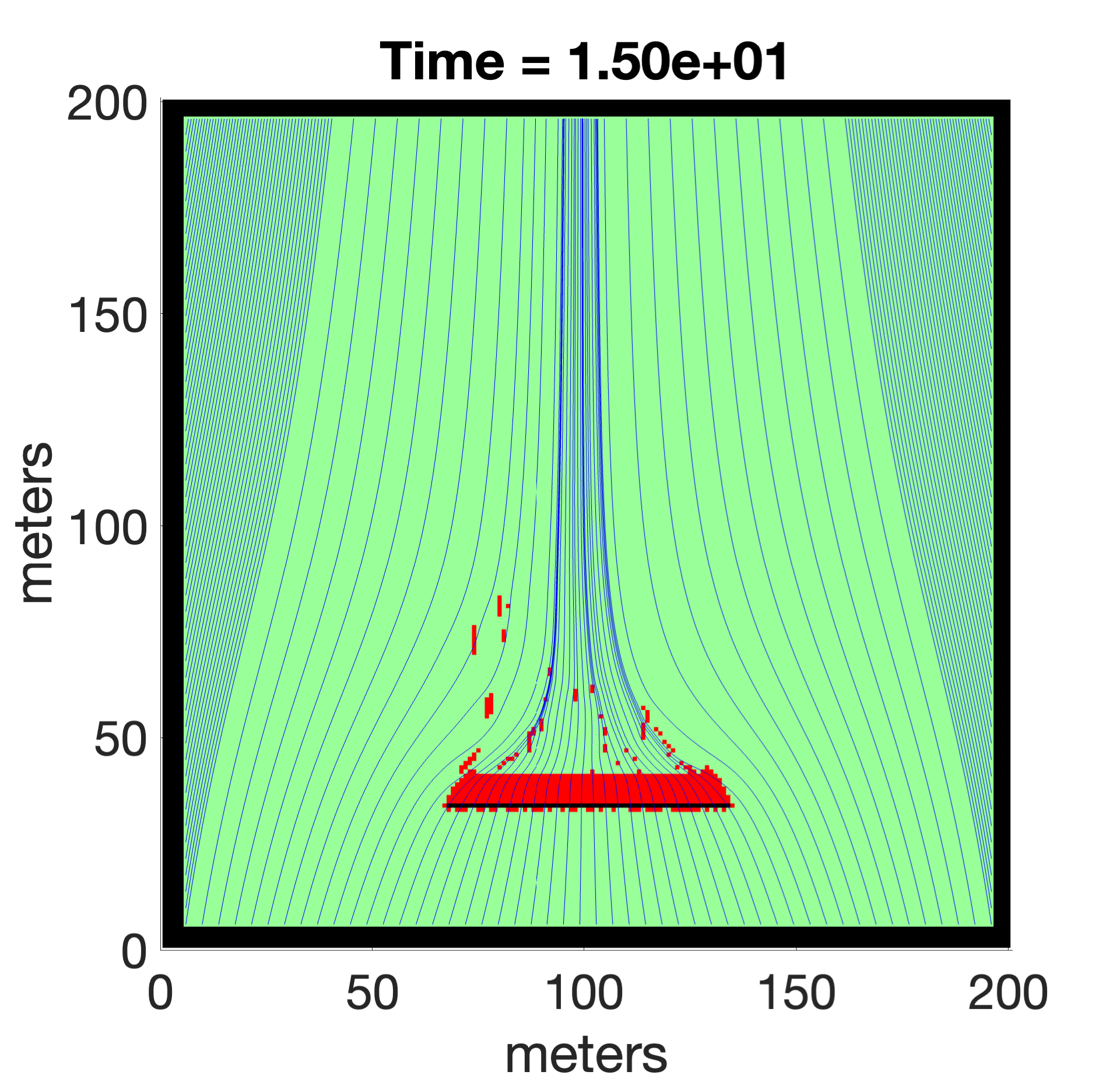}
  \includegraphics[width=0.243\textwidth,trim=0.5cm 0.5cm 1.5cm
  1.0cm, clip=true]{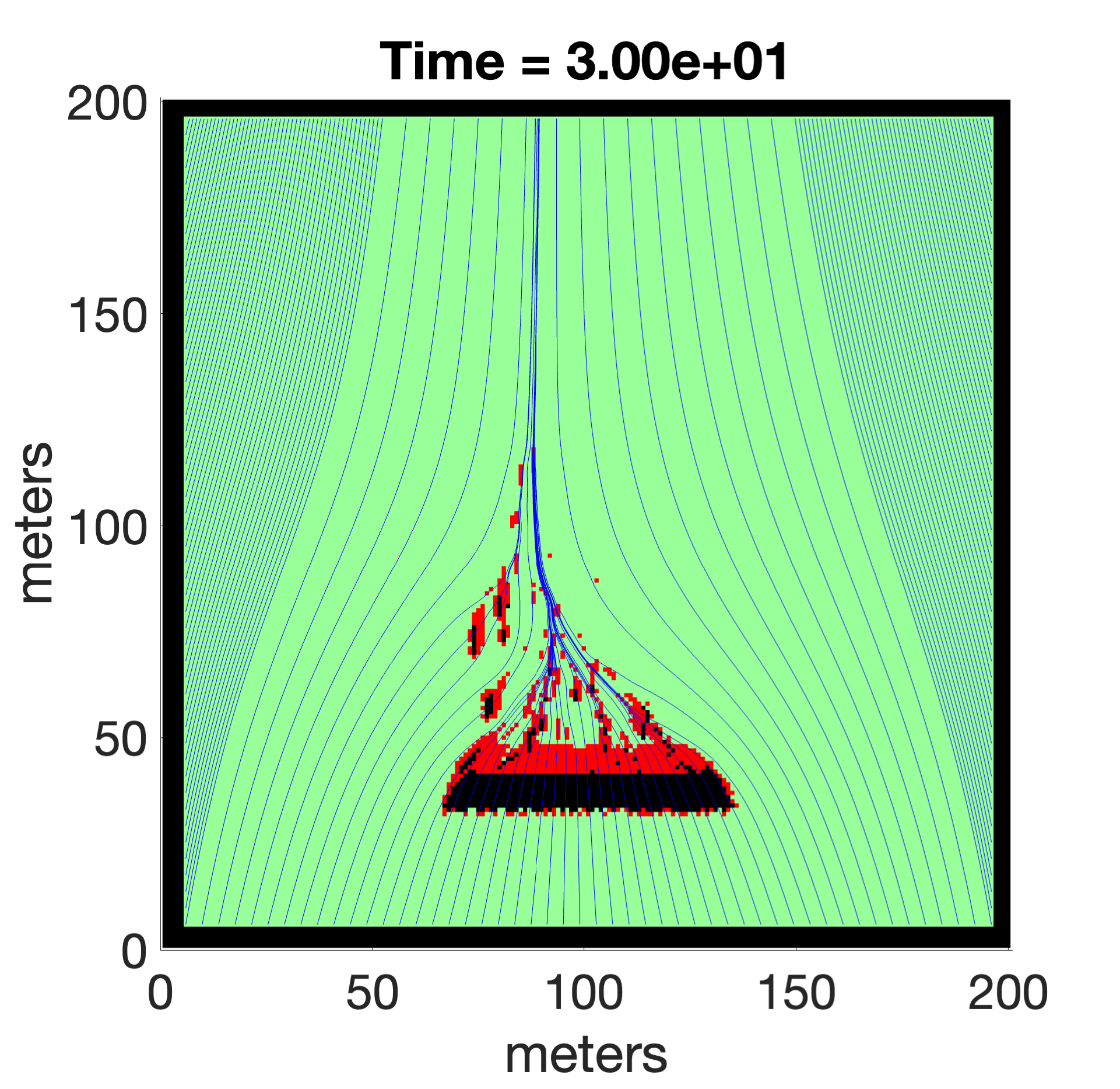}
  \includegraphics[width=0.243\textwidth,trim=0.5cm 0.5cm 1.5cm
  1.0cm, clip=true]{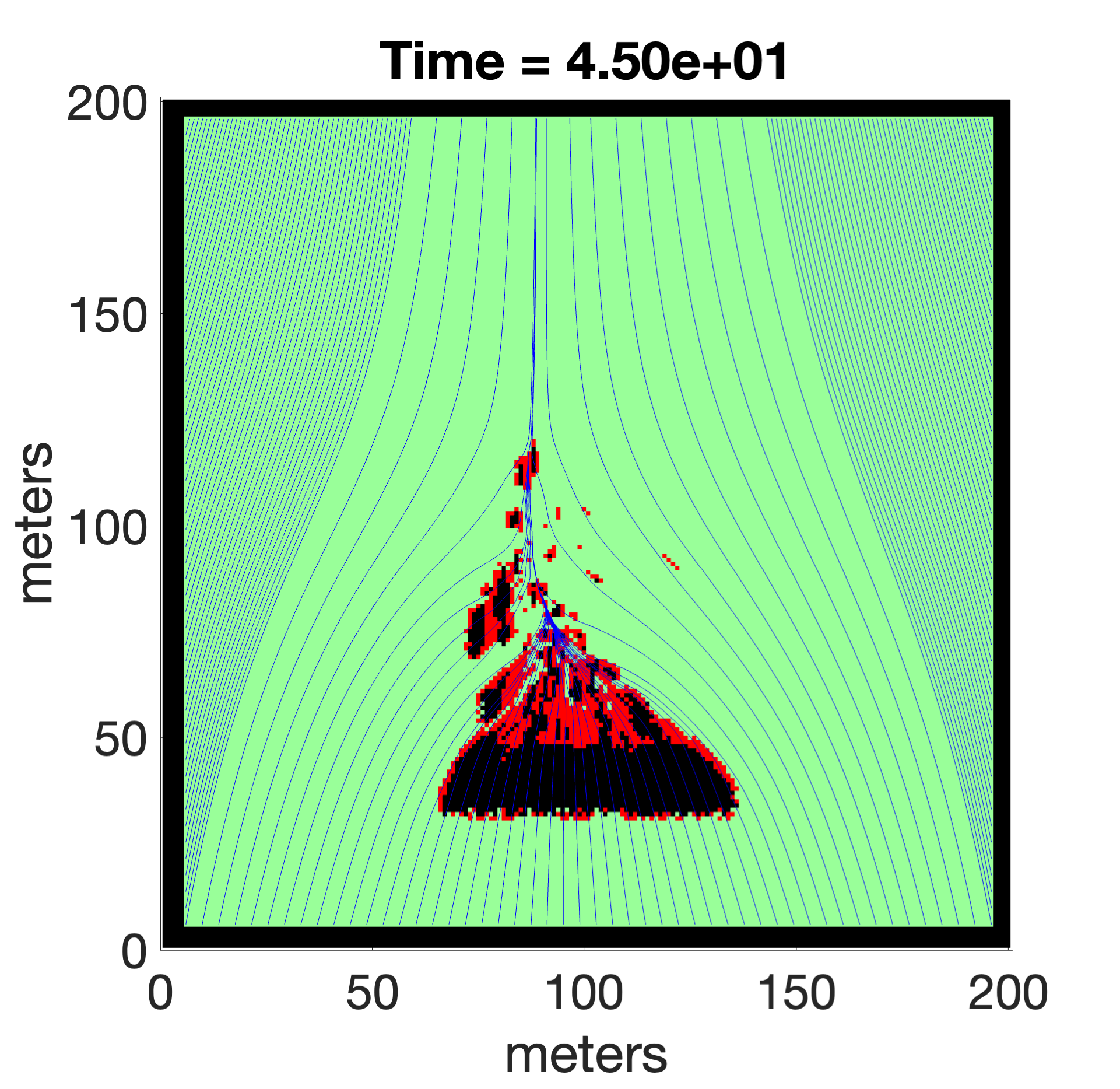}
  \includegraphics[width=0.243\textwidth,trim=0.5cm 0.5cm 1.5cm
  1.0cm, clip=true]{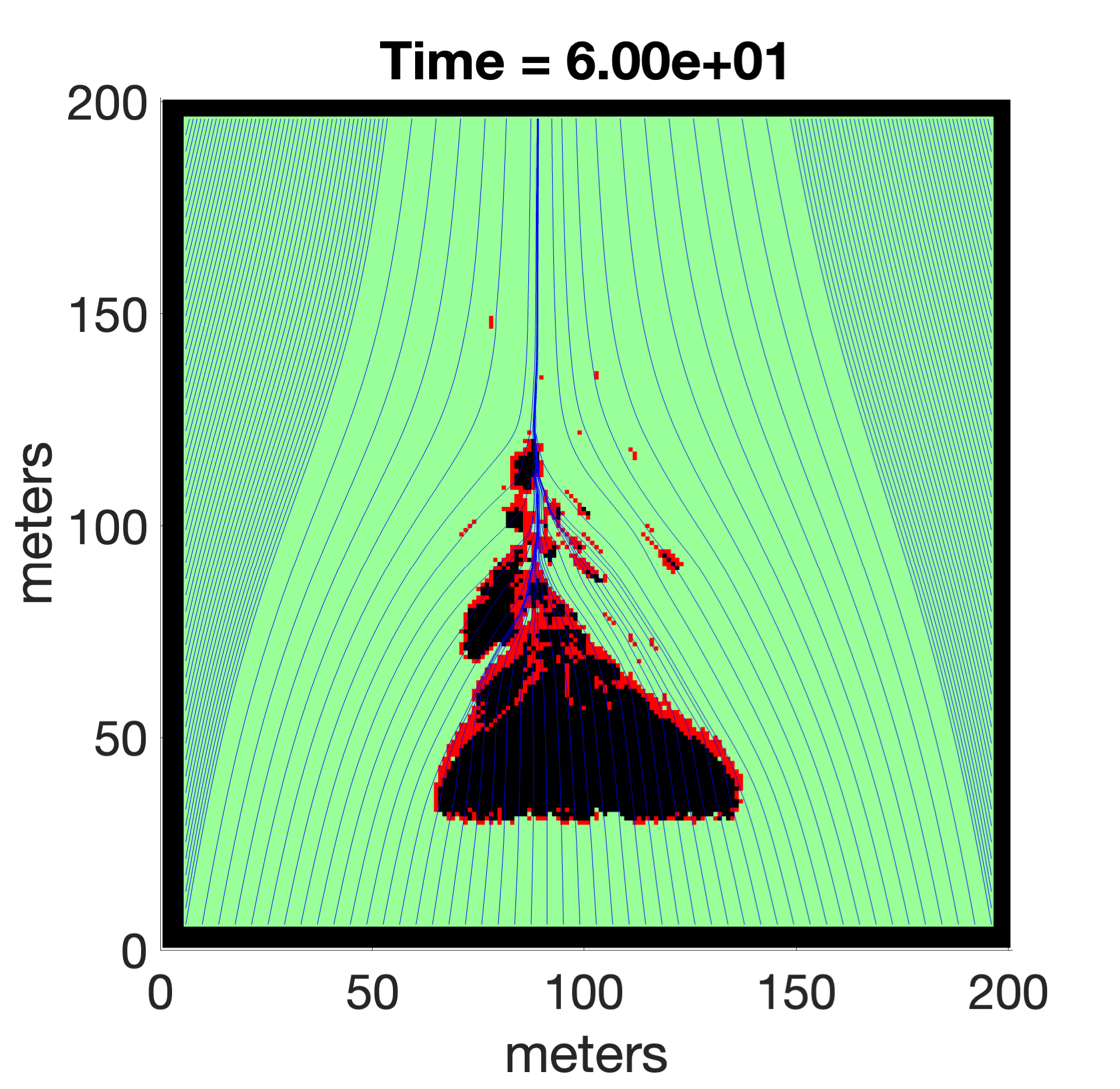}
  \\
  \includegraphics[width=0.243\textwidth,trim=0.5cm 0.5cm 1.5cm
  1.0cm, clip=true]{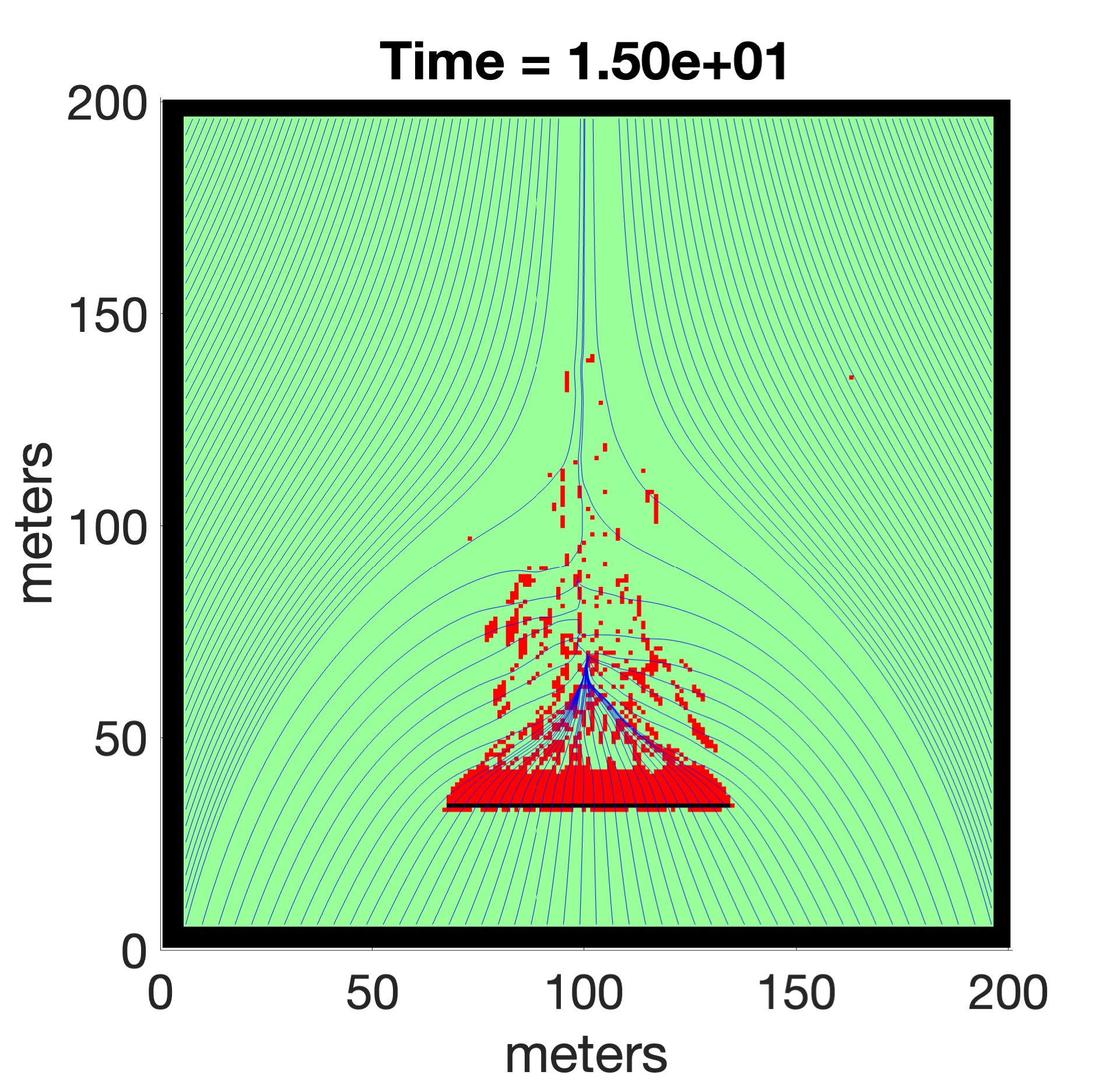}
  \includegraphics[width=0.243\textwidth,trim=0.5cm 0.5cm 1.5cm
  1.0cm, clip=true]{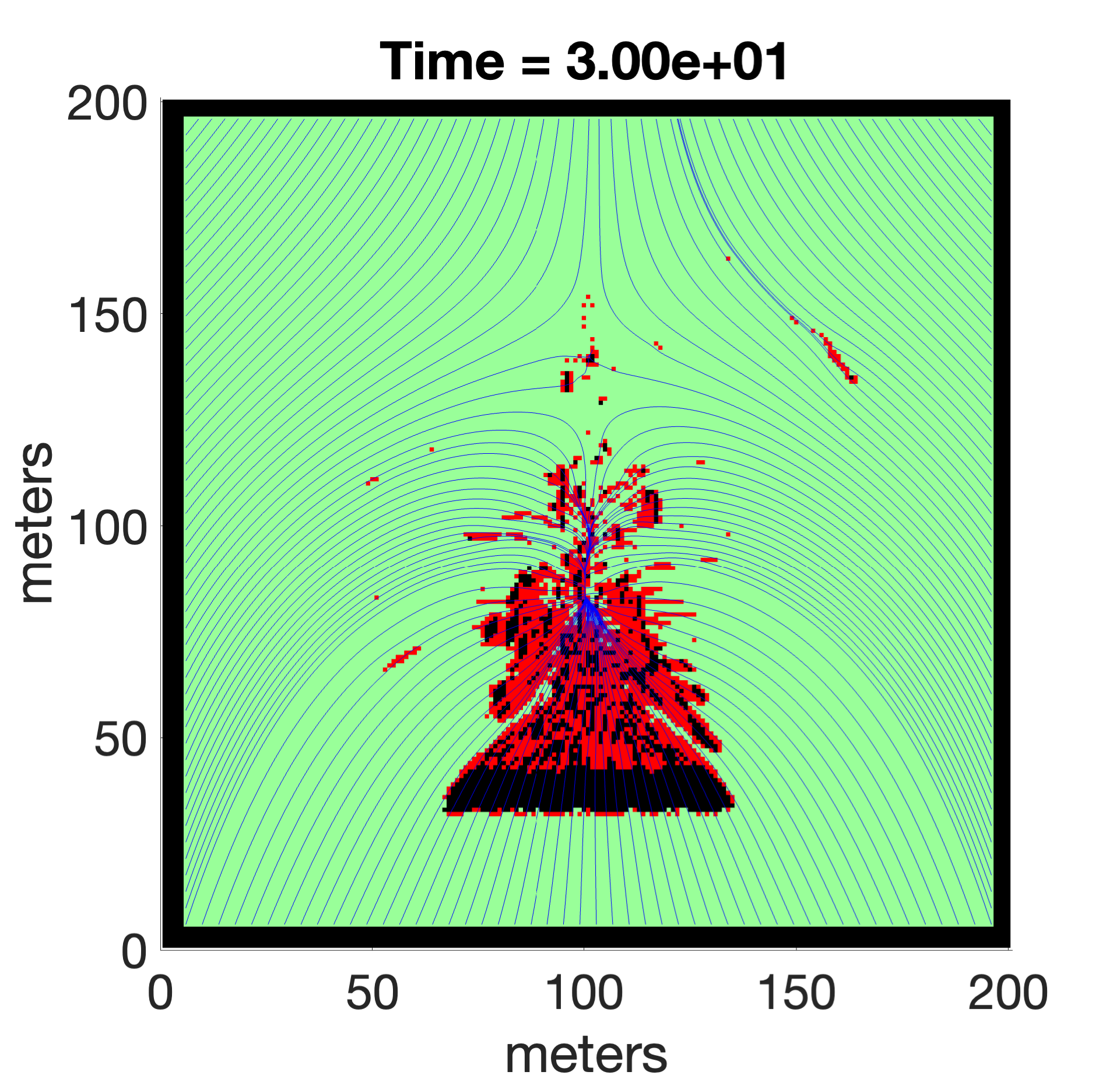}
  \includegraphics[width=0.243\textwidth,trim=0.5cm 0.5cm 1.5cm
  1.0cm, clip=true]{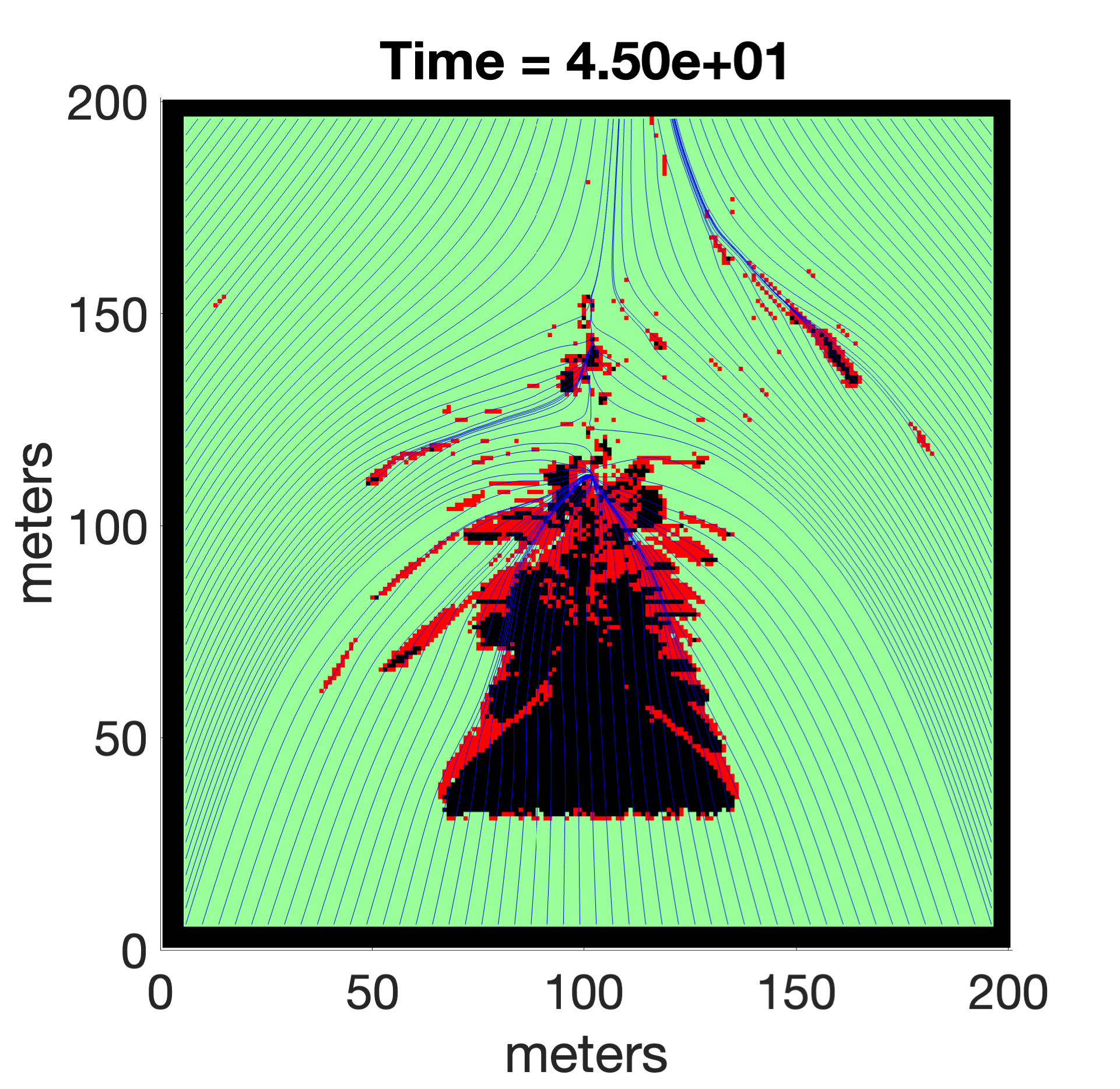}
  \includegraphics[width=0.243\textwidth,trim=0.5cm 0.5cm 1.5cm
  1.0cm, clip=true]{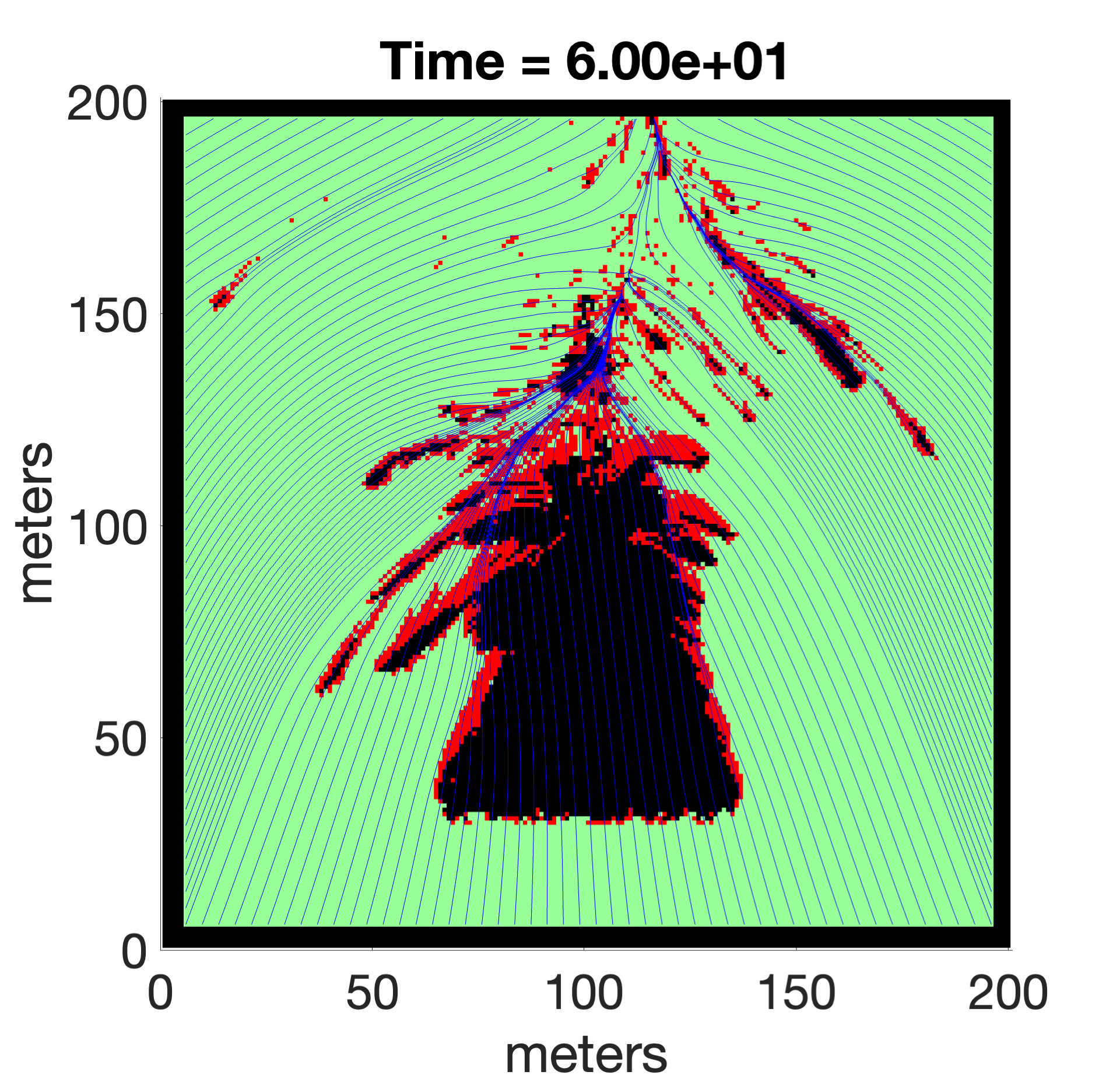}
  \caption{\label{fig:visualsegment1} \em An idealized example of fire
  spread due to ember wash. The ember flight time is exponentially
  distributed with a mean of $\mu = 20~s$. The probability of ignition
  is $\pig = 0.1$ (top) and $\pig = 0.5$ (bottom). Streamlines of the
  wind (solid blue curves) indicate the interaction of a constant
  background wind of 4~m/s from the south with the fire-induced wind.}
\end{figure}

Using our simplified model~\citep{qua-spe2021}, we vary the parameters
$\mu$ and $\pig$ to characterize the effects of mean ember flight time
and probability of ignition. For the flight time, we use exponential
distributions with means 5~s $\leq \mu \leq$ 50~s, and use probability
of ignition values of $\pig$ = 0.1 and 0.5. We considered intermediate
probability of ignition values and find that the rate of area growth is
smooth with respect to $\pig$. Figure~\ref{fig:visualsegment1} shows
simulation results from the simplified model coupled with the ember wash
model. The background wind is 4~m/s, and the ember flight time is
exponentially distributed with a mean value of $\mu = 20~s$. The
probability of ignition is $\pig = 0.1$ (top) and $\pig = 0.5$ (bottom).
The solid blue lines are streamlines of the wind which affect the
trajectory of embers. Wind is the sum of both background wind and
fire-generated wind, thus, as the embers create fires ahead of the main
fire front the wind field evolves. More specifically, as the fire
evolves, the ember flight time distribution is fixed, but the surface
velocity changes, thereby changing the flight distance of the embers.

For large values of $\pig$, enough new fire is generated to form
stagnation points downwind of the main fire. These regions can increase
local ember residence time and concentrate subsequent ignitions, creating a
positive feedback between fire geometry and the near-surface flow. As
expected, the presence of embers, especially for large probabilities of
ignition, results in a faster spread. Similarly, the flight distance
distribution is affected by the presence of different fuels, landscapes,
and the presence of structures. Section~\ref{sec:WUI} provides an example
that an array of simple rectangular structures has on the ember
distributions.

\begin{figure}[htp]
  \centering
  \includegraphics[width=0.243\textwidth,trim=0.5cm 0.5cm 1.5cm
  1.0cm, clip=true]{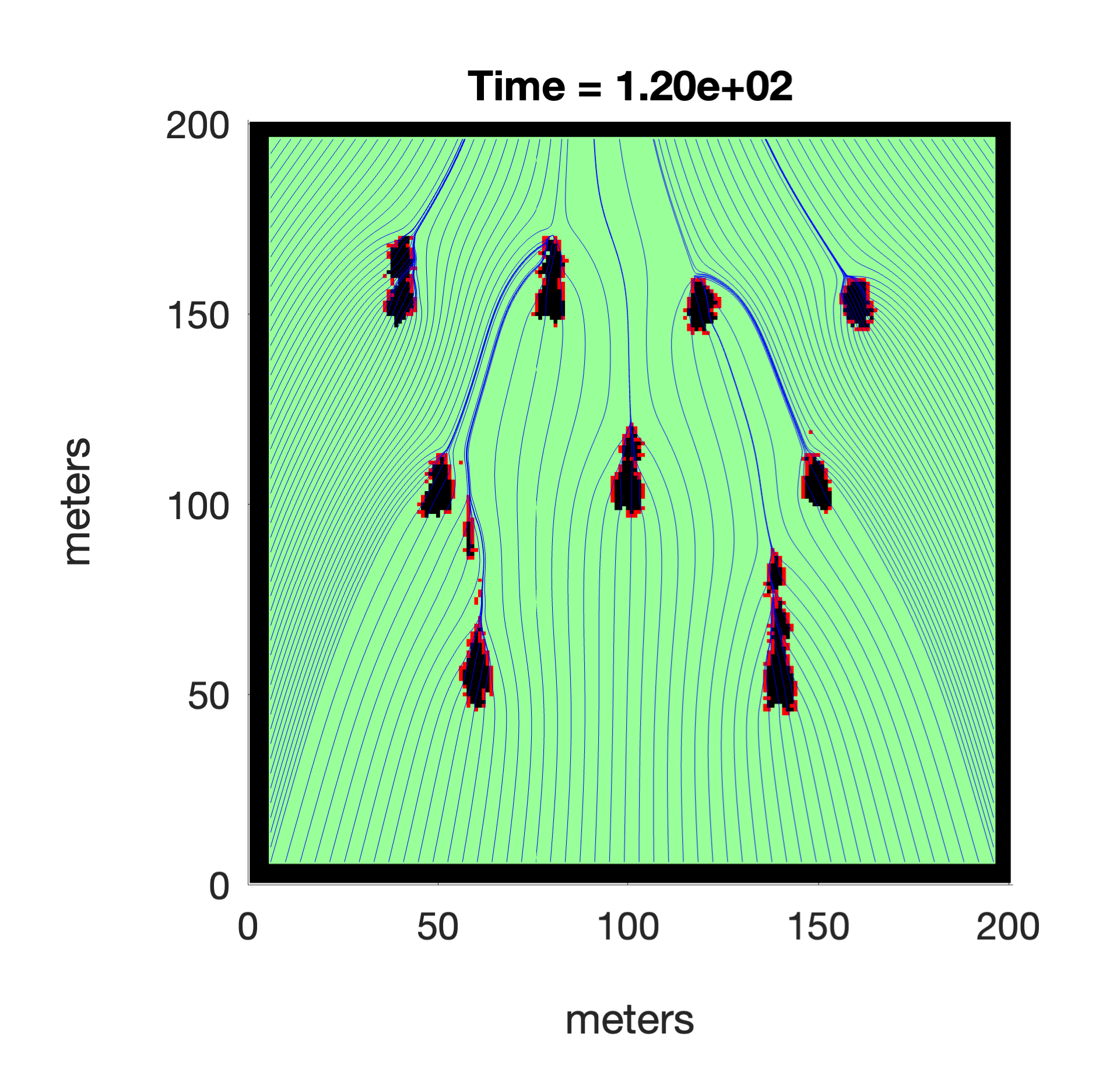}
  \includegraphics[width=0.243\textwidth,trim=0.5cm 0.5cm 1.5cm
  1.0cm, clip=true]{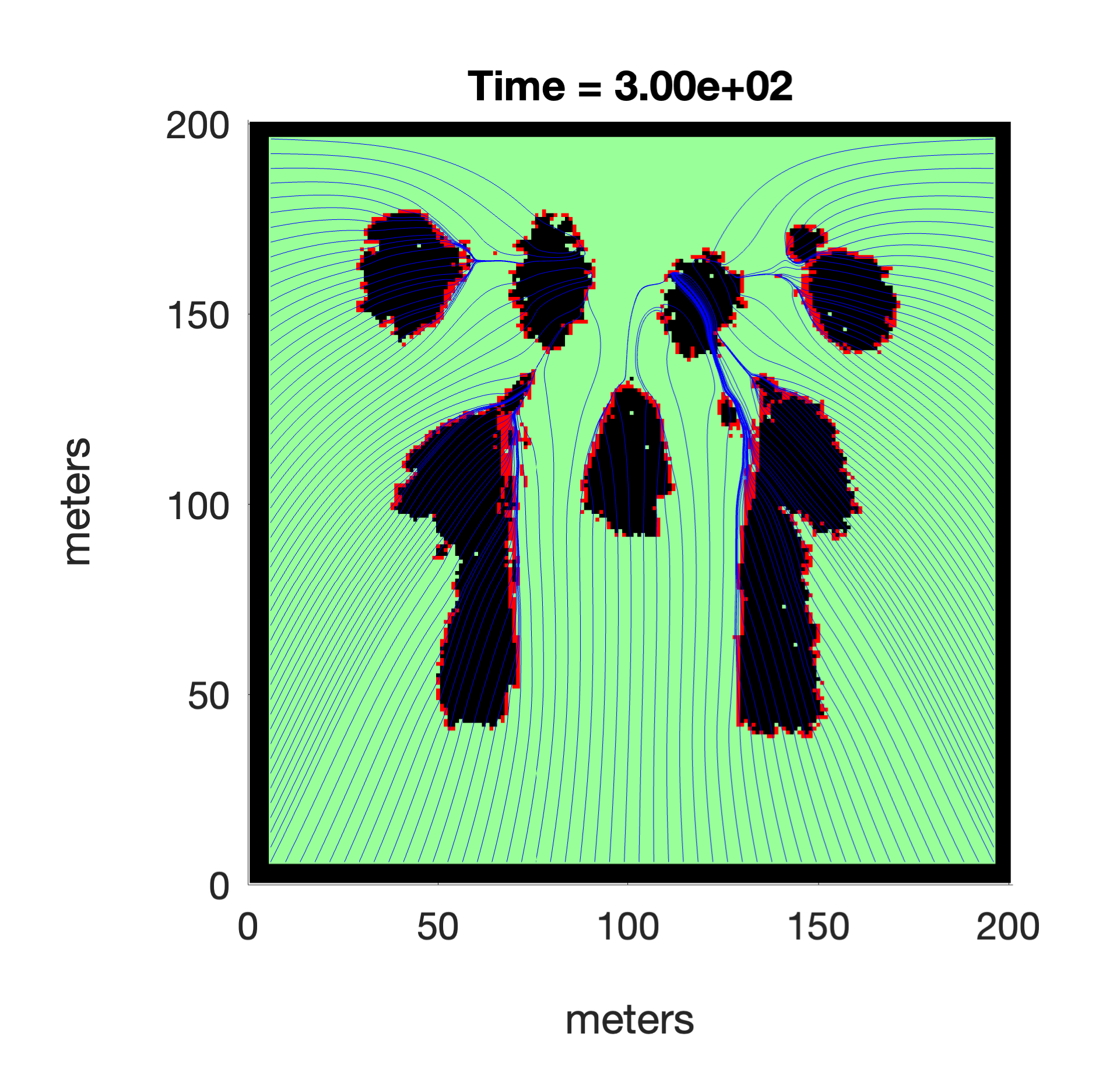}
  \includegraphics[width=0.243\textwidth,trim=0.5cm 0.5cm 1.5cm
  1.0cm, clip=true]{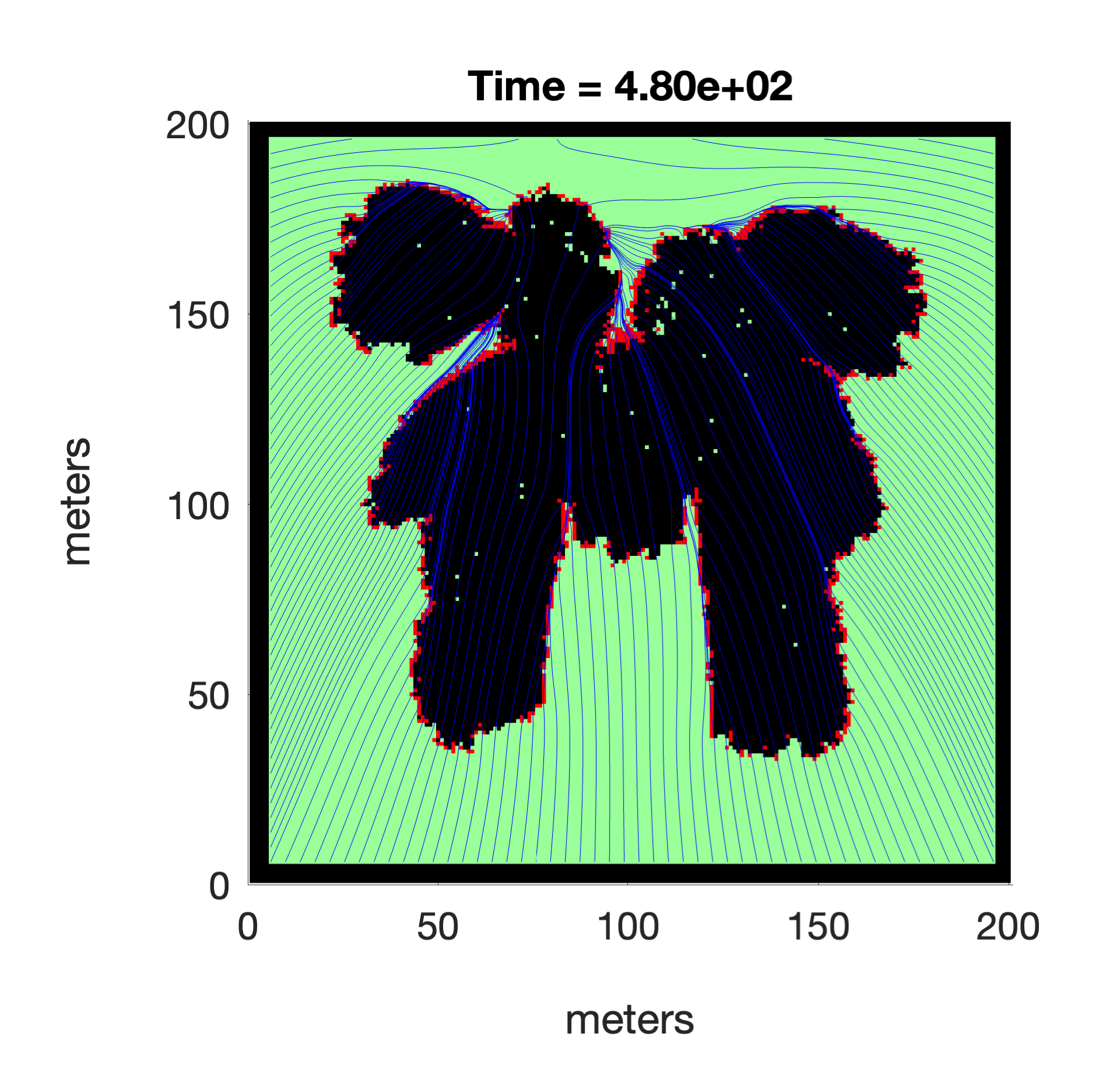}
  \includegraphics[width=0.243\textwidth,trim=0.5cm 0.5cm 1.5cm
  1.0cm, clip=true]{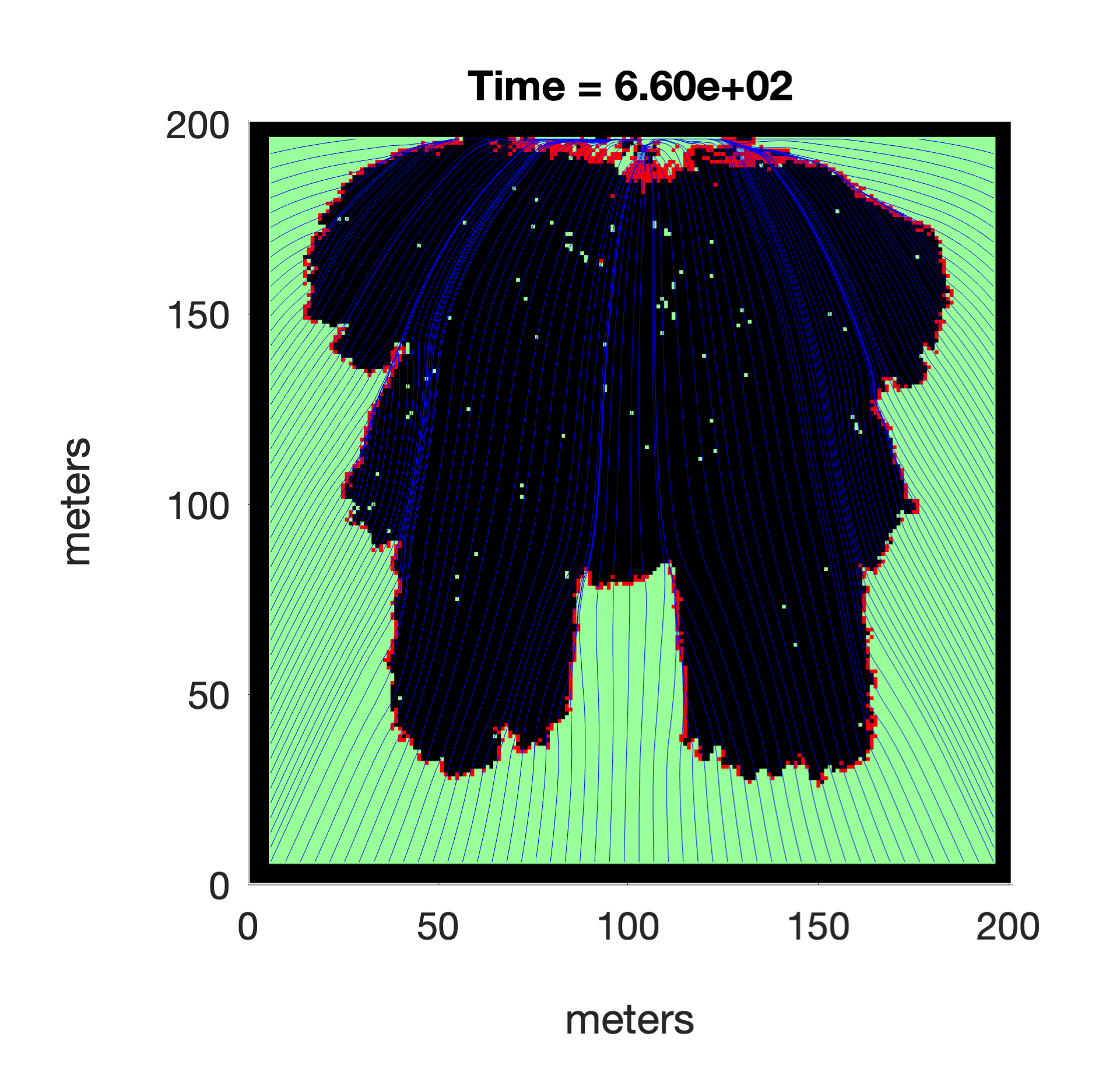}
  \includegraphics[width=0.243\textwidth,trim=0.5cm 0.5cm 1.5cm
  1.0cm, clip=true]{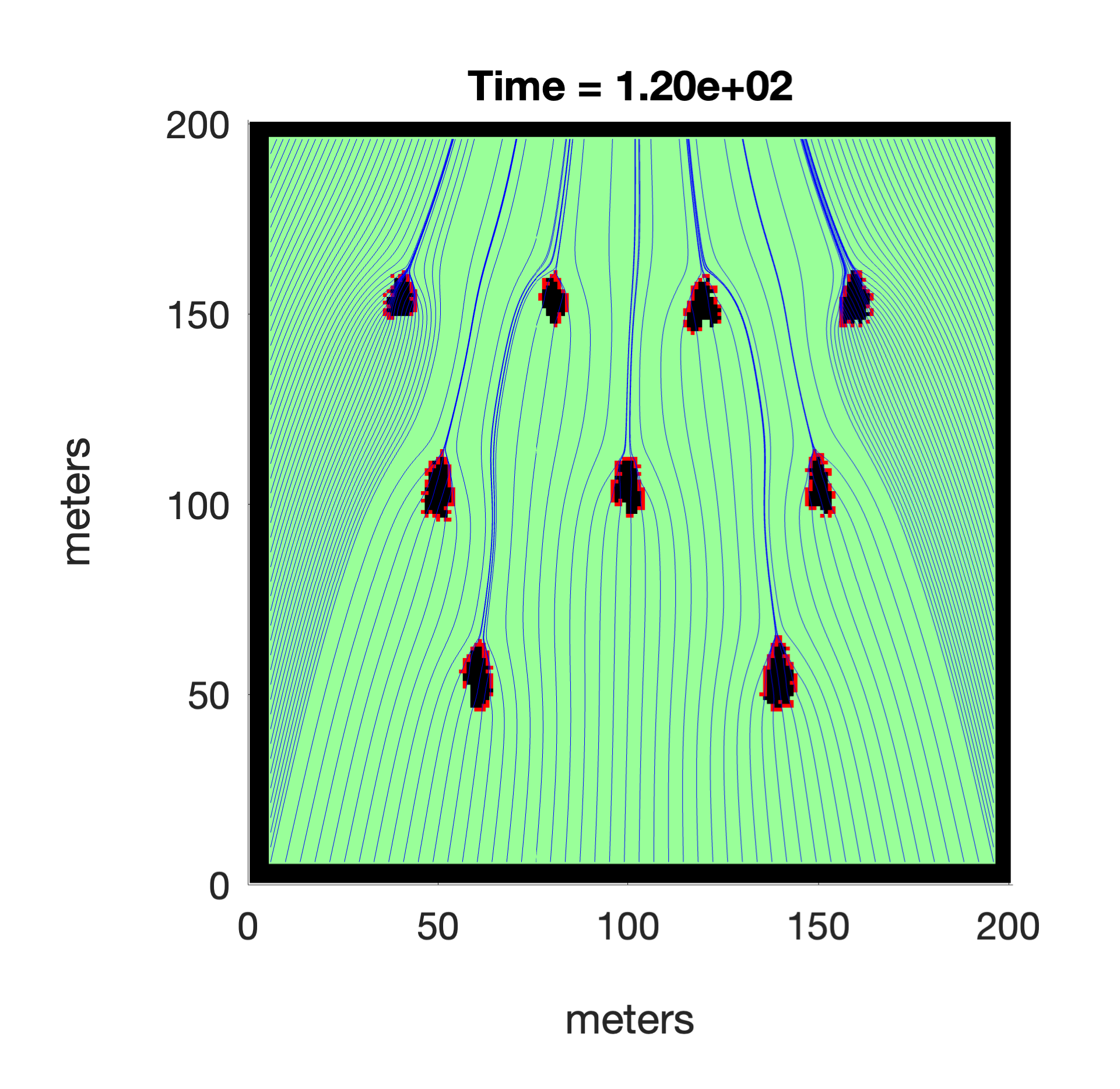}
  \includegraphics[width=0.243\textwidth,trim=0.5cm 0.5cm 1.5cm
  1.0cm, clip=true]{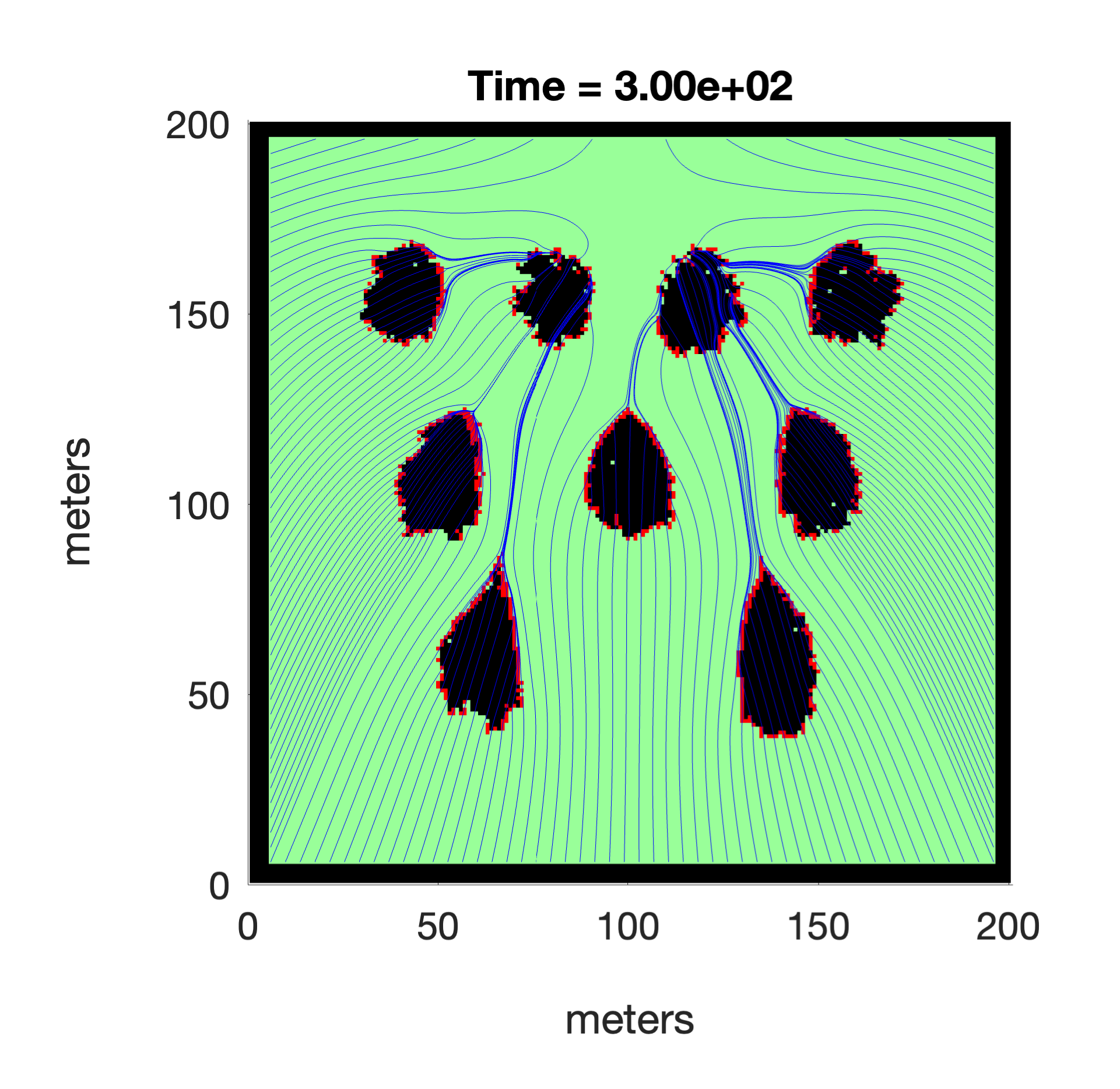}
  \includegraphics[width=0.243\textwidth,trim=0.5cm 0.5cm 1.5cm
  1.0cm, clip=true]{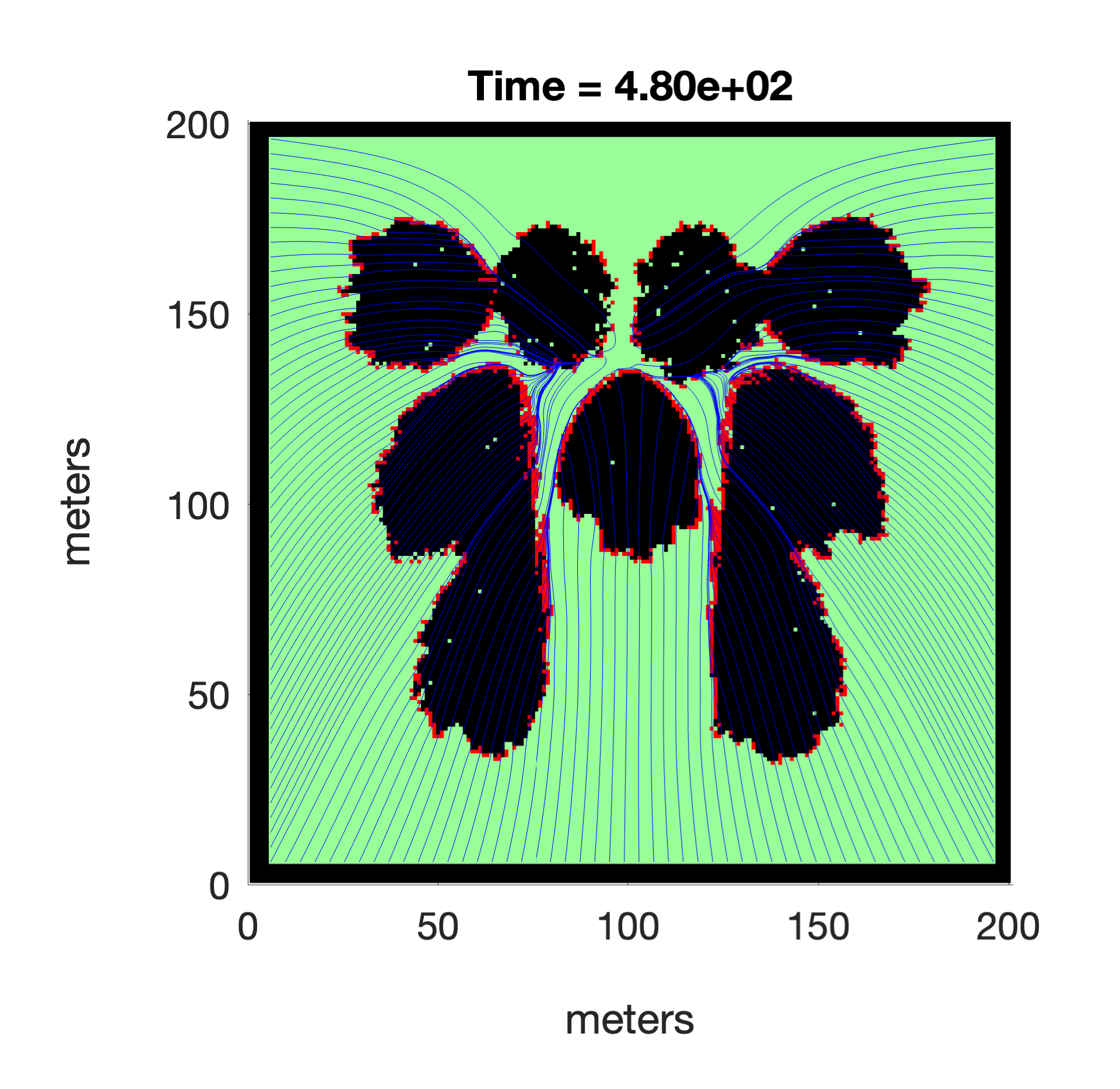}
  \includegraphics[width=0.243\textwidth,trim=0.5cm 0.5cm 1.5cm
  1.0cm, clip=true]{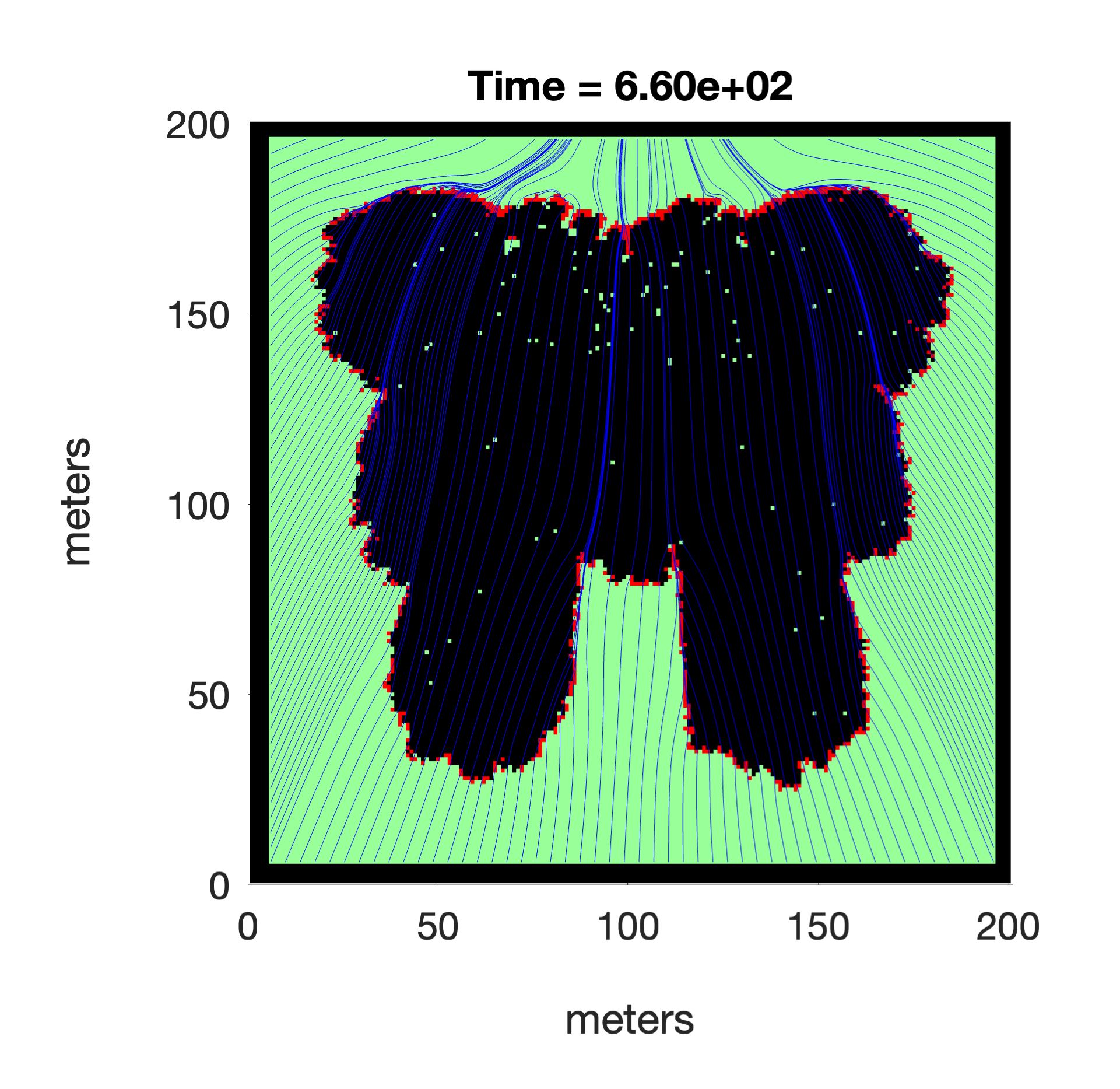}
\caption{\label{fig:Wash} \em An idealized example of fire spread due to
ember wash (top) and without ember-driven spread (bottom). The ember
flight time is exponentially distributed with a mean of $\mu = 20~s$ and
the probability of ignition is $\pig = 0.5$. The ignition pattern is
nine isolated spots. With ember wash, a large region of ignition appears
following the direction of the coupled fire-induced wind field.
Streamlines of the wind (solid curves) indicate the interaction of a
constant background wind from the south at 1~m/s with the fire-induced
wind. Note that the ember velocity is controlled by the surface wind which includes both the background and fire-induced velocities.} 
\end{figure}

At sufficiently strong ember wash, the model no longer produces only
isolated downwind ignitions. Instead, repeated near-surface transport
and ignition broaden the active burning region, so that gaps between
burning patches are progressively filled. A simulation designed to
illustrate conditions near an area of multiple spot fires takes the
ember wash effect on spread to a more extreme behavior
(Figure~\ref{fig:Wash}). Wind streamlines show the very different effect
of fire-atmosphere interactions in the case of an ember wash that
produces a much stronger perturbation to the ambient wind due to the
larger area on fire at any given time compared to a simpler, roughly
continuous, front with a relatively narrow width.

The combination of ember wash and fire-induced wind generates a spread geometry not easily described by a simple distribution in distance from the main front, but rather a ``mass ignition" process filling in the areas of unburned fuel. Mass ignition is a well known example of extreme fire behavior. In cases with strong inflow, the local wind can overwhelm the background flow and reverse the direction of influence, thus eliminating ember effects from that source. But the complex evolution of the fire, even in the simplified 2D model setting, produces openings, breakthrough jets, and
wind pulses which can carry embers away from the main front. Moreover,
if background vorticity is included, intense fingering due to vorticity
dynamics together with ember wash can also enhance total spread rates in
unexpected directions~\citep{qua-spe2021, cla-jen-coe-pac1996a}.

\subsection{Simulated Growth Regimes}
\label{sec:growth}

Finally we are able to return to the observed fire area growth rates and
apparent regimes of growth. Lateral fire spread emerges from the coupled
interaction between fire and background winds, stochastic ember
transport and ignitions of new fire which, in turn, modify the wind. The
fire growth regimes discussed here have been designed to illustrate
coupled wind, fire, and ember transport effects. They correspond to
relatively simple background flow conditions and fuel distributions, but
provide benchmark cases for interpreting growth rates.

\begin{figure}[htp]
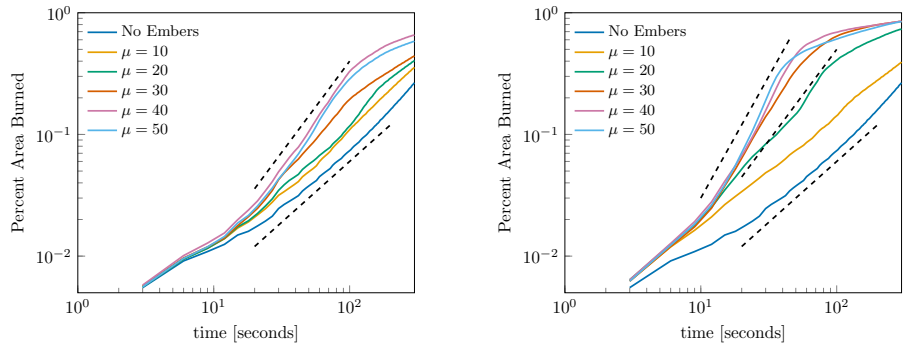

  \begin{center}
  \begin{tabular}{cc}
    \scalebox{0.65}{\input{area_vs_time_exponential_PIG0p1.tikz}} &
    \scalebox{0.65}{\input{area_vs_time_exponential_PIG0p5.tikz}}
  \end{tabular}
  \end{center}
  \caption{\label{fig:exponential1} \em The percentage of the available
  fuels that are actively burning or have been burnt. The probability of
  ignition is $\pig = 0.1$ (left) and $\pig = 0.5$ (right). The flight
  time of the embers follows an exponential distribution with the mean
  value displayed in the legends. For reference, slopes 1 and 1.5
  (left), and 1, 1.5, and 2 (right) are shown (dashed lines).}
\end{figure}

Figure~\ref{fig:exponential1} shows percent area burned of a line
ignition parallel to the wind direction with $\pig = 0.1$ (left) and
$\pig = 0.5$ (right). Each colored line represents a different a mean
flight time for embers, and snapshots of the simulations for $\mu =
20$~s are shown in Figure~\ref{fig:visualsegment1}. The dashed black
lines of slopes 1, 1.5, and 2, can be used to show how the rate of area
growth depends on $\mu$ and $\pig$. For the smaller probability of
ignition (Figure~\ref{fig:exponential1}, left panel), we see a clear
transition in the asymptotic scaling of the percent area
burned---without embers, it grows linearly, and as $\mu$ increases, it
transitions to the scaling $A(t) \sim t^{1.5}$. With the larger
probability of ignition $\pig = 0.5$, the area growth covers all regimes
and scales all the way to $A(t) \sim t^{2}$.

These results emphasize the importance of feedbacks between ember combustion and flight properties, fuels and probability of ignition, wind and transport. The complex geometry of fire spread emerges from this interaction, leading to growth rate scaling behavior similar to a large body of data from real wildland fires. This suggests that the first-order effects are captured by the idealized model, although more realistic landscape scale models are required to represent the broader range of dynamics and fire behavior with the interaction of the atmosphere with topography.

To bring this Chapter to a close we return to the high impact of wildland fires in the Wildland-Urban Interface or WUI. Again, a physics-based context helps to understand the role of the near-surface wind and the influence of structures on the flow patterns and speed, with areas of concentration and stagnation modifying the transport of embers.

\section{Embers in the Wildland Urban Interface}
\label{sec:WUI}
Previous sections emphasized some physical and statistical core models
for spread rate and fire area growth, motivated by the notion that
growth rate is a dominant predictor of spotting and related extreme
behavior. Hence, ember transport and spotting models must get the basic
growth right, or must pull observational data for fire area
distribution, to be able to represent or parameterize spotting in a
simplified, possibly operational, context. Consequently, ember-driven
fires and spread rate are tightly linked. Following that discussion,
idealized modeling for both the fire front spread and ember-driven
ignitions were provided. A very brief view of the impact of idealized
structural elements on flow dynamics and ember transport is given next,
and we end with a discussion of key concepts and future directions.

The presence of structural elements near wildland fire constitutes the
wildland-urban interface or WUI, and as has been noted earlier, bears
the brunt of ember threats and societal impact from wildland fires, and
much of the attention from public and private fire suppression forces as
well as insurance risk analyses. Despite our overall emphasis on ember
modeling in the wildland environment, it is important to note that the
same dynamical considerations come into play in the WUI. At great
distance, the WUI is just another fuel bed to catch fire, but as the
fire approaches and more structures are ignited a whole new fire
dynamics emerges with a modified boundary layer, complex surface flow
patterns, and long-lived intense fires. State-of-the-art combined
wildland and WUI fire spread modeling is just beginning to account for
embers. For example, \citet{purnomo2024reconstructing} used downwind
lognormal and lateral normal landing distributions to represent ember
landing. In addition to the flight and landing processes, a variety of
ember production mechanisms from building construction materials, and,
not to be neglected, fuels and materials from the interior of the
structures, for instance furniture, paper, etc., create new, and
difficult to predict, ember sources. 

\begin{figure}
  \centering
  \includegraphics[width=0.5\linewidth]{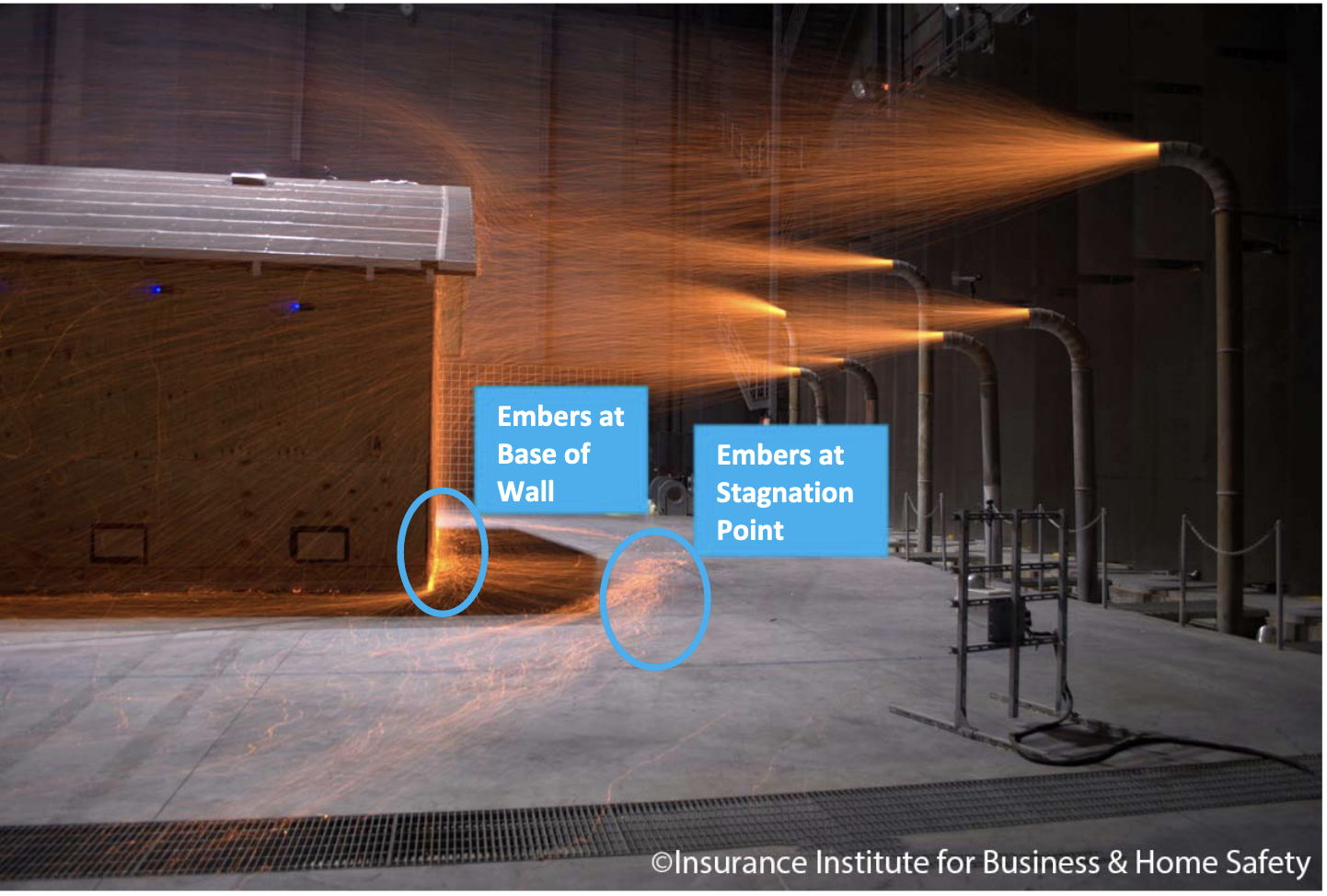}
  \caption{\label{fig:IBHS} \em Ember generators inject burning embers
  at several heights in a large-scale wind tunnel experiment to
  illustrate effect of flow on ember concentration patterns. Multiple
  stagnation points occur in flow around a simple cube-like structure. Credit: Insurance Institute for Business and Home Safety, used by permission.}
\end{figure}

We continue with a brief mention and acknowledgment of the insight
provided by extensive experimental work on ember transport near
structures by the Fire Safety Research Institute (FSRI),National Fire
Research Lab at NIST, Institute for Business and Home Safety (IBHS) and
other groups and agencies (Figure~\ref{fig:IBHS}). Similar to the
results of long-standing research on the atmospheric boundary layer and
canopies, structures modify the wind field, create drag and wakes that
guide fire spread, ember transport, and smoke concentrations.

The notion of flow patterns, wakes, and stagnation points around
structural elements extends to full 3D simulations of ember transport
and landing. Landing distributions and ember spotting risk in built
environments is the subject of much new research with sophisticated 3D
CFD models and both ember flight dynamics and combustion. In these
models the evolving embers influence the air flow around them as they
are advected by the turbulent boundary layer flow. Moreover, in 3D
simulations, the building structures themselves have an effect of flow
above the ground that can modify the dispersion characteristics of
embers.

\begin{figure}[htp]
\centering
  \includegraphics[width=1.0\textwidth]{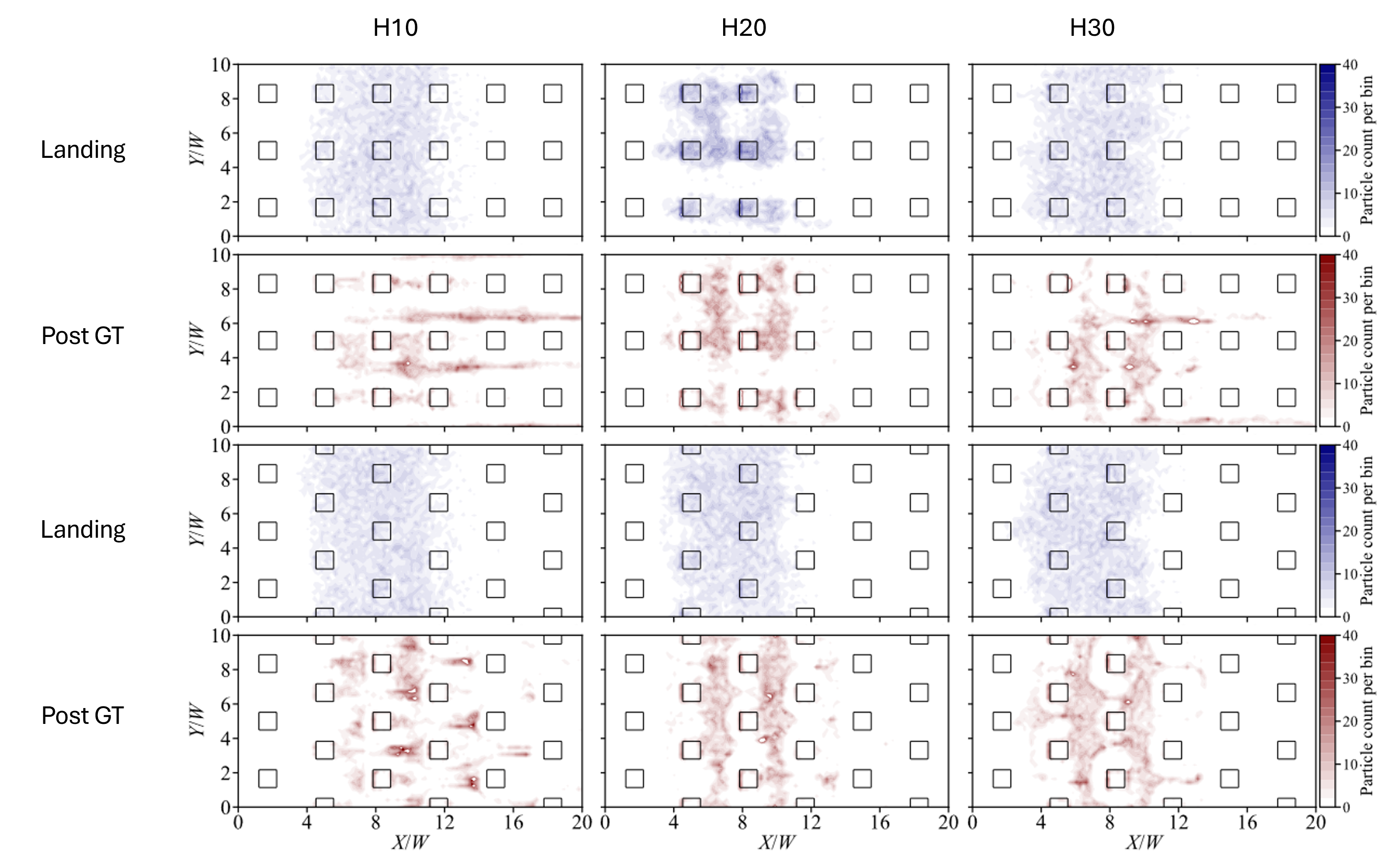}
  \caption{\label{fig:NedaModel} \em Surface view of ember landing
  distribution in a 3D simulation of ember transport in a WUI scenario.
  Embers are represented by combusting particles, with their mass, size,
  and temperature changing during flight. Shown is their landing
  distribution (Landing) and subsequent rolling motion on the ground due to the interaction with the near-surface flow Post GT). Two building configurations (inline and staggered) and three different building heights (10~m, 20~m, and 30~m) are modeled.} 
\end{figure}

We illustrate these effects with results from a recent 3D simulation
study with unsteady turbulent flow field over a range of idealized
structures to quantify its influence on ember transport, landing
patterns, and post-landing dispersal~\citep{eva-dos-yag2025a,
eva-dos-yag2025b}. The computational domain of this simulation consists
of canonical urban-like arrays of surface-mounted cubic obstacles
representing idealized built environments. The example displayed in
Figure~\ref{fig:NedaModel} shows two arrangements, inline and staggered,
at low packing densities, with uniform building heights of 10~m and
30~m, where the width $W$ is 20~m, the characteristic building width.
Embers are modeled as spherical particles with diameters ranging from
2--5~mm and are released from a horizontal source plane covering
approximately one-third of the domain area. Ember releases plane occur
at a height of 58~m, to simulate a source within a near-field plume or
lofting region, with a specified atmospheric wind condition.

The inflow wind condition corresponds to a neutrally stratified boundary
layer with a bulk wind speed of approximately 20~m/s, producing a
high-Reynolds-number flow regime characteristic of strong wind-driven
fire scenarios. Each trajectory of ember positions is obtained via a
particle-tracking model that calculates instantaneous positions and
velocity of individual particles by solving the conservation of linear
momentum in a Lagrangian frame of reference. More results from the model
are described in~\citet{dos-yag2023}. 

These simulations identify a number of features related to ember
transport, and bring many of the the processes described above into the WUI context. \citet{dos-yag2023} found that ember dispersion in the
presence of buildings produced a shorter flight time and more complex
settling patterns. With structures, embers pile up at stagnation points
(Figure~\ref{fig:NedaModel}) and create a higher risk of ignitions,
similar to the wind tunnel observations (Figure~\ref{fig:IBHS}). Post-landing distributions depend on both the arrangement and the height of the structures. Clearly, the risk associated with ember fallout and surface spread in the WUI varies with the geometry of the structures as well as the layout of the developed area. This suggests that some structures may be at a much greater risk than others, and also that some degree of control of risk may be possible by arranging or re-arranging urban development when ember transport taken into account during the planning, or post-fire recovery, stages.

%
%

\section{Discussion}
\label{sec:discussion}

Observational analyses of spot fires show diverse patterns of near and
long distance ignition, and using a variety of models we find that
localized but highly intermittent modes of transport can create spot
fires, fingers, but also widen the fire front without producing clearly
separated spot fires. Ember transport near the surface, with repeated
deposition and re-suspension, plays a large role in fire spread, yet it
is not a component of standard fire spread models. Models with surface
ember transport show that an extended fire width produced by residual
burning or by ember ignitions is associated with more heat and
fire-generated wind, inducing complex time-dependent fire behavior, even
with highly simplified dynamics \citep{qua-spe2021}. The ember wash
effect produces larger burning area hence greater fire-induced component
of wind, as well as broader fires hence an area-dependent fire spread.

With all the complexities of ember transport, the last factor in the
Chain of Models~\eqref{eqn:Chain}, probability of ignition, is also one
of the most difficult to determine. In our idealized model of ember
spread the probability of ignition along the Bresenham ignition line
plays several roles. One of the roles is of ember flight and combustion
lifetime; not all embers in motion are burning, and not all that are
burning will ignite the fuel bed. Another role is the ignition
probability of the fuel bed itself, since embers may gather on
nonflammable surfaces or other locations with no ignition, as well as on
fuels easily ignited. A third is the time the ember spends at rest
before resuspension, burning and releasing heat all the while. All of
these parts have been examined in different studies and contexts, and
need to be considered in a model of fire spread. The nonlinear nature of
the interaction with local winds makes the fire spread sensitive to this
parameter.


Fire spread in highly constrained experimental pine straw fuel bed
conditions without ember effects suggests an exponential distribution of
fine-scale spread rate or front
displacements~\citep{sag-spe-pok-qua2021}. Here, the random nature of
turbulence in the wind, though weak compared to larger prescribed fires
and wildland fires, both at the fuel surface, plume and within the fuel
bed, is at play in the convection of the hot gases and associated
diffusive-like spread of fire close to the
front~\citep{beb-oli-qua-sko-hei-spe2020}.

We expect a fundamental role for boundary layer turbulence in the ember
wash phenomenon with complex coupled interactions between surface shear
layers, buoyancy, and roughness elements. We have left out a discussion
of the role of canopies in fire spread and ember transport, a clearly
crucial component of boundary layer structure and ember dynamics in the
near-surface regime~\cite{ban2026}. Similarly, topography, neglected
here, plays a role both locally and for the more general atmospheric
flow near the surface, focusing flow and engendering various dynamical
effects (see for example~\citet{coen2018generation}).

A Bayesian framework for rate of spread has been
developed~\citep{sto-bed-pri-bra-sha2021} which could provide a model
for similar approach to ember driven fire spread. We note that many of
the processes at work with ember transport consists of multiplicative
terms similar to surface erosion processes
that may be interpreted as being sampled from an appropriate
distribution, forming a product of random variables. For example, wind
velocity components are typically Gaussian while the wind speed
distribution is often assumed to be Weibull~\citep{drobinski2015surface}
or Rayleigh distributed, with higher order tendencies under intermittent
processes associated with turbulence~\citep{monahan2011probability}.
Size distributions of firebrands from trees appear to be exponentially
distributed~\citep{adu-blu2021}.

The product of estimated distributions for these terms together with
other factors produces results that appear to approach approximate
exponential decay for ember transport velocity. We conducted Monte-Carlo
simulations with simulated data that also suggests an exponential-like
form, primarily due to wind effects. Theoretical studies point to an
exponential distribution for sediment bed load velocities
\citep{furbish2013probabilistic} lending some support to this framework.
To this spatial structure, a simple Bernoulli law for probability of
ignition can be added as a starting point since not all embers actually
ignite the fuel. Such an elementary model cannot be expected to
encompass the entire range of possible fire spread outcomes, but does a
surprisingly good job of illustrating fundamental effects of coupled
fire-atmosphere transport processes.

We have emphasized the exponential and also more extreme, heavy-tailed
Pareto nature of some of the key processes at work generating ember
transport and fire spread. Do these processes lead naturally to
realistic longer range landing patterns and spotting or fire risk? Is
the appearance of lognormal spotting distributions a trivial result or
fundamental? It may be that this is a more fundamental consequence of
the underlying source-sink processes including diffusion~\cite{and2021}.
It remains open whether there is enough flexibility in the lognormal
framework to describe the different regimes of fire spread.

Finally, we return to the underlying component of all fires, that is,
fuel. In the same weather conditions, a practitioner of prescribed fire
approaches a short understory grassy fuel bed very differently from a
multi-year heavy fuel load in a location that has not seen fire for many
years. An open grassland fire exhibits very different behavior depending
on the height of the vegetation. These differences reflect real physical
principles acting in a coupled fire-atmosphere system that naturally
will have different statistical behaviors as well. Incorporating these
into operational models will help to anticipate extreme behavior.
Ultimately we cannot neglect the importance of relaying the results of
models and observations in a predictive, operational context to the
people that need the information~\citep{coen2024framework}.

The possible range of turbulent surface and boundary layer fire spread
dynamics and effects is vast and only a limited set of conditions is
explored here. The applicability of physically constrained stochastic
process models is promising. More judicious and quantitative comparisons with experimental results and wildland fire observations are needed to
determine suitable explicit model parameterizations of all modes of
ember transport.

\backmatter



\bmhead{Acknowledgments}

We would like to acknowledge the extensive effort by Craig Anderson for
his analysis of the historical fire perimeters from GEOMACS, Daryn Sagel
for her experimental work on fire spread in controlled conditions, and
the many contributions from colleagues investigating wildland fire
dynamics in a observational and modeling context. Partial support for
this work came from National Science Foundation grants AGS-2427321 and
AGS-2427323, and from the Department of Defense SERDP contract
RC20-1298.

This is a preprint of the following chapter: [Speer, Quaife, Sun], [Models of Ember Transport], published in [Wildland Fire], edited by K. Bhaganagar and C. Wikle, [2027], [Springer (as it appears on the cover of the book)]. It is the version of the author’s manuscript prior to acceptance for publication and has not undergone editorial and/or peer review on behalf of the Publisher (where applicable). The final  authenticated version is available online at: http://dx.doi.org/[insert DOI].




\bibliographystyle{plainnat}

\end{document}